\documentclass[a4paper,fleqn,usenatbib]{mnras}

\usepackage{newtxtext,newtxmath}

\usepackage[T1]{fontenc}
\usepackage{ae,aecompl}

\usepackage{graphicx}	
\usepackage{amsmath}	
\usepackage{amssymb}	
\usepackage{multicol}        
\usepackage{bm}		
\usepackage{pdflscape}	
\usepackage{csquotes}
\usepackage{subfig}
\usepackage{float}
\usepackage{nccmath}
\newcommand{\angstrom}{\mbox{\normalfont\AA}}

\title[]{ERQs are the BOSS of Quasar Samples: The Highest-Velocity [OIII] Quasar Outflows}

\author[]{
S. Perrotta,$^{1}$\thanks{E-mail: serenap@ucr.edu}
F. Hamann,$^{1}$
N. L. Zakamska,$^{2}$
R. M. Alexandroff,$^{3,4}$
D. Rupke,$^{5}$
\newauthor
D. Wylezalek$^{6}$
\\
\\
$^{1}$Department of Physics and Astronomy, University of California, 900 University Avenue, Riverside, CA 92521, USA\\
$^{2}$Department of Physics and Astronomy, Johns Hopkins University, Baltimore, MD 21218, USA\\
$^{3}$Canadian Institute for Theoretical Astrophysics, The University of Toronto, Toronto, ON M5S 3H8, Canada\\
$^{4}$Dunlap Institute for Astronomy and Astrophysics, The University of Toronto, Toronto, ON M5S 3H4, Canada\\
$^{5}$Department of Physics, Rhodes College, Memphis, TN 38112, USA\\
$^{6}$European Southern Observatory, Karl-Schwarzschildstr 2, D-85748 Garching bei M\"{u}nchen, Germany
}

\date{Accepted XXX. Received YYY; in original form ZZZ}

\pubyear{2019}

\begin{document}
\label{firstpage}
\pagerange{\pageref{firstpage}--\pageref{lastpage}}
\maketitle

\begin{abstract}

We investigate Extremely Red Quasars (ERQs), a remarkable population of heavily-reddened quasars at redshift $z$ $\sim$2 $-$ 3 that might  be caught during a short-lived ``blow-out'' phase of quasar/galaxy evolution. We perform a near-IR observational campaign using Keck/NIRSPEC, VLT/X-shooter and Gemini/GNIRS to measure rest-frame optical spectra of 28 ERQs with median infrared luminosity $\langle$log $L$(erg/s)$\rangle$ $\sim$46.2. They exhibit the broadest and most blue-shifted [O{\sevensize III}] $\lambda$4959,5007 emission lines ever reported, with widths ($w_{90}$) ranging between 2053 and 7227 km s$^{-1}$, and maximum outflow speeds (v$_{98}$) up to 6702 km s$^{-1}$.
ERQs on average have [O{\sevensize III}] outflows velocities about 3 times larger than those of luminosity-matched blue quasar samples. 
We show that the faster [O{\sevensize III}] outflows in ERQs are strongly correlated with their extreme red colors and {\it{not}} with radio-loudness, larger quasar luminosities, nor higher Eddington ratios. We estimate for these objects that at least 3$-$5 per cent of their bolometric luminosity is being converted into the kinetic power of the observed wind. Our results reveal that ERQs have the potential to strongly affect the evolution of host galaxies.

\end{abstract}




\section{Introduction}
\label{intro}

A key problem in galaxy formation and evolution is understanding how active galactic nuclei (AGN) interact with their host galaxies.
The potential impact of galaxy-scale outflows driven by quasars on their environment has become widely recognized (e.g., \citealp{Blandford2004, Scannapieco2004, Vernaleo2006, Kormendy2013}). These outflows provide a mechanism that might regulate and quench star formation activity in the host galaxy by dispersing or expelling the gas that feeds star formation and black hole growth.

Theoretical studies show that this so-called AGN feedback can provide an explanation for a variety of observations, e.g., the chemical enrichment of the intergalactic medium (IGM), the self-regulation of the growth of the supermassive black hole (SMBH) and of the galactic bulge, the steep slope of the high end of the stellar mass function, the existence of the red sequence of massive passive galaxies (e.g., \citealp{Silk1998, Granato2004, DiMatteo2005, Hopkins2010}). In some theoretical scenarios, quasar feedback occurs at a particular critical stage in galaxy evolution. It begins with a major merger or accretion event triggers a massive burst of star formation and rapid accretion onto the central SMBH. These starburst galaxies are shrouded in gas and dust and appear observationally as sub-mm galaxies or ultra-luminous infrared galaxies (e.g., \citealp{Sanders1988, Hopkins2005, Hopkins2008, Veilleux2009, Simpson2014}). AGN feedback occurs when energy and momentum liberated by the accreting SMBH couple to the surrounding interstellar medium (ISM) to produce a ``blowout'' of gas and dust that quenches the star formation, thus revealing a visibly luminous quasar in the galactic nucleus (e.g. \citealp{Sanders1988, DiMatteo2005, Hopkins2006, Hopkins2008, Hopkins2016, Rupke2011, Rupke2013, Liu2013}). 

Dust-reddened quasars are valuable to test this evolution picture because they are candidate young objects participating in this early dusty stage of massive galaxy formation. In particular, they might have more common or more powerful accretion-disk outflows that drive feedback/blowouts during the brief transition phase from dusty starburst to normal blue quasar (e.g., \citealp{Canalizo2001, Hopkins2005, Urrutia2008, Glikman2012, Glikman2015, Wu2014, Banerji2015, Assef2015}). 

It is at the peak epoch of quasar and star formation activity ($z \sim$ 2 $-$ 3) that AGN feedback should have had the greatest impact on massive galaxy evolution. Our team recently discovered a remarkable population of extremely red quasars (ERQs; \citealp{Ross2015, Hamann2017}) at redshifts $z \sim$ 2.0 to 3.4 in Data Release 12 (DR12) of the Baryon Oscillation Sky Survey (BOSS, \citealp{Dawson2013, Ross2012}) in the Sloan Digital Sky Survey-III (SDSS-III, \citealp{2011Eisenstein}). 
ERQs are defined simply by extreme red colors in the rest-frame ultraviolet (UV) to mid-IR, namely i$-$W3 $>$ 4.6 (AB) from SDSS and the Wide-field Infrared Survey Explorer (WISE; \citealp{Wright2010, Lang2016}). They have sky densities a few percent of luminous blue quasars consistent with a short obscured phase of quasar activity. X-ray observations of 11 ERQs showed high obscuration, with typical intervening column densities $N_H$ $\approx$ 10$^{24}$ cm$^{-2}$ \citep{Goulding2018}. 
However, the remarkable aspect of ERQs is the suite of exotic rest-frame UV spectral properties that accompany their red colors, including (1) exceptionally large broad emission-line equivalent widths, (2) peculiar ``wingless'' broad emission-line profiles with frequent highly blueshifted centroids, e.g., in C{\sevensize IV} $\lambda$1549, (3) unusual emission-line flux ratios such as N{\sevensize V} $\lambda$1240 $\gg$ C{\sevensize IV} $\lambda$1549 and in some cases N{\sevensize V} $\lambda$1240/Ly$\alpha$ $>$ 1, and (4) an unusually high incidence of outflows identified by broad absorption lines (BAL). 
Several studies of highly reddened quasars at $z \sim$ 2 \citep{Banerji2013, Banerji2015, Assef2015, Wu2012} find objects with generally normal UV line properties consistent with normal quasars behind a dust reddening screen. ERQs are fundamentally different.
They are heavily-reddened, but their exotic line properties also require extreme physical conditions that all could be linked to powerful outflows \citep{Hamann2017}.

Another remarkable property of ERQs is extremely broad and blueshifted [O{\sevensize III}] $\lambda$4959,5007 emission lines. This was first reported by \citet{Zakamska2016} from near-IR observations of four of the reddest ERQs in the \citealp{Ross2015} and \citealp{Hamann2017} samples. Their data include the broadest and most blueshifted [O{\sevensize III}] $\lambda$4959,5007 ever recorded, with full widths at half maximum (FWHMs) and blueshifted wings both reaching $\sim$5000 km s$^{-1}$. These features clearly identify high-speed outflows in ionized gas, consistent with the evidence for prodigious outflows found in the rest-frame UV data mentioned above. 

However, the [O{\sevensize III}] $\lambda$4959,5007 lines are important because broad and blueshifted [O{\sevensize III}] profiles (at speeds larger than expected from galaxy dynamics) are common tracers of outflows on large (galactic) scales. As forbidden transitions, the [O{\sevensize III}] emission lines arise from relatively low density ($n \lesssim 7 \times 10^5$ cm$^{-3}$) warm ($T \sim 10^{4}$ K) clouds \citep{Baskin2005}. In particular, [O{\sevensize III}] cannot be produced in the high-density subparsec scales of the AGN broad-line region (BLR) making it a good tracer of the kinematics in the narrow-line region (NLR) on parsecs to tens of kiloparsec scales (e.g. \citealp{Wampler1975, Boroson1985, Stockton1987}). Therefore, large velocity shifts in the [O{\sevensize III}] lines provide strong evidence for high-velocity outflows on galactic scales (e.g. \citealp{Spoon2009, Mullaney2013, Rupke2013, Veilleux2013, Zakamska2014, Harrison2014, Cresci2015, Brusa2015, Brusa2016, Carniani2015, Carniani2016}). 

In this paper, we present the results of new near-IR observations to determine if the extreme [O{\sevensize III}] kinematics discovered by \citet{Zakamska2016} are common in ERQs and/or are related to some particular property of the quasars such as their luminosities or accretion rates. To do that, we observed 24 more ERQs and ERQ-like quasars that span a wide range of reddenings and rest-frame UV line properties.

The paper is organized as follows: Section~\ref{obs} describes the sample, observations, and data reductions; Section~\ref{analysis} presents our measurements and analysis of the [O{\sevensize III}] kinematics; Section~\ref{results} presents our main results with  comparisons to other significant quasar samples; and Section~\ref{discussion} discusses the broader implications of our study. Our conclusions are summarized in Section~\ref{summary}. 
We adopt a $\Lambda$CDM cosmology throughout this manuscript, with $\Omega_M$ = 0.315, $\Omega_{\Lambda}$ = 0.685, and H$_0$ = 67.3 km s$^{-1}$ Mpc$^{-1}$ \citep{planck14}.

\begin{table*}
\caption[]{Properties of the targets presented in this work. $z_{best}$ is the best estimate of the emission redshift taking into account the centroid of H${\beta}$, H${\alpha}$ (when available), [O{\sevensize III}] and low-ion emission lines in the corresponding BOSS spectra. $\Delta$v(H${\beta}$) represents the H${\beta}$ velocity shift with respect to $z_{best}$ (negative values indicate blueshift and positive values indicate redshift). R is the resolving power. i magnitude and i$-$W3 color have been corrected for Galactic extinction. F$_{20cm}$ is the 20 cm radio flux from FIRST where no entry means the source was not covered by FIRST, 0.0 indicates a non-detection with 5$\sigma$ upper limit $\sim$1 mJy \citep{Becker1995, Helfand2015}, non-zero entries are measurements with SNR $>$ 3 as recorded in DR12Q.}
\begin{center}
\label{tab_data}
\begin{tabular}{ c c c c c c c c c c}
\hline
\hline

 Object Name  & $z_{best}$ & Reference$^a$ & $\Delta$v(H${\beta}$) & Instrument & Obs. Date & $\lambda$ coverage &  R  & i$-$W3 & F$_{20cm}$  \\
 &  & line & [km s$^{-1}$] & & & [$\mu$m] &  &  [mag] &  [mJy] \\

 \hline
 
J000610.67+121501.2  &  2.3198   & [O{\sevensize III}] & $-$1664& NIRSPEC & Aug 13 2017 &1.413 - 1.808 & 2000 & 8.0 & 0.0 \\
J001120.22+260109.2  &  2.2807   & H${\beta}$ & $-$ &NIRSPEC & Nov 10 2016 &1.413 - 1.808  & 2000 & 4.9 & $-$ \\       
J013413.22$-$023409.7  &  2.3834    & H${\beta}$ & $-$ &X-shooter & Oct 31 2016 & 0.3 - 2.48  & 5100 & 3.3 & 0.0 \\       
J020932.15+312202.7  &  2.3595   & H${\beta}$ & $-$ &NIRSPEC & Gen 22 2017 &1.413 - 1.808 & 2000 & 5.1 & $-$ \\        
J080547.66+454159.0  &  2.3147   & [O{\sevensize III}] & $-$127 &NIRSPEC & Gen 22 2017 &1.413 - 1.808  & 2000 & 6.3 & 0.0 \\       
J082618.04+565345.9  &  2.3347   & H${\beta}$ & $-$ &NIRSPEC & Feb 16 2017 &1.413 - 1.808  & 2000 & 4.6 & 0.0 \\        
J082653.42+054247.3$^{c}$  &  2.5767   & H${\beta}$ & $-$ &GNIRS & Feb 22 2015 & 0.9 - 2.5  & 1700 & 6.0 & 1.1 \\       
J083200.20+161500.3  &  2.4252   & H${\beta}$ & $-$ & NIRSPEC & Feb 16 2017 &1.413 - 1.808 & 2000 & 6.7 & 1.0 \\
J083448.48+015921.1$^{b}$  &  2.5928 & low-ions  & $-$203 &X-shooter & Apr 03 2014 & 0.3 - 2.48  & 5100 & 6.0 & 0.0 \\  
J091303.90+234435.2  &  2.4356   & [O{\sevensize III}] & +462 &GNIRS & Mar 01/25 2016 & 0.9 - 2.5 & 1700 & 5.3 & 0.0 \\
J093226.93+461442.8  &  2.3130   & H${\beta}$ & $-$ &GNIRS & Mar 18 2016 & 0.9 - 2.5 & 1700 & 5.7 & 0.0 \\
J095823.14+500018.1  &  2.3626   & low-ions & $-$200 &GNIRS & Feb 02 2016 & 0.9 - 2.5  & 1700 & 5.2 & 10.3 \\
J101324.53+342702.6  &  2.4609   & H${\beta}$ & $-$ &GNIRS & Apr 20 2016 & 0.9 - 2.5  & 1700 & 4.7 & 0.0 \\
J102541.78+245424.2  &  2.3994   & H${\beta}$ &$-$ & NIRSPEC & Mar 13 2017 &1.413 - 1.808  & 2000 & 4.8 & 0.0 \\
J103146.53+290324.1  &  2.2955   & [O{\sevensize III}] & +379 &GNIRS & Feb 26 2016 & 0.9 - 2.5  & 1700 & 5.7 & 0.0 \\
J113834.68+473250.0  &  2.3154   & H${\beta}$ & $-$ &NIRSPEC & Mag 3 2017 &1.413 - 1.808  & 2000 & 6.1 & 0.0 \\
J121704.70+023417.1  &  2.4266   & [O{\sevensize III}] & $-$66 &X-shooter & Feb 18 2017 & 0.3 - 2.48 & 5100 & 5.6 & 0.0 \\
J123241.73+091209.3$^{b}$  &  2.3886 &  low-ions  & +310 &X-shooter & Apr 03 2014 & 0.3 - 2.48  & 5100 & 6.8 & 0.0 \\
J134254.45+093059.3  &  2.3451   & H${\beta}$ & $-$ & NIRSPEC & Mar 13 2017 &1.413 - 1.808  & 2000 & 4.9 & 0.0 \\
J134800.13$-$025006.4  &  2.2382    & [O{\sevensize III}] & $-$122 &NIRSPEC & Mag 3 2017 &1.413 - 1.808  & 2000 & 5.7 & 0.0 \\
J135608.32+073017.2$^{c}$  &  2.2751   & H${\beta}$ & $-$ &GNIRS & Feb 25 2015 & 0.9 - 2.5 & 1700 & 5.1 & 0.0 \\
J155057.71+080652.1  &  2.5087   & H${\beta}$ & $-$ &GNIRS & Mar 22/24 2016 & 0.9 - 2.5 & 1700 & 3.8 & 1.3 \\
J160431.55+563354.2  &  2.4914   &  low-ions  & +677 &GNIRS & Feb 21/22 2016 & 0.9 - 2.5 & 1700 & 5.7 & 0.0 \\
J165202.64+172852.3$^{c}$  &  2.9482   & H${\beta}$ & $-$ &GNIRS & Apr 6 2015 & 0.9 - 2.5 & 1700 & 5.4 & 1.6 \\
J215855.10$-$014717.9  &  2.3068    &  low-ions  & $-$300 &NIRSPEC & Nov 10 2016 &1.413 - 1.808  & 2000 & 4.1 & 0.0 \\
J221524.00$-$005643.8$^{b}$  &  2.4975    &  low-ions  & $-$413 &X-shooter & Jun 2 2014 & 0.3 - 2.48  & 5100 & 6.2 & 0.0 \\
J232326.17$-$010033.1$^{b}$  &  2.3695    & H${\beta}$ & $-$ & X-shooter & Jun 10 2014 & 0.3 - 2.48  & 5100 & 7.2 & 0.0 \\
J232611.97+244905.7  &  2.3784   & H${\beta}$ & $-$ & NIRSPEC & Nov 11 2016 &1.413 - 1.808 & 2000 & 4.5 & $-$ \\
\hline
\\
\end{tabular}
\end{center}
\begin{flushleft}
$^{a}$  Redshift reference line with respect to which the kinematics of the emission lines is measured: [O{\sevensize III}] = centroid of the [O{\sevensize III}] emission line; H${\beta}$ = centroid of the H${\beta}$ emission line; low-ions = centroid of the low-ions (e.g. O{\sevensize I}, C{\sevensize II} and Mg{\sevensize II}) in the BOSS spectra.\\ 
$^{b}$ From \citet{Zakamska2016}\\
$^{c}$ From \citet{Alexandroff2018}\\
\end{flushleft}
\end{table*}

\begin{figure*}
 \includegraphics[width=\textwidth]{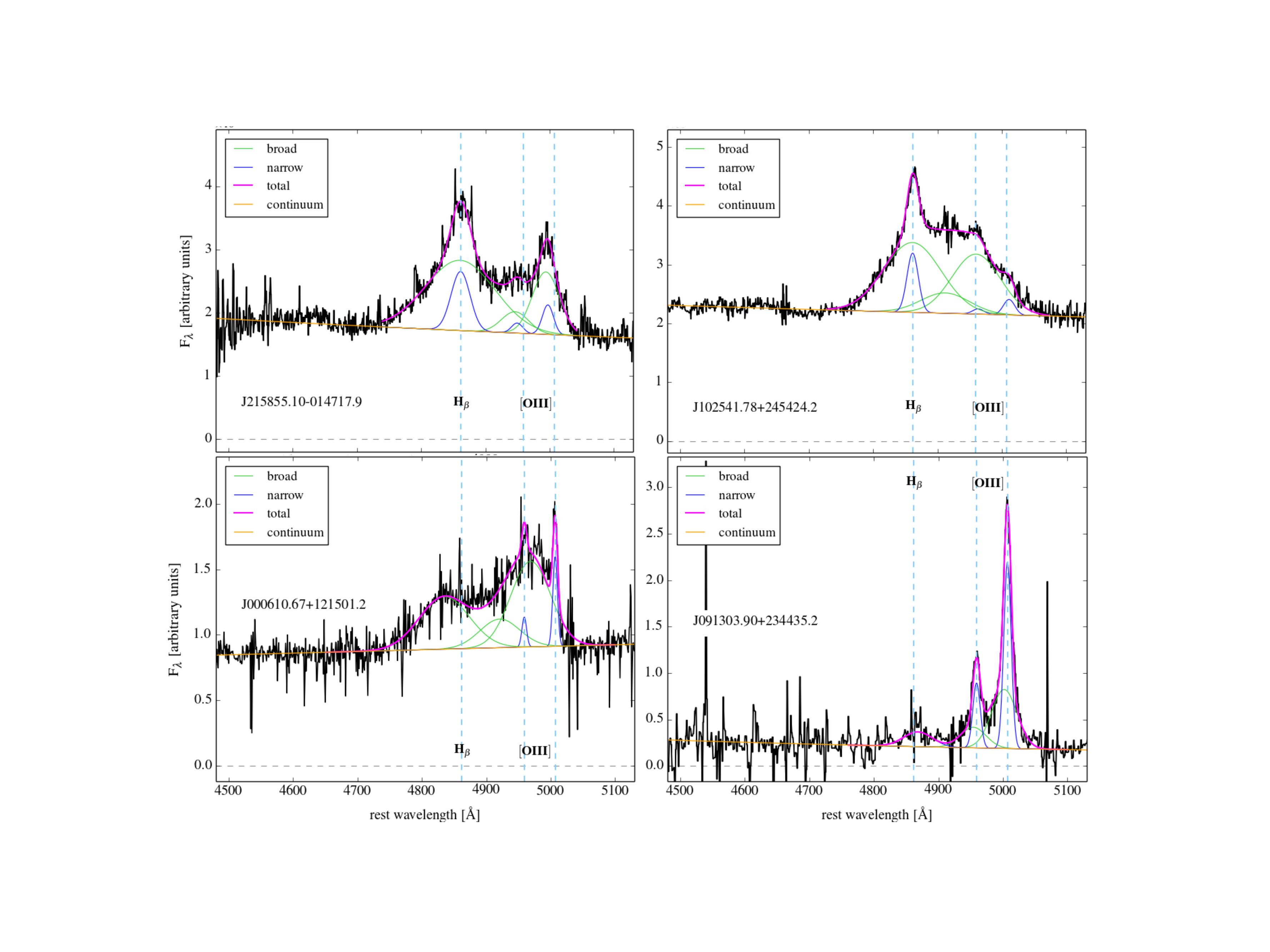}
 \caption{Fits to the H$\beta$+[O{\sevensize III}] blend in four extremely red quasars in our sample. The iron fits have been subtracted from the original spectra to give visibility to the components of the [O{\sevensize III}] fits. The orange line represents the continuum. The magenta solid line shows the best fit to the entire complex, two Gaussian components for [O{\sevensize III}] and one or two kinematically independent Gaussians for H$\beta$. The dashed light blue vertical lines mark the locations of H$\beta$ $\lambda$ 4861, [O{\sevensize III}] $\lambda$4959 and [O{\sevensize III}] $\lambda$5007 in the frame associated with our best estimate of the systemic redshift, $z_{best}$. }
 \label{ERQ_fits}
\end{figure*}

\begin{figure}
 \includegraphics[width=\columnwidth]{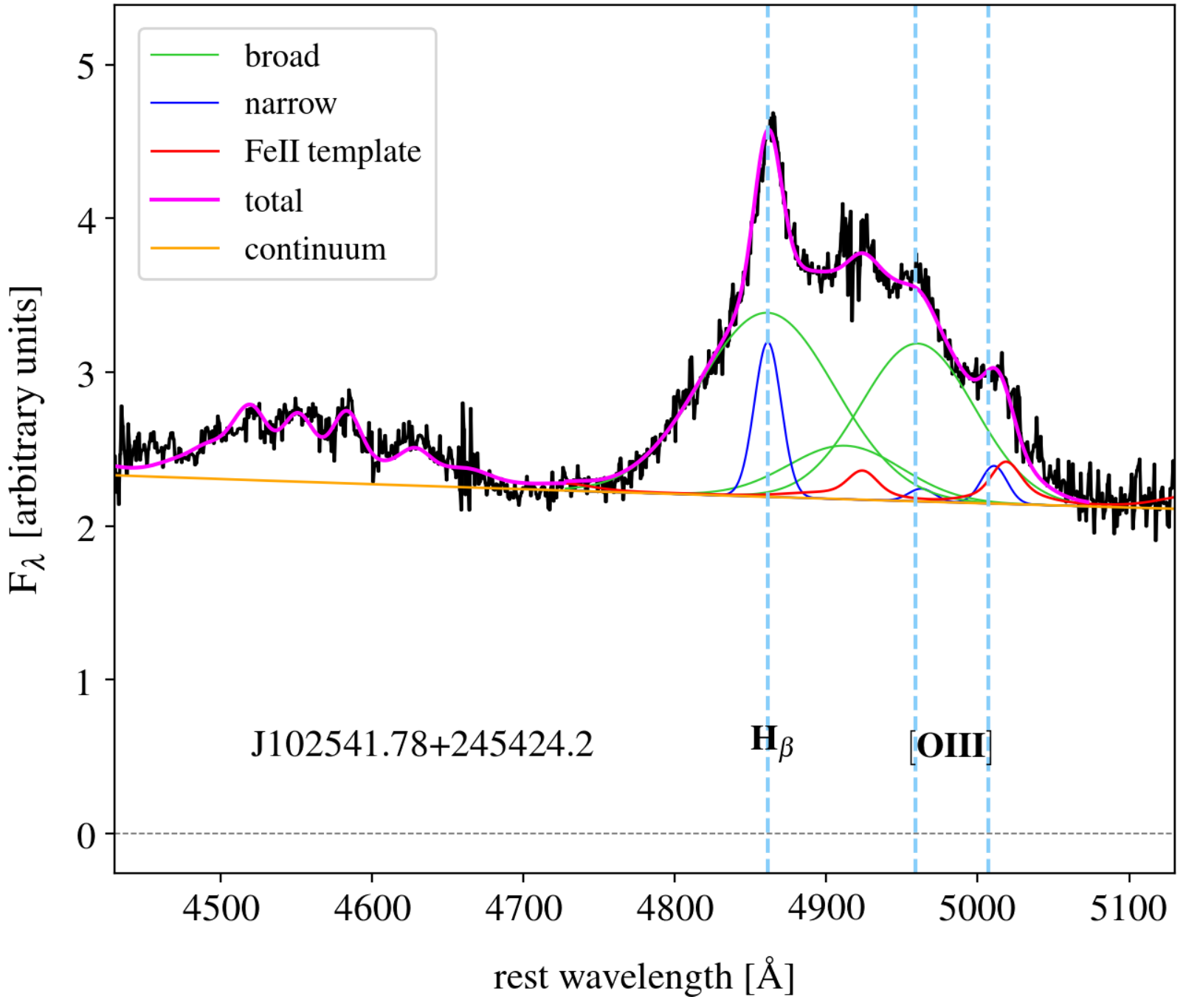}
 \caption{Fit to the H$\beta$+[O{\sevensize III}]+Fe{\sevensize II} blend in one extremely red quasar in our sample. The magenta solid line shows the best fit to the entire complex. Green and blue curves refer to the H$\beta$+[O{\sevensize III}] line fitting decomposition. The red solid line represents the iron line template from \citet{Veron2004} convolved with a Gaussian kernel with a velocity dispersion similar to the H$\beta$ line.}
 \label{ERQ_FeII}
\end{figure}

\section{Observations and Data Reduction}
\label{obs}

We obtained near-IR spectra of 28 broad emission-line (type 1) quasars selected from the lists of ERQs and ``ERQ-like'' quasars in \cite{Hamann2017}. Table~\ref{tab_data} provides some basic information about the quasars and our new observations.

We select only broad-line quasars based on FWHM(C{\sevensize IV}) $>$ 2000 km s$^{-1}$ \citep{Alexandroff2013} to exclude type 2 sources where the obscuration is generally attributed to orientation effects, e.g., in the so-called unified model of AGN \citep{Padovani1993,Urry1995, Netzer2015}. Then our highest priority was to observe ERQs in the ``core'' sample defined by \cite{Hamann2017} to have i$-$W3 $>$ 4.6 and rest-frame equivalent width of C{\sevensize IV}, REW(C{\sevensize IV}) $>$ 100\AA. The additional requirement for large REW(C{\sevensize IV}) in the core sample helps to isolate ERQs with both the reddest colors and the most extreme UV emission-line properties (that might be related to an early active evolution stage, Section 1). We also observed several quasars outside of the core ERQ sample to explore the relationship of the [O{\sevensize III}] kinematics to a wider range of quasar properties. 

The final quasar sample listed in Table~\ref{tab_data} includes 20 type 1 core ERQs, 4 ERQs (with i$-$W3 $>$ 4.6) not in the core sample due to REW(C{\sevensize IV}) $<$ 100\AA, and 4 ``ERQ-like'' quasars that are not ERQs due i$-$W3 $<$ 4.6 but they are still redder than the median for W3-detected BOSS quasars at these redshifts and they have emission-line properties similar to the core ERQs (see \citealp{Hamann2017}). For convenience, we will refer to this entire sample as ``ERQs'' throughout the remainder of this article.

Near-IR spectra were obtained for twelve of the targets using the NIRSPEC instrument \citep{McLean1998} on the Keck II telescope. 
We used the NIRSPEC-5 filter covering 1.413-1.808 $\mu$m, corresponding to the photometric H band. All targets were observed with a 0\farcs76 $\times$ 42\arcsec\,long slit for a spectral resolution of R = $\lambda/\Delta\lambda \approx$ 2000. We employed the standard ABBA slit-nodding pattern. Individual exposures were 360 s, with total integration times of $\sim$1.5 hr per object.

We reduced the data using an IDL custom pipeline developed by George Becker \citep[see][]{Becker2009}. The exposures were dark subtracted and flat-fielded using an internal flat-field calibration lamp. We did not subtract pairs of A--B exposures because that multiplies the noise from the sky and the dark current by a factor of $\sqrt{2}$. Instead, we built a low-noise dark frame by taking a large set of dark exposures with the same exposure time as our science frames. This is used to remove dark current features and other blemishes prior to sky subtraction. The sky was then modeled along the slit on each 2-dimensional exposure frame using a b-spline fit and removed from each exposure using optimal sky subtraction techniques for long-slit spectra \citep{Kelson2003}. All the steps were applied to the two-dimensional frames before the data are rectified. A single one-dimensional spectrum was extracted simultaneously across all the orders and all exposures of a given object. We performed relative flux calibrations and telluric absorption corrections using spectra of standard stars observed the same night. The effect of telluric corrections is small for most of our targets. The wavelengths are calibrated to the vacuum heliocentric system using spectra of internal arc lamps.

Ten ERQs were observed with Gemini Near-IR Spectrograph (GNIRS; \citealp{Elias1998}) on Gemini-North. For these observations, we selected the cross-dispersed mode, using the short camera, the 32 lines mm$^{-1}$ grating centred at 1.65 $\mu$m and the 0\farcs45 slit width, which gives a resolution R $\sim$1700. In this observing mode, the entire near-IR region from 0.9 to $\approx$ 2.5 $\mu$m is covered in a single observation. The Gemini observations were conducted in service mode in a series of nodded 330 sec exposures along the slit, giving a total exposure time of $\sim$45 min. A telluric standard was also observed both before and after the target, at a similar airmass. 

The data from Gemini-GNIRS were reduced using the GNIRS sub-package in the Gemini IRAF software package (v1.13.1). Briefly, a correction was first applied to the raw science, standard star and arc lamp spectral images for the s-distortion in the orders. The data were then flat-fielded, taking care to flat-field each order with the corresponding correctly exposed flat. Subsequently, difference pairs were assembled from the science and standard star images and any significant remaining sky background removed by subtracting linear functions, fitted in the spatial direction, from the data. The spectral orders of the objects and the standard stars were then extracted and assigned the wavelength solution derived from the relevant arc spectrum. Then, the science spectral orders were divided by the corresponding standard star spectral orders and multiplied by a blackbody of appropriate temperature. 

The spectra of six objects, have been obtained with the echelle spectrograph X-shooter on the European Southern Observatory (ESO) Very Large Telescope (VLT), a medium resolution spectrograph allowing simultaneous observations over the wavelength range from 0.3 to 2.48 $\mu$m \citep{Vernet2011}. The spectra of these objects have been reduced as described in \citet{Zakamska2016}.

Seven sources are part of previous X-shooter and GNIRS programs (\citealp{Zakamska2016} and \citealp{Alexandroff2018}; see Table~\ref{tab_data}). In particular, the spectrum of J135608.32+073017.2 is unpublished from the GNIRS program by \citet{Alexandroff2018}.

\section{Analysis of Spectra}
\label{analysis}

\subsection{Line Fitting}
\label{fitting}

Extreme kinematics in the [O{\sevensize III}] $\lambda$4959,5007 emission lines are a common characteristic of ERQs. Figure~\ref{ERQ_fits} shows the H$\beta$$-$[O{\sevensize III}] spectral region for four ERQs that illustrate the range of properties in our sample. (Similar spectra of the remaining quasars in our study are shown in Appendix \ref{Appendix1}). The [O{\sevensize III}] lines are blended together and they often have broad blueshifted wings that blend with the nearby H$\beta$ $\lambda$4861 line. The line widths and blueshifts that cause this blending imply line-of-sight velocities of order several thousand km s$^{-1}$ in the [O{\sevensize III}]-emitting gas. We quantify the [O{\sevensize III}] kinematics for each ERQ in our sample by fitting the H$\beta$ and [O{\sevensize III}] emission lines as follows.

First, we fit a local power-law continuum ($f_{\lambda} \propto \lambda^{\alpha}$) constrained by wavelength regions that avoid the H$\beta$ and [O{\sevensize III}] emission lines as well as possible contributions from blended broad Fe{\sevensize II} emission. To estimate the strength of the Fe{\sevensize II} emission, we convolve the iron line template from \cite{Veron2004} with a Gaussian kernel that has a velocity dispersion similar to the H$\beta$ line (as derived from a preliminary fit or direct measurements). We use two strong emission features in the Fe{\sevensize II} template at $\sim$4600\AA\ to estimate the iron contributions to the observed spectra. Figure~\ref{ERQ_FeII} shows an example of the Fe{\sevensize II} spectral line fitting. We find that Fe{\sevensize II} does not critically change the measured [O{\sevensize III}] kinematics. The [O{\sevensize III}] line widths decrease by a factor 2 to 8 per cent.
 The pseudo-continuum and iron fits are then subtracted from the original spectrum. We then model H$\beta$ and the two [O{\sevensize III}] lines with one or two Gaussian functions each, depending on the complexity of the emission profiles and the signal-to-noise ratio (S/N) of the spectrum. We assume both lines in the [O{\sevensize III}] doublet always the same kinematics (i.e., the same velocity widths and shifts in the Gaussian fit components), and we fix their amplitude ratio [O{\sevensize III}]$\lambda$4959/[O{\sevensize III}]$\lambda$5007 to 0.337 to match the transition strengths \citep{Storey2000}. The H$\beta$ profile is allowed to have a different kinematic structure than the [O{\sevensize III}] lines. 

Figure~\ref{ERQ_fits} shows fits to the H$\beta$ and [O{\sevensize III}] lines in four ERQs that illustrate the range of line properties in our data. H$\beta$ is well described by a single Gaussian in 19 objects (see the bottom panels of Figure~\ref{ERQ_fits} for examples). In the remaining 9 cases, we add a second Gaussian component at the same redshift as the first one because residuals from the fit of a single component turn out to be large (upper panels of Figure~\ref{ERQ_fits}). Forcing the two components to have the same redshift helps the fitting procedure by eliminating a degree freedom and it is justified by the generally symmetric appearance of the H$\beta$ emission lines in our data. It is also consistent with the fact that our quasar sample is mostly radio-quiet \citep{Hwang2018}, and previous studies have shown that H$\beta$ is typically symmetric in radio-quiet quasars \citep[while radio-loud quasars often have more asymmetric profiles][]{Boroson1992, Marziani1996, Zamfir2010}. 

The [O{\sevensize III}] lines generally require two Gaussian components to fit their broad asymmetric profiles. The only exceptions are in J013413.22-023409.7 and J135608.32+073017.2 (see Appendix \ref{Appendix1}) where a second Gaussian components yields no statistical improvement due to the low S/N ratios across the [O{\sevensize III}] lines in the data. Many ERQs require a strong broad blueshifted component to fit the [O{\sevensize III}] lines. This can lead to complex blends where the blueshifted component in [O{\sevensize III}] $\lambda$5007 substantially boosts the flux near the peak of the $\lambda$4959 line (e.g., J000610.67+121501.2 and J102541.78+245424.2 in Figure~\ref{ERQ_fits}). 
Table~\ref{tab:fits} lists the centroid wavelength of each Gaussian component in our fits, plus the FWHM and the REW measured from the full fitted profiles.  

\begin{figure*}
 \includegraphics[width=\textwidth]{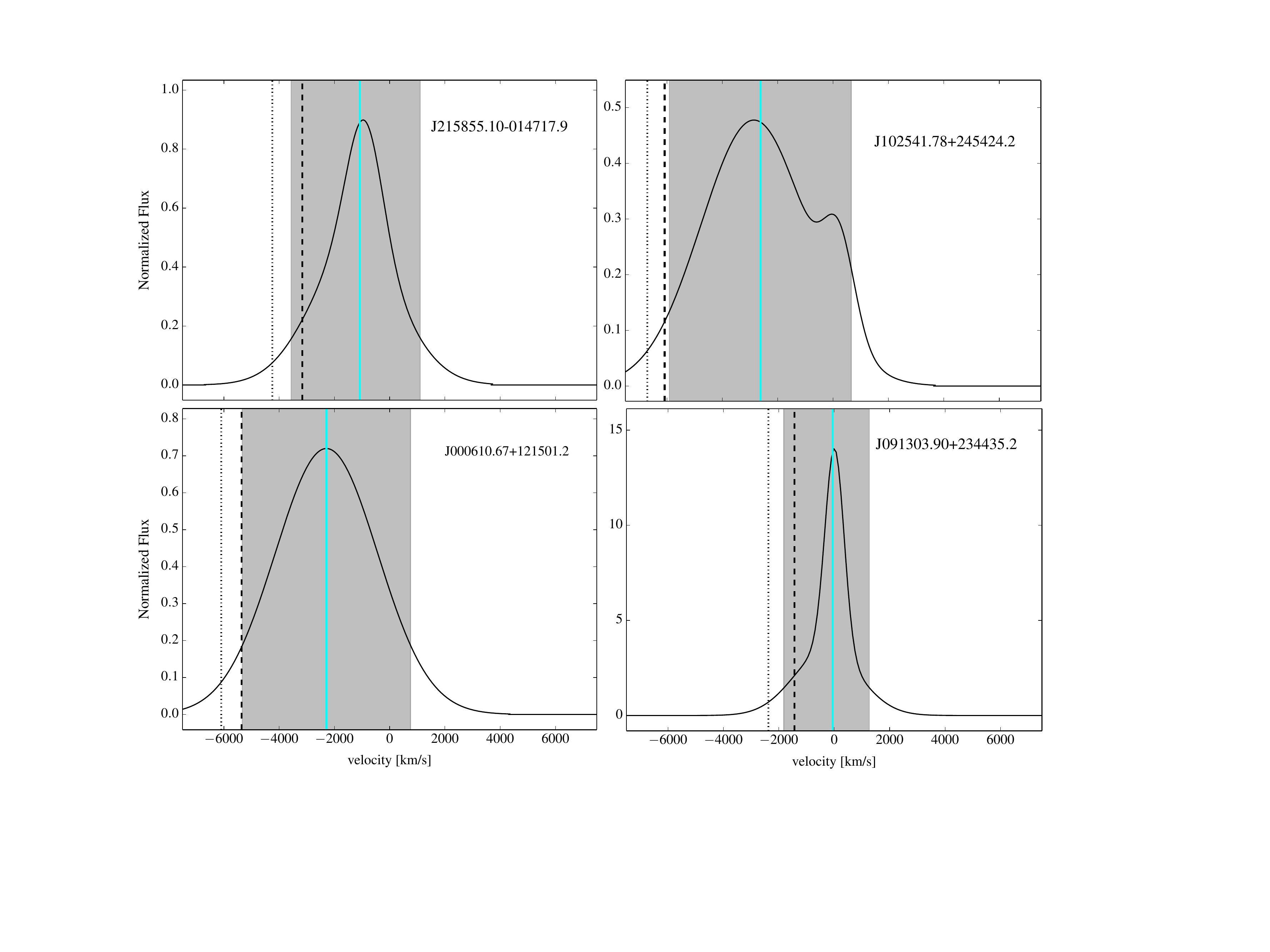}

 \caption{The best-fit profiles of [O{\sevensize III}] $\lambda$5007 relative to our best estimate of the systemic redshift, $z_{best}$, shown in velocity space. The pseudo-continuum is subtracted from the spectrum, leaving an emission-line-only spectrum. The grey area marks the part of the profile containing 90\% of the line power ($w_{90}$), whereas vertical lines mark v$_{98}$ (black dotted), the median velocity of the profile v$_{50}$ (cyan solid) and the flux-weighted mean outflow velocity $\langle {\rm v}_r\rangle$ (black dashed).}
 \label{O3profiles}
\end{figure*}

\subsection{Redshift Estimates}

Accurate redshifts are important to determine line shifts and outflow velocities. Table~\ref{tab_data} lists our best estimates of the systemic redshifts ($z_{best}$) for each ERQ along with the reference line used for these estimates.

Stellar absorption lines are not detected in any of the ERQ spectra. Therefore we must use quasar emission lines to estimate the systemic redshifts. Accurate redshifts are notoriously difficult to measure from rest-frame UV broad emission lines in quasars due to blueshifts and asymmetries in the line profiles \citep{Richards2011, Denney2016}. Previous studies have shown that the redshifts derived from narrow forbidden [O{\sevensize III}] lines are consistent to within $\sim$50 km s$^{-1}$ with those based on stellar absorption lines and H{\sevensize I} 21 cm emission in the host galaxies of AGN \citep{Gaskell1982, McIntosh1999, Hewett2010, Bae2014}. On the other hand, [O{\sevensize III}] lines in ERQs are often extremely broad and blueshifted (Section~\ref{fitting}).

Following \citealp{Zakamska2016}, we adopt H$\beta$ as the nominal redshift indicator for the ERQs. However, we also examine other available emission lines to choose a final best-guess redshift, $z_{best}$, for each ERQ. These checks against other lines are important because i) H$\beta$ can be weak and/or severely blended with [O{\sevensize III}] in ERQ spectra (see Figure~\ref{ERQ_fits}), and ii) H$\beta$ can be significantly shifted from low-ionization broad emission lines. Low ionisation permitted lines, such as Mg{\sevensize II} $\lambda$2800, are often used when narrow [O{\sevensize III}] lines are  not available \citep{Hewett2010, Shen2014, Shen2016}. \citet{Shen2016} showed that Mg{\sevensize II} has a mean offset of only $\sim$100 km s$^{-1}$ from narrow [O{\sevensize III}] lines, with an object-to-object dispersion of $\sim$280 km s$^{-1}$. \citet{Shen2016} also reported that H$\beta$ is typically blueshifted by $-$100 km s$^{-1}$ with respect to low-ionization Ca{\sevensize II} lines with object-to-object dispersion of 418 km s$^{-1}$. 

For the ERQs in our study, we compare the redshifts determined from the H$\beta$ centroids (Section \ref{fitting}) to other lines as available in existing spectra, namely i) H$\alpha$, ii) narrow components in [O{\sevensize III}] that resemble traditional NLR features, and iii) any low-ionization emission lines (such as Mg{\sevensize II} $\lambda$2800, O{\sevensize I} $\lambda$1334 and C{\sevensize II} $\lambda$1335) available in the rest-UV spectra. We adopt the redshift derived from H$\beta$ for 16 objects in our sample, as they well match those inferred from H$\alpha$ and low-ionization ions (including J102541.78+245424.2 in Figure~\ref{ERQ_fits}). We use the narrow [O{\sevensize III}] component as the best redshift reference for 6 ERQs where it is clearly distinct from the rest of the broad H$\beta$-[O{\sevensize III}] emission-line blend (e.g., J000610.67+121501.2 and J091303.90+234435.2 in Figure~\ref{ERQ_fits}), and we use low-ionisation lines in the remaining 6 objects (see Table~\ref{tab_data}). For example, in J215855.10-014717.9, we use the low-ionisation lines because H$\beta$ is blueshifted by $\sim$300 km s$^{-1}$ with respect to the well measured O{\sevensize I}, C{\sevensize II} and Mg{\sevensize II} lines. The other 5 cases in which we adopt low-ion lines as reference for the redshift have some of the most noisy near-IR spectra in our sample (e.g. J221524.00-005643.8 in Appendix~\ref{Appendix1}).

\begin{landscape}
 \begin{table}
\begin{center}
  \caption{Properties of the H${\beta}$ and [O{\sevensize III}] emission lines derived from parametric model fits. $ \lambda_n$ [O{\sevensize III}] and $ \lambda_b$ [O{\sevensize III}] represent the centroid of the narrow and broad [O{\sevensize III}] emission line components, respectively. $ \lambda$ H${\beta}$ is the centroid of the H${\beta}$ emission line.}
  \label{tab:fits}
\begin{tabular}{ c c c c c c c c c c }
\hline
\hline

 Object Name & $ \lambda_n$ [O{\sevensize III}] & $ \lambda_b$ [O{\sevensize III}] &  FWHM[O{\sevensize III}]  & REW[O{\sevensize III}] & $ \lambda$ H${\beta}$  &  FWHM H${\beta}$ & REW H${\beta}$  & REW C{\sevensize IV} & FWHM C{\sevensize IV} \\
     & [\AA] & [\AA] & [km s$^{-1}$] & [\AA] & [\AA] & [km s$^{-1}$] & [\AA] &   [\AA] & [km s$^{-1}$]  \\

 \hline
J000610.67$+$121501.2  & 5006.84 & 4968.60 & 5132 $\pm$ 488 & 63 $\pm$ 3 & 4834.43 & 5733 $\pm$ 512& 44 $\pm$ 2& 107 &4540 \\
J001120.22+260109.2  & 5001.04 &  4982.15 & 1475 $\pm$ 96 & 61 $\pm$ 2 & 4861.34 & 2963 $\pm$ 136 & 100 $\pm$ 3 &98 & 3804  \\
J013413.22$-$023409.7& $-$ & 4996.99  & 2912 $\pm$ 106& 46 $\pm$ 2& 4862.69 & 3646 $\pm$ 693& 92 $\pm$ 13&56 &4534   \\
J020932.15+312202.7  & 5003.47 & 5000.36 & 788 $\pm$ 182 & 157 $\pm$ 16 & 4861.34 & 1557 $\pm$ 157& 38 $\pm$ 5&  108 &2180 \\
J080547.66+454159.0  & 5006.84 & 4976.77 & 3909 $\pm$ 253& 104 $\pm$ 7& 4859.28 & 2914 $\pm$ 222& 45 $\pm$ 5&  109 &2667  \\
J082618.04+565345.9  & 4999.32 & 4979.76 & 1834 $\pm$ 419& 89 $\pm$ 11& 4861.34 & 4626 $\pm$ 294& 64 $\pm$ 5&     82 &3508  \\
J082653.42+054247.3 & 5007.51 & 4986.37 & 1541 $\pm$ 303& 441 $\pm$ 36& 4861.34 & 2140 $\pm$ 326& 72 $\pm$ 11&    205 &2434   \\
J083200.20+161500.3 & 5001.22 & 4977.16 & 3938 $\pm$ 148 & 206 $\pm$ 7& 4861.34 & 5428 $\pm$ 275& 78 $\pm$ 4&    300 &3082  \\
J083448.48+015921.1 & 4986.48 & 4979.02 & 2204 $\pm$ 92& 493 $\pm$ 12& 4859.4 & 4525 $\pm$ 42& 220 $\pm$ 3&    209 &2863   \\
J091303.90+234435.2 & 5006.84 & 5004.10 & 996 $\pm$ 183& 316 $\pm$ 21& 4868.85 & 3193 $\pm$ 495& 50 $\pm$ 9&    145& 2190 \\
J093226.93+461442.8 & 5005.66 & 4988.13 & 1633 $\pm$ 129& 646 $\pm$ 39& 4861.34 & 3581 $\pm$ 736& 231 $\pm$ 60&  443 &1960 \\
J095823.14+500018.1 & 4974.64 & 4967.25 & 1451 $\pm$ 230& 105 $\pm$ 6& 4858.09 & 4052 $\pm$ 389& 111 $\pm$ 8&   263 &4345 \\
J101324.53+342702.6 & 5004.27 & 5003.88  & 1045 $\pm$ 206& 271 $\pm$ 19& 4861.34 & 1716 $\pm$ 212& 145 $\pm$ 8&   204 &4157 \\
J102541.78+245424.2 &  5010.63 & 4959.14 & 5751 $\pm$ 195& 42 $\pm$ 1& 4861.34 & 2753 $\pm$ 94& 71 $\pm$ 2&    114& 5324 \\
J103146.53+290324.1 & 5006.84 & 4989.28 & 2590 $\pm$ 187& 51 $\pm$ 4& 4867.5 & 7456 $\pm$ 512& 66 $\pm$ 5&      121 &4364 \\
J113834.68+473250.0 & 5003.20 & 4983.35 & 1738 $\pm$ 169& 197 $\pm$ 10& 4861.34 & 4290 $\pm$ 136& 54 $\pm$ 5&    177 &3296 \\
J121704.70+023417.1 & 5008.24 & 4995.29 & 689 $\pm$ 68& 67 $\pm$ 2& 4861.62 & 5160 $\pm$ 367& 122 $\pm$ 5&    181 &2604\\
J123241.73+091209.3 & 5011.52 & 4970.66 & 4876 $\pm$ 209& 311 $\pm$ 12& 4867.73 & 5412 $\pm$ 116& 111 $\pm$ 3&   224& 4787 \\
J134254.45+093059.3 & 4999.51 & 4974.46 & 1654 $\pm$ 224& 90 $\pm$ 4& 4861.34 & 4098 $\pm$ 87& 78 $\pm$ 2&    66 &3246 \\
J134800.13$-$025006.4 & 5006.84 & 4983.05 & 3330 $\pm$ 130& 67 $\pm$ 3& 4859.36 & 6013 $\pm$ 199& 70 $\pm$ 3&   87 &3654 \\
J135608.32+073017.2 & $-$ & 5001.18 & 2017 $\pm$ 198& 66 $\pm$ 7& 4861.34 & 4162  $\pm$ 326& 73 $\pm$ 4& 110 & 2043 \\
J155057.71+080652.1 & 4977.29 & 4967.12 & 1089 $\pm$ 185& 130 $\pm$ 13& 4861.34 & 3550 $\pm$ 635& 162 $\pm$ 18&  149 &4446 \\      
J160431.55+563354.2 & 5008.99 & 4981.11 & 1733 $\pm$ 412& 382 $\pm$ 68& 4872.33 & 4036 $\pm$ 437& 151 $\pm$ 20&  205 &4221 \\  
J165202.64+172852.3 & 5008.28 & 4993.92 & 1330 $\pm$ 220& 153 $\pm$ 11& 4861.34 & 2662 $\pm$ 342& 57 $\pm$ 12&    124 &2403 \\
J215855.10$-$014717.9  &  4993.82 & 4986.67 & 2346 $\pm$ 262 & 38 $\pm$ 4& 4856.47 & 4001 $\pm$ 183 & 110 $\pm$ 3 & 73 &4735 \\
J221524.040$-$005643.8& 5012.56 & 4988.31 & 2870 $\pm$ 415 & 255 $\pm$ 8& 4855.99 & 6485 $\pm$ 532& 116 $\pm$ 12& 153 &4280 \\
J232326.17$-$010033.1& 5009.01& 4956.49  & 3065 $\pm$ 800& 96 $\pm$ 7& 4862.69 & 4508 $\pm$ 160& 46 $\pm$ 2& 256 &3989 \\
J232611.97+244905.7  & 5009.36& 4990.62  & 1103 $\pm$ 275& 67 $\pm$ 5& 4861.34 & 4295 $\pm$ 91& 98 $\pm$ 3& 131 &2402 \\
      
    \hline      
  \end{tabular}      
\end{center}    

 \end{table}
\end{landscape}

\subsection{[OIII] kinematics}
\label{O3kinematics}

We measure the [O{\sevensize III}] kinematics from our line profile fits in three ways, following \citet{Zakamska2016}, \citet{Zakamska2014} and other authors (e.g. \citealp{Veilleux1991}). We report the non-parametric measures for every ERQ in Table~\ref{tab_derived}. We note that the [O{\sevensize III}] widths and velocity shifts probed by the ERQ sample are much larger and significant than the uncertainties of the redshift determination. 

For each fit to the [O{\sevensize III}] $\lambda$5007 profile, we calculate the velocity width comprising 90\% of the flux, $w_{90}$, by rejecting the most extreme 5\% blue-shifted and red-shifted parts of the line profile. $w_{90}$ is a measure of the profile width that for a gaussian profile is 1.397FWHM. The measured values of $w_{90}$ range from 2053 to 7227 km s$^{-1}$ in our sample. Such [O{\sevensize III}] widths are much larger than those found in type 1 and type 2 quasars at lower redshifts. We will come back to this point in Section~\ref{results}. We also measure the line velocity shifts relative to the systemic (based on $z_{best}$) at which 98 (v$_{98}$) and 50 (v$_{50}$) per cent of the line flux accumulates moving from red (positive velocities) to blue (negative velocities) across the line profile. v$_{50}$ is thus the median velocity in the profile while v$_{98}$ measures the blueshifted wing near its maximum extent. 

Figure~\ref{O3profiles} shows examples of these measurements for the same four ERQs plotted in Figure~\ref{ERQ_fits}. The best-fit [O{\sevensize III}] $\lambda$5007 profiles are plotted on a velocity scale relative to $z_{best}$. The grey shaded area marks the part of the profile containing 90\% of the line power. The velocities v$_{98}$ and v$_{50}$ are marked by dotted and solid vertical lines, respectively. 

We also estimate a mean radial outflow velocity, $\rm\langle v_r\rangle$, following \citet{Zakamska2016}. The strong blue asymmetry in most of the measured [O{\sevensize III}] $\lambda$5007 profiles, e.g., with most of the flux appearing on the blue side of the lines profiles (v$_{50} < 0$, Table~\ref{tab_derived}), strongly suggests that the [O{\sevensize III}] emission is affected by extinction. We assume that 1) extinction affects only the redshifted side of the profiles (v$_z$ $<$ 0), 2) the outflow is spherically symmetric in the hemisphere facing the observer, and 3) the gas moves radially away from the quasar with radial velocity distribution, f(v$_r$)dv$_r$. Then, regardless of the shape of the distribution function f(v$_r$), the flux-weighted mean outflow velocity has a simple relation to the observable line-of-sight value, v$_z$, given by
\begin{ceqn}
\begin{equation}
\rm   \langle v_r\rangle = 2 \,\langle\vert v_z\vert \rangle.
\label{vr}
\end{equation}
\end{ceqn}
where $\langle\vert v_z\vert \rangle$ is the flux-weighted average \textit{observed} velocity on the blueshifted side of the line profile. $\rm\langle v_r\rangle$ derived this way measures a typical radial speed for (spherically-symmetric) outflows that might have a range of radial speeds. The derived values in our sample range 218 to 6141 km s$^{-1}$. The mean value across the whole sample is 2548 km s$^{-1}$. 

In contrast to $\langle\vert v_z\vert \rangle$, v$_{98}$ yields a direct measure of the maximum speeds reached in the outflows (or a lower limit to the maximum speed if the gas producing the blueshifted line wings is not moving directly toward the observer). The values of v$_{98}$ in our ERQ sample range from $-$1993 to $-$6702 km s$^{-1}$ (Table~\ref{tab_derived}). To estimate errors on v$_{98}$ due to uncertainties in the fits, we consider the best-fit parameters uncorrelated, vary them in a range of $\pm$1$\sigma$ and calculate the resulting change in v$_{98}$. We adopt the maximal variation of v$_{98}$ as upper limit error, with typical values of 200$-$500 km s$^{-1}$ for our sample. The median [O{\sevensize III}] line profile velocities, v$_{50}$, are also blueshifted in all cases, with values ranging from $-$36 to $-$2613 km s$^{-1}$.

\subsection{Luminosities and Black Hole Mass}
\label{luminosity}

Table~\ref{tab_derived} lists the bolometric luminosity ($L_{bol}$) and [O{\sevensize III}] luminosity ($L$([O{\sevensize III}])) for every target in our sample.
$L_{bol}$ values are difficult to determine for ERQs because they have severe (but uncertain) amounts of extinction across the rest-frame visible and UV. We adopt the procedure outlined in  \cite{Hamann2017}, who use the bolometric correction $L_{bol} = 8\, \lambda\,$$L_{\lambda}$ at $\lambda$ = 3.45 $\mu$m in the rest frame. WISE W3 photometry measures $\sim$3.45 $\mu$m in the rest-frame at the typical redshift of ERQs in our study. We adjust all of the observed W3 fluxes to this fixed rest wavelength by assuming the spectra have slope $L_{\lambda} \sim \lambda^{-0.65}$ near this wavelength \citep{Polletta2007}.

We estimate the luminosity at $\lambda$ = 5007 and 5100\AA~multiplying the fluxes at 5007 and 5100\AA\, by 4$ \pi D_L^2$, where $D_L$ is the luminosity distance of the quasar. To obtain the fluxes at 5007 and 5100\AA, we linearly interpolate the WISE W1(3.6$\mu$) and SDSS i or z magnitudes and then convert the derived values at 5007 and 5100\AA\, to fluxes.

$L$([O{\sevensize III}]) is estimated for each ERQ, multiplying the [O{\sevensize III}] REW by the luminosity at 5007\AA. The median ERQ spectral energy distribution (SED) is suppressed at $\sim$5000\AA\, by about 2 magnitudes relative to normal/blue quasars \citep[see Figure 16 in][]{Hamann2017} and, therefore, we expect the values of $L$([O{\sevensize III}]) in Table~\ref{tab_derived} to be too low typically by a factor of $\sim$6.3. 

We estimate the black hole mass ($M_{ BH}$) of each ERQ using the virial relation
\begin{ceqn}
\begin{equation}
 M_{BH} = f \frac{\Delta {\rm V}^2 \, R_{BLR}}{G}
\label{M_BH}
\end{equation}
\end{ceqn}
where $\Delta$V is the velocity of the broad-line region (BLR) gas, $R_{BLR}$ is the radial size of the BLR, $G$ is the gravitational constant and $f$ is a calibration factor. Usually, the FWHM of the H$\beta$ line (FWHM$_{{\rm H}\beta}$) or the second moment of the line profile (i.e., the line dispersion; $\sigma_{ {\rm H}\beta}$) is used for the velocity of the BLR gas \citep{Peterson2004}. We adopt log $f$ = 0.05 $\pm$ 0.12 ($f$ = 1.12) for FWHM-based mass ($M_{BH}^{FWHM}$), while we used log $f$ = 0.65 $\pm$ 0.12 ($f$ = 4.47) for $\sigma$-based mass ($M_{ BH}^\sigma$). These average $f$ values are derived in \citet{Woo2015} by comparing the reverberation-mapped AGN and quiescent galaxies in the $M_{ BH}$-$\sigma^*$ plane, where $\sigma^*$ is the stellar velocity dispersion of the host galaxy. \citet{Woo2015} combine classical and pseudo bulges in determining the best-fit $M_{ BH}$-$\sigma^*$ relation. The use of the different calibration factors, $f$, derived by \citet{Ho2015} for the two bulge types, yields similar results within $\sim$0.15 dex.

We estimate $R_{ BLR}$ from
\begin{ceqn}
\begin{equation}
 {\rm log} (R_{BLR}) = K + \alpha {\rm log} (\lambda L_{\lambda}(5100\, \angstrom ))
\label{R_BLR}
\end{equation}
\end{ceqn}
where $\alpha$ = 0.519 is the slope of the power-law relationship between $R_{BLR}$ and $\lambda$$L_{\lambda}$ (5100\AA) and $K$ = $-$21.3 is the zero point \citep{Bentz2009}.

Table~\ref{tab_derived} presents black hole masses derived using both the H$\beta$ line width, $M_{BH}^{FWHM}$, and velocity dispersion, $M_{ BH}^\sigma$ for our ERQ sample. Also listed are Eddington ratios (i.e. $L_{bol} / L_{Edd}$, where $L_{Edd}$ is the Eddington luminosity).
Given the uncertainties on the black hole masses, we adopt the average value of $L_{Edd}$ calculated using $M_{BH}^{FWHM}$ and $M_{ BH}^\sigma$.
We use measured values of $L_{\lambda}$(5100\AA) uncorrected for extinction. If the typical extinction in ERQs is $\sim$2 magnitudes (see above), then the corrected black hole masses should be typically $\sim$2.5 times larger than listed in Table~\ref{tab_derived}. The Eddington luminosities scale directly with black hole mass and, therefore, the corrected Eddington ratios should be typically $\sim$2.5 times smaller than the vales in Table~\ref{tab_derived}.

\subsection{Radio Properties}
\label{radio}

The radio properties of our sample and many more ERQs are discussed extensively by \citet{Hwang2018}. 
The radio-loudness of ERQs cannot be measured by the standard ratio of observed radio to blue-visible fluxes \citep{Kellermann1989} because ERQs have large but uncertain amounts of blue-visible extinction. Therefore, \citet{Hwang2018} use a definition for radio-loudness based on radio luminosity $\nu$$L_{\nu}$[5 GHz] $>$ 10$^{41.8}$ erg s$^{-1}$ for quasars with [O{\sevensize III}] luminosities of order $\sim$10$^{10}$ L$_{\odot}$ \citep[from][]{Xu1999}. The radio luminosities at 5 GHz are computed using the following equation:
\begin{ceqn}
\begin{equation}
 \nu L_{\nu} = 4 \pi D_L^2 (1 + z)^{-1-\alpha} (\nu/\nu_{obs})^{1+\alpha} \nu_{obs}F_{\nu_{obs}}
\end{equation}
\end{ceqn}
where $D_L^2$ is the luminosity distance at redshift $z$, $\alpha$ is the spectral index and $F_{\nu_{obs}}$ is the observed flux density. This analysis indicates that ERQs are mostly radio-quiet (RQ) objects, with a radio-loud (RL) fraction of $\sim$8 per cent, comparable to the general quasar populations at similar $z$ \citep{Jiang2007}.

With their extreme [O{\sevensize III}] kinematics, ERQs are found to lie on the extension of the $w_{90}$-radio luminosity relationship of low-redshift, less extreme quasars \citep{Zakamska2014}. While the nature of radio emission from radio-quiet quasars remains poorly understood \citep[for a review see][]{Panessa2019}, this relationship is consistent with radio emission being a bi-product of shocked winds \citep{Hwang2018}.

\begin{table*}
\caption[]{Derived quantities, including [O{\sevensize III}] kinematic parameters $w_{90}$, v$_{50}$ v$_{98}$, and v$_r $ (Section~\ref{O3kinematics}); quasar bolometric luminosities, $L_{bol}$, derived from observed rest-frame 3.45 micron flux; observed [O{\sevensize III}] luminosities, $L$([O{\sevensize III}]), uncorrected for extinction; BH masses derived from the H$\beta$ FWHM, $M^{ FWHM}_{BH}$, and $\sigma$, $M^{\sigma}_{BH}$, using measured $L$(5100\AA) values uncorrected for extinction; Eddington ratios, $L_{bol}$/$L_{edd}$, using the average value of $L_{edd}$ derived using $M^{ FWHM}_{BH}$ and $M^{\sigma}_{BH}$ uncorrected for extinction. Correcting for a typical extinction of $\sim$2 mags at 5100\AA\,in ERQs would increase the black hole masses by a factor of $\sim$2.5 and decrease the Eddington ratios by the same $\sim$2.5 compared to the values listed here (see Section~\ref{luminosity}).}
\begin{center}
\label{tab_derived}
\begin{tabular}{ c c c c c c c c c c c}
\hline
\hline

 Object Name  & $w_{90}$ & v$_{50}$ & v$_{98}$ & $\langle {\rm v}_r\rangle$ & log $L_{bol}$ & log $L$([O{\sevensize III}]) & $M^{ FWHM}_{BH}$ & $M^{\sigma}_{BH}$ & $L_{bol}/ L_{Edd}$ \\
     & [km s$^{-1}$] & [km s$^{-1}$] & [km s$^{-1}$] & [km s$^{-1}$] & [erg s$^{-1}$] & [erg s$^{-1}$] & [10$^9$M$\odot$] & [10$^9$M$\odot$] & \\

 \hline
J000610.67$+$121501.2 & 6206.38 & $-$2289.67 &$-$6092.24 & 5363.62 & 47.86 & 43.66 & 4.30 & 3.19& 1.55\\
J001120.22+260109.2  & 3669.45  & $-$833.88 &$-$3749.36 & 2680.58 & 47.36 & 43.79 & 1.62 & 4.45 & 0.76\\
J013413.22$-$023409.7& 4072.42 & $-$674.11 &$-$3192.29 & 2546.47 & 47.08 & 43.40 & 3.32 & 6.29 & 0.22 \\
J020932.15+312202.7  & 2445.93 & $-$263.05 &$-$1992.9 & 1247.67 & 46.98 & 43.77 & 0.25 & 0.18 &  3.53 \\
J080547.66+454159.0  & 5471.48 & $-$1801.32 &$-$5165.58 & 4455.31 & 47.29 & 43.52 & 0.82 & 0.60 &  2.22 \\
J082618.04+565345.9  & 4123.97 & $-$870.32 &$-$4142.75 & 2865.04 & 46.76 & 43.57 & 2.39 & 1.77 & 0.22\\
J082653.42+054247.3 & 3645.44 & $-$315.19 &$-$3307.6 & 2258.40 & 47.51 & 44.53 & 0.76 & 0.55 & 4.03  \\
J083200.20+161500.3  & 5649.39 & $-$1562.12 &$-$5283.29 & 4240.18 & 47.54 & 43.96 & 3.34 & 2.40 & 0.98\\
J083448.48+015921.1 & 5378.64 & $-$1270.74 &$-$5079.2 & 3595.42 & 47.64 & 44.38 & 3.06 & 2.29 & 1.32  \\
J091303.90$+$234435.2 & 3100.83 & $-$36.01 &$-$2283.67 & 1336.64 & 46.97 & 43.93 & 0.99 & 0.75 & 0.87 \\
J093226.93+461442.8 & 2724.34 & $-$480.07 &$-$2552.3 & 1891.83 & 47.09 & 44.30 & 1.22 & 2.53 &  0.59\\
J095823.14+500018.1 & 5434.86 & $-$2137.1 &$-$5781.66 & 5046.84 & 47.34 & 43.89 & 2.62 & 1.94 & 0.79 \\
J101324.53+342702.6 & 3240.37 & $-$163.39 &$-$2278.94 & 1443.25 & 47.42 & 44.40 & 0.59 & 2.68 &  2.17 \\
J102541.78$+$245424.2 & 6703.55 & $-$2613.82 &$-$6702.18 & 6079.92 & 47.84 & 44.08 & 2.59 & 8.89 & 1.38 \\
J103146.53+290324.1 & 3621.70 & $-$1052.04 &$-$3290.2 & 2769.58 & 47.17 & 43.22 & 7.17 & 5.31 & 0.24 \\
J113834.68+473250.0 & 4135.56 & $-$777.47 &$-$4002.6 & 2871.40 & 47.15 & 43.95 & 1.65 & 1.20 & 0.81 \\
J121704.70+023417.1 & 2661.69 & $-$181.88 &$-$2518.88 & 1618.88 & 47.36 & 43.82 & 4.60 & 11.4 & 0.28 \\
J123241.73+091209.3 & 6336.17 & $-$1084.06 &$-$5579.6 & 4419.53 & 47.76 & 44.28 & 4.16 & 3.22 & 1.26  \\
J134254.45+093059.3  & 4479.28 & $-$882.75 &$-$4609.54 & 3084.26 & 47.13  & 43.71 & 2.38  & 1.75 & 0.53 \\
J134800.13$-$025006.4 & 4659.62 & $-$1424.80 &$-$4296.34 & 3650.95 & 47.09 & 43.33 & 3.62 & 2.71 & 0.32 \\
J135608.32+073017.2 & 2819.69 & $-$339.47 &$-$2089.49 & 1642.32 & 46.87 & 43.41 & 1.83 & 1.35 & 0.38 \\
J155057.71+080652.1 & 2979.31 & $-$2027.31 &$-$4286.54 & 4405.48 & 47.02 & 44.09 & 2.54 & 5.65 & 0.24 \\
J160431.55+563354.2 & 4028.07 & $-$342.49 &$-$3006.38 & 2595.56 & 47.24 & 44.17 & 1.93 & 1.16 & 0.96 \\
J165202.64+172852.3 & 2749.33 & $-$277.68 &$-$2033.68 & 1780.10 & 47.73 & 44.57 & 2.46 & 5.62 & 1.24 \\
J215855.10$-$014717.9 & 3751.88 & $-$1083.88 &$-$3060.08 & 2858.23 & 47.19 & 43.72 & 3.52 & 8.75 & 0.25 \\
J221524.00$-$005643.8 & 4195.56 & $-$399.29 &$-$2872.28 & 2615.98 & 47.13 & 43.96 & 4.38 & 4.71 & 0.23 \\  
J232326.17$-$010033.1 & 7227.95 & $-$714.75 &$-$5479.84 & 4612.67 & 47.08 & 43.47 & 2.01 & 1.45 & 1.25 \\ 
J232611.97+244905.7 & 3079.82 & $-$165.08 &$-$2288.39 & 1979.63 & 46.85 & 43.54 & 2.34 & 1.75 & 0.28\\

\hline
\\
\end{tabular}
\end{center}
\end{table*}

\begin{figure*}

 \includegraphics[width=\textwidth]{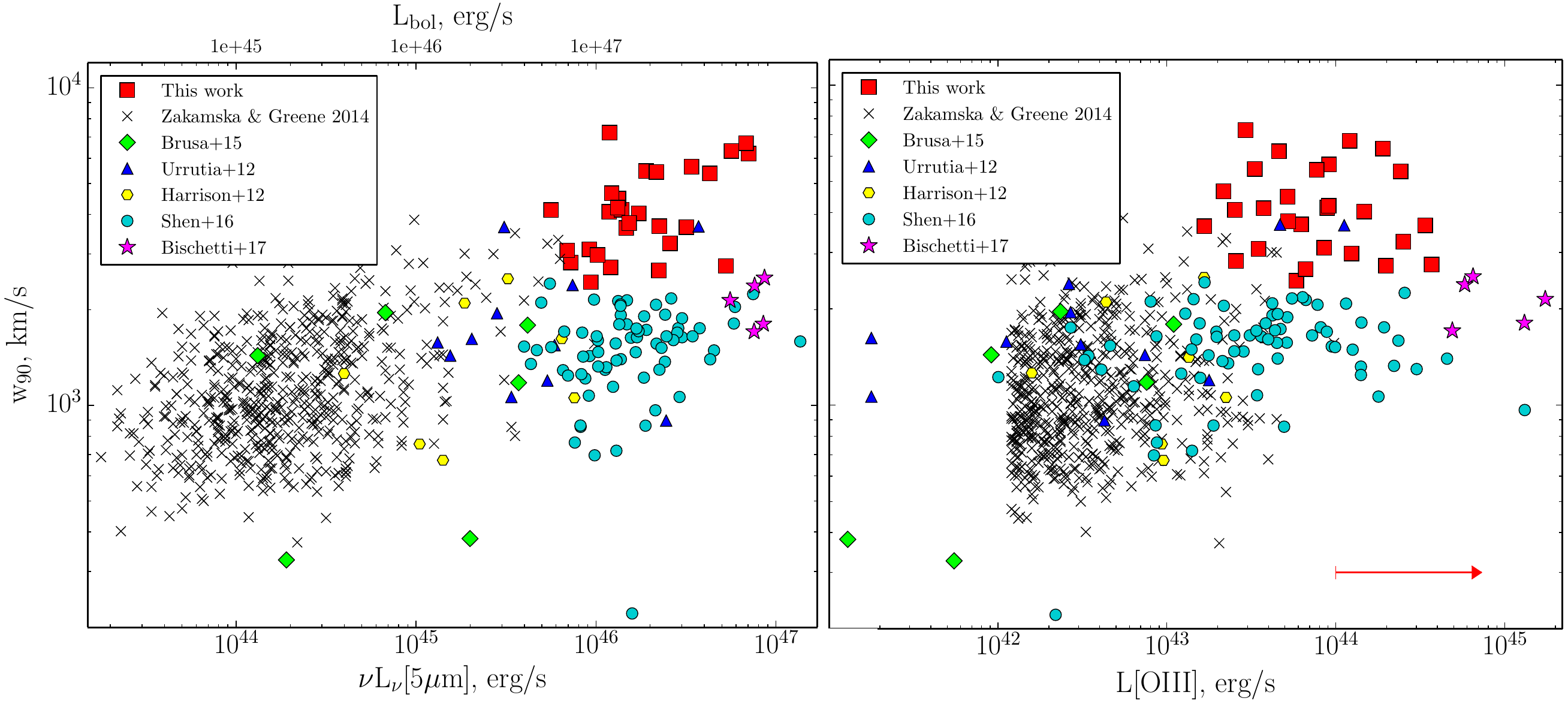}

 \caption{[O{\sevensize III}] kinematics as a function of mid-infrared luminosities (left) and [O{\sevensize III}] luminosities (right) for the objects presented in this paper
(red squares, median $w_{90}$ = 4050 km s$^{-1}$) compared with those of type 1 quasar samples: 1.5 $< z <$ 3.5 quasars from \citet{Shen2016} shown with cyan circles (median $w_{90}$ = 1568 km s$^{-1}$); five $z \approx$ 2.3 - 3.5 quasars from \citet{Bischetti2017} shown with magenta stars (median $w_{90}$ = 2143 km s$^{-1}$). We also show red and type 2 quasar samples: $z <$ 1 type 2 quasars from \citet{Zakamska2014} shown with black crosses (median $w_{90}$ = 1060 km s$^{-1}$); X-ray selected obscured quasars at $z \sim$ 1.0 - 1.5 from \citet{Brusa2015} shown with green diamonds (median $w_{90}$ = 1300 km s$^{-1}$); infrared-selected red quasars at $z <$ 1 from \citet{Urrutia2012} shown with blue triangles (fitting parameters for [O{\sevensize III}] emission for these
objects are published by \citet{Brusa2015}; median $w_{90}$ = 1580 km s$^{-1}$); submm-selected $z \sim$ 2 active galaxies from \citet{Harrison2012} shown with yellow hexagons (median $w_{90}$ = 1330 km s$^{-1}$).  The sharp cutoff for black points in the right panel
is a selection effect, as only objects with $L$([O{\sevensize III}]) $>$ 10$^{8.5}$ L$\odot$ were analyzed by \citet{Zakamska2014}. The red arrow (bottom right of the right panel) represents the typical extinction correction for ERQs.}
 \label{compare_others}
\end{figure*}

\section{Results \& Comparison to Other Studies}  
\label{results}

Extremely broad and blueshifted [O{\sevensize III}] emission lines, indicative of high velocity outflows, are an important common feature of ERQs. Our new observations reveal a previously unexplored range of [O{\sevensize III}] line velocities, with FWHM[O{\sevensize III}]  reaching $\sim$5750 km s$^{-1}$, $w_{90}$ up to $\sim$7200 km s$^{-1}$, and maximum outflow speeds v$_{98}$ up to 6700 km s$^{-1}$. 

Figure~\ref{compare_others} plots [O{\sevensize III}] width, represented by $w_{90}$, as a function of mid-infrared luminosity (left panel) and [O{\sevensize III}] luminosity (right panel). Figure~\ref{compare_others} resembles Figure 9 in \citet{Zakamska2016}, with some of the same data from \citet{Zakamska2014}, \citet{Harrison2012}, \citet{Urrutia2012} and \citet{Brusa2015}. Sources from the sample by \citet{Zakamska2014} are type 2 quasars, which are a useful for comparison to type 1s because traditional NLRs that produce [O{\sevensize III}] emission are extended far beyond the dusty torus (or other nuclear obscurer). Figure~\ref{compare_others} adds our new data for ERQs, including re-measurements of the 4 ERQs in \citet{Zakamska2016}, plus luminous type 1 quasars from \citet{Shen2016} and \citet{Bischetti2017}. The latter two samples are primarily blue quasars with redshifts and luminosities similar to the ERQs; only one of them from \citet{Bischetti2017} is an ERQ according to its i$-$W3 color. We estimate their 5 $\mu$m luminosities using W3 photometry and extrapolate to the rest-frame 5 $\mu$m assuming a $L_{\lambda} \sim \lambda^{-0.65}$ slope (from the \citealp{Polletta2007} QSO1 SED). We recreate their [O{\sevensize III}] emission line profiles using the published fit parameters to measure the kinematic parameters consistently in all samples, as described in Section~\ref{O3kinematics}. We also calculate their $L$([O{\sevensize III}]) consistently with our targets (see Section~\ref{luminosity}). To this aim, we estimate the luminosity at 5007\AA\ from the published luminosity at 5100\AA, and assuming a $L_{\lambda} \sim \lambda^{-0.65}$ slope. Then, we multiply $L$ at 5007\AA\ by the [O{\sevensize III}] REW measured from the recreated [O{\sevensize III}] emission line profiles.

Another useful data set for comparison to ERQs is the low redshift ($z <$ 0.85) quasar sample drawn from SDSS DR7 by \citet{Shen2011}. Their spectral fits are not published so we cannot reproduce their [O{\sevensize III}] emission line profiles and measure the $w_{90}$. However, those objects would populate lower-left corner of the $w_{90} - \nu L_{\nu}$[5$\mu$m] plot at low luminosities (typically $\sim$10$^{44}$-10$^{45}$ erg/s) and narrow line profile widths ($\lesssim$1000 km s$^{-1}$). They follow the general trend described by \citet{Shen2016} for increasing [O{\sevensize III}] line widths in more luminous quasars. This trend is also evident in Figure~\ref{compare_others} \citep[see also][]{Zakamska2016}. However, it is important to note that the [O{\sevensize III}] line widths in ERQs are substantially offset toward larger values (typically by factors of 2 to 4) compared to blue quasars with similar luminosities. 

\begin{figure*}
 \includegraphics[width=\textwidth]{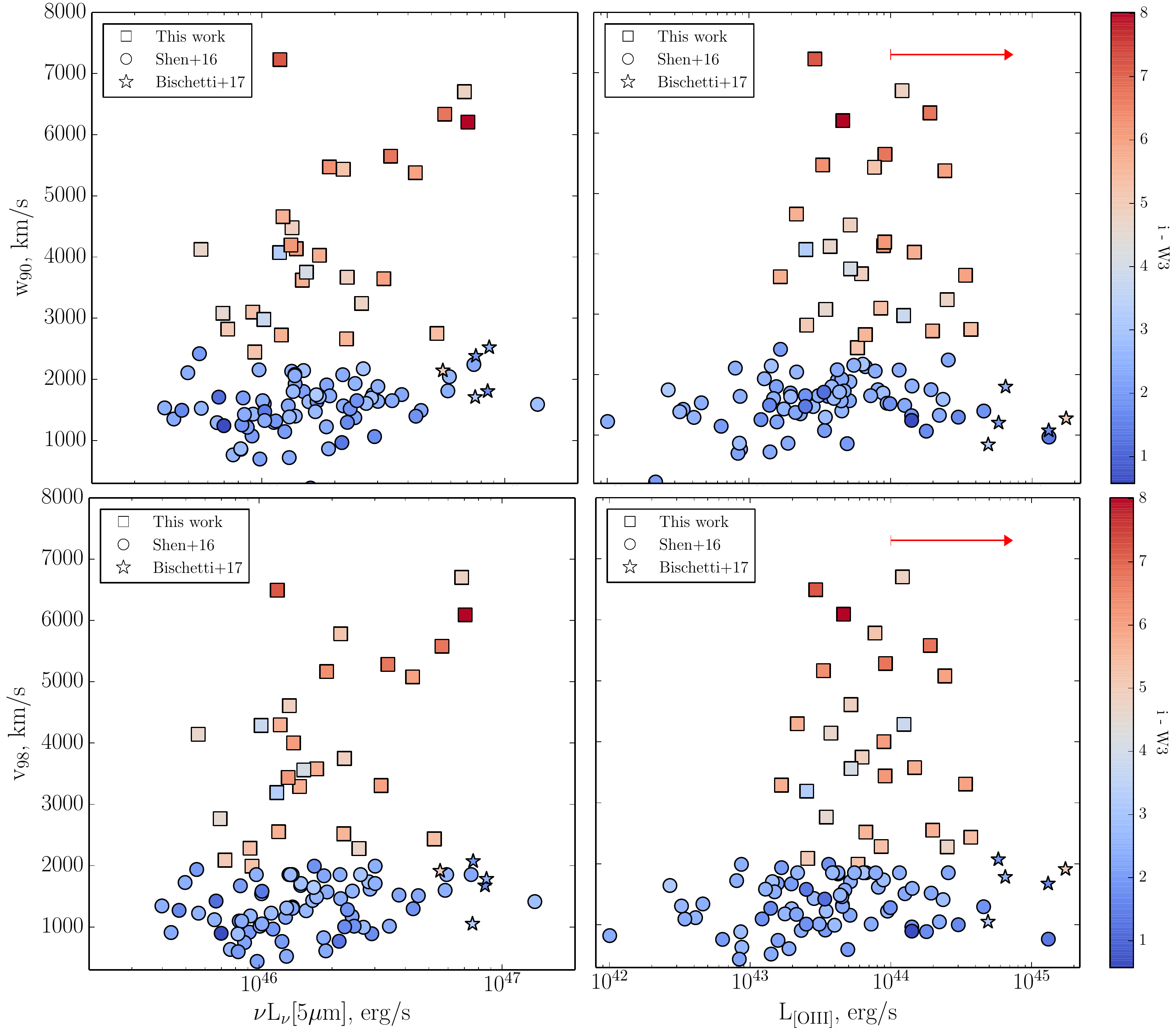}

 \caption{[O{\sevensize III}] kinematics, as represented by $w_{90}$ (top row) and v$_{98}$ (bottom row), as a function of mid-infrared luminosities (left) and [O{\sevensize III}] luminosities (right) for the quasars in this study (squares), 1.5 $< z <$ 3.5 luminous quasars from \citet{Shen2016} (circles) and the five $z \approx$ 2.3 - 3.5 luminous quasars from \citet{Bischetti2017} (stars). All the symbols are color-coded indicating their i$-$W3. The red arrows (top right of the right panels) represent the typical extinction correction for ERQs.}
 \label{L5_LO3_vel}
\end{figure*}

The top row Figure~\ref{L5_LO3_vel} presents an expanded view of Figure~\ref{compare_others} for just the ERQs in our study and the luminous (mostly) blue quasars from \citet{Shen2016} and \citet{Bischetti2017}. The symbols are color-coded to denote the i$-$W3 color of each quasar. The bottom row of Figure~\ref{L5_LO3_vel} shows similar results for v$_{98}$, which is a more direct indicator of the outflow speeds. If we consider only the normal blue quasars in this figure, e.g., from \citet{Shen2016} and \citet{Bischetti2017}, we see that $w_{90}$ and v$_{98}$ both display a weak dependence on rest-frame 5 $\mu$m quasar luminosity and [O{\sevensize III}] line luminosity. However, there is a much stronger relationship to i$-$W3 color, such that ERQs typically have $\sim$2 to $\sim$4 times broader and more blueshifted [O{\sevensize III}] lines than blue quasars at the same luminosity. The quasars in our sample that are not ERQs because they have i$-$W3 $<$ 4.6 (light blue squares in Figure~\ref{L5_LO3_vel}) are very rare quasars with ERQ-like emission-line properties \citep[Section 2 and][]{Hamann2017}. Figure~\ref{L5_LO3_vel} shows that they also have [O{\sevensize III}] kinematics like the ERQs, strongly suggesting that they are physically similar to ERQs and unlike normal blue quasars. 

\begin{figure*}
\includegraphics[width=\textwidth]{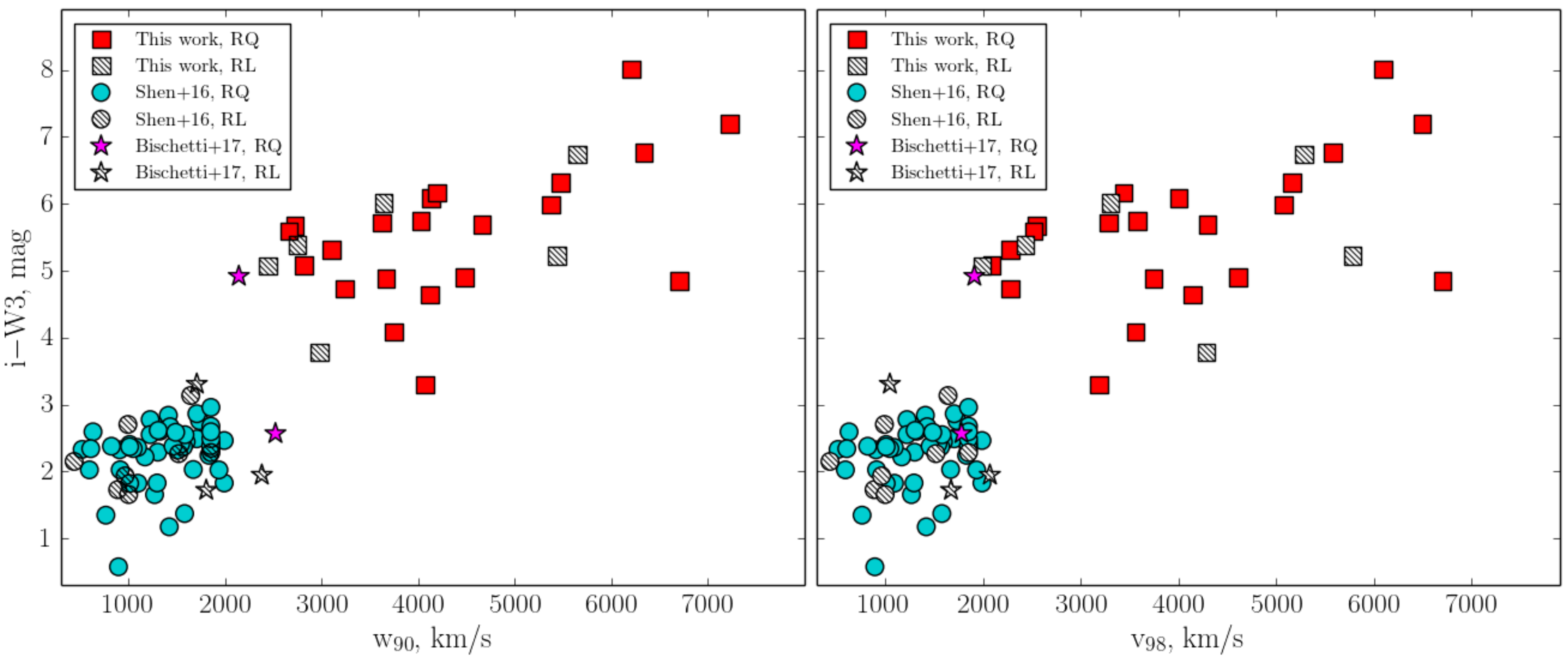}

 \caption{i$-$W3 color as a function of [O{\sevensize III}] kinematics, as represented by $w_{90}$ (left) and v$_{98}$ (right). Filled and open hashed symbols represent radio quiet (RQ) and radio loud (RL) object, respectively. We omit from the plot 11 quasars from \citet{Shen2016} that are not covered by FIRST and J232611.97+244905.7 from our sample, for which the radio flux is unknown. The exclusion of these targets does not alter the results.}
 \label{scatter_vel}
\end{figure*}

Figure~\ref{scatter_vel} plots directly the relationships of $w_{90}$ and v$_{98}$ to i$-$W3 color for the same three quasar samples as Figure~\ref{L5_LO3_vel}, but now with separate filled and open hashed symbols representing RQ and RL objects, respectively. We determine the radio-loudness of the quasars from \citet{Shen2016} and \citet{Bischetti2017} in the same manner as our ERQ sample, as described in Section \ref{radio}. 
In particular, we compute their radio luminosities at 5 GHz using the observed integrated flux density at 20 cm from FIRST\footnote{The version of the FIRST source catalog used is http://sundog.stsci.edu/first/catalogs/readme.html.} and assuming $\alpha$ = $-$0.5 \citep{Richards2006}. 
We also compute the radio-loudness of these blue quasars using the ratio between the rest-frame flux densities at 5 GHz and 2500\AA, i.e., R = $f_{\nu}$(5 GHz)/$f_{\nu}$(2500\AA) \citep{Sramek1980}, for comparison. The use of the two different methods to identify RL targets yields similar results. For example, 13\% of the sample from \citet{Shen2016} are found to be RL using the threshold $\nu L_{\nu}$[5 GHz] $>$ 10$^{41.8}$ erg s$^{-1}$, while 8\% adopting the traditional R $>$ 10 \citep{Kellermann1989} as criterion for radio-loudness. 

Two results are immediately evident from Figure~\ref{scatter_vel}. First, the outflow kinematics are strongly dependent on i$-$W3 color (much more so than the weak luminosity trend among blue quasars in Figure~\ref{L5_LO3_vel}). Second, we do not see any relationship to the radio properties in the different samples. In particular, radio-loudness does not favor larger $w_{90}$ nor v$_{98}$ values and it cannot account for the very different [O{\sevensize III}] kinematics between ERQs and blue quasars at similar luminosities. 

\begin{figure*}
\includegraphics[width=\textwidth]{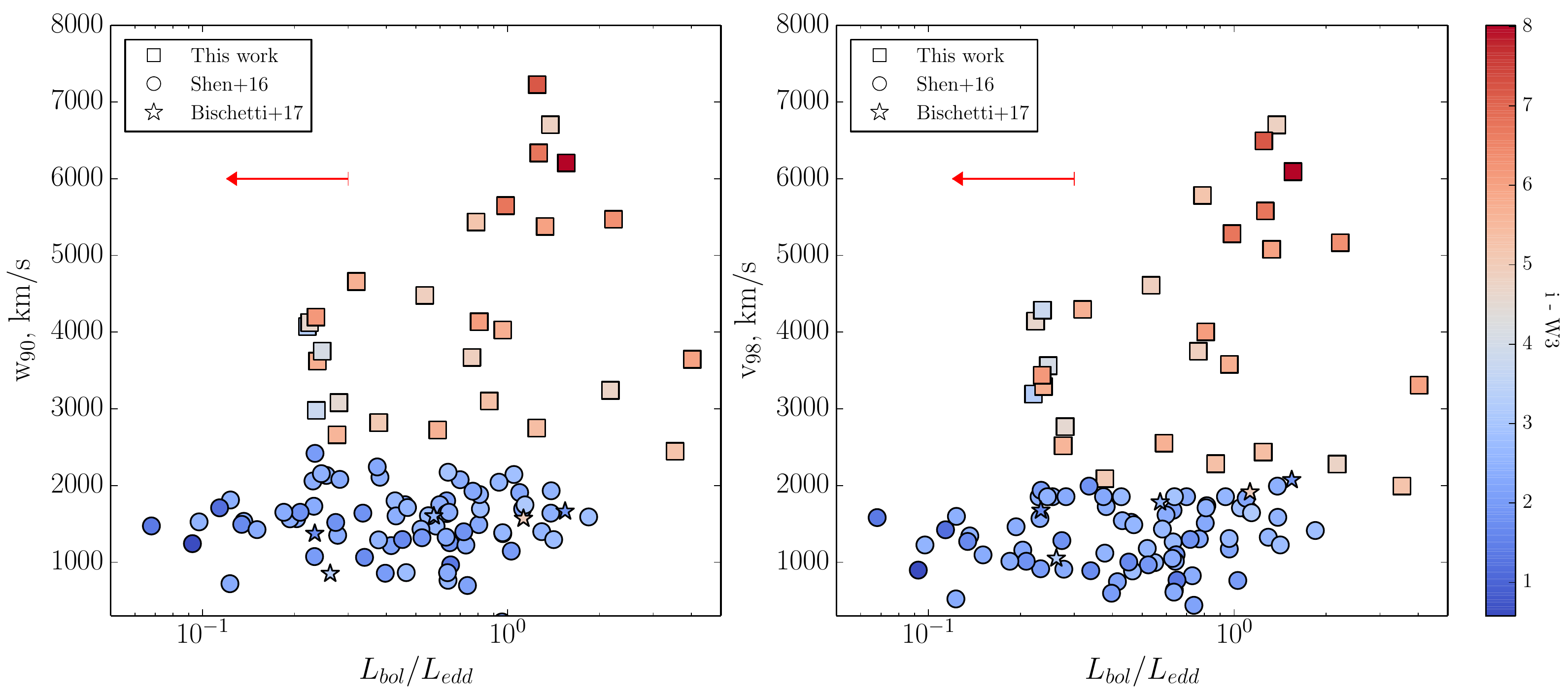}

 \caption{[O{\sevensize III}] kinematics, as represented by $w_{90}$ (left) and v$_{98}$ (right) as a function of the Eddington ratio for the quasars in this study (squares), 1.5 $< z <$ 3.5 luminous quasars from \citet{Shen2016} (circles) and the five $z \approx$ 2.3 - 3.5 luminous quasars from \citet{Bischetti2017} (stars). All the symbols are color-coded indicating their i$-$W3. The red arrows represent the typical extinction correction for ERQs.}
 \label{vel_Edd}
\end{figure*}

Figure~\ref{vel_Edd} explores the relationship of the quasar Eddington ratios (Table~\ref{tab_derived}) to the [O{\sevensize III}] outflow kinematics measured by $w_{90}$ and v$_{98}$, again for the same three quasar samples in Figures~\ref{L5_LO3_vel}$-$\ref{scatter_vel} roughly matched in luminosity. We estimate the Eddington ratios for the luminosity-matched blue quasars consistently to our targets (see Section~\ref{luminosity}). The red arrows represent the typical extinction correction for ERQs estimated as follows.
The median ERQ SED is suppressed at $\sim$5000\AA\ by about 2 magnitudes relative to normal/blue quasars (Section~\ref{luminosity}) and, therefore, we expect the values of the luminosity at 5100\AA\ to be too low by a factor of $\sim$6.3. If the typical extinction in ERQs is $\sim$2 magnitudes, then the corrected $R_{BLR}$ (Eq.~\ref{R_BLR}) is $\sim$2.5 times larger, and therefore $M_{BH}$ in Eq.~\ref{M_BH} is $\sim$2.5 times larger as well. 
Considering such correction, both panels show no significant relationship between [O{\sevensize III}] kinematics and the Eddington ratio. In addition, the median value of the Eddington ratio corrected for extinction in ERQs, 0.32, is similar to the median ratio 0.52 in the combined blue quasar samples in Figure~\ref{vel_Edd}. 

The results in this figure do not support speculation elsewhere (\citealt{Hamann2017}, Zakamska et al., submitted) that ERQs have higher accretion rates (larger $L_{bol}/L_{edd}$) that could naturally produce faster and more powerful outflows compared to normal/blue quasars. However, there are significant uncertainties in both the measurements and assumptions, and there might be systematic offsets between the ERQs and normal blue quasar samples. For example, the ERQs might have \textit{intrinsically} different SEDs than blue quasars that would affect the $L_{bol}$ estimates. One possibility is that the mid-IR emissions are optically thick and thereby suppressed in ERQs \citep[e.g., as in][]{Pier1992}. In our study, this would lead to underestimates of $L_{bol}$ and $L_{bol}/L_{edd}$ in the ERQs. Another possibility is that the H$\beta$ kinematics we measure in ERQs are affected by outflows in the broad emission line regions (Hamann et al., in prep.). We do not expect this to be a serious problem because the median FWHM of H$\beta$ in our ERQ sample is 4075 km s$^{-1}$ (e.g., in the normal range of type 1 AGN and quasars \citealt{Hao2005,Steinhardt2013}); however, we cannot exclude the possibility that some of the measured H$\beta$ line widths come from outflows. This could again lead us overestimated black hole masses and underestimated Eddington ratios. 

Conversely, the analysis in Zakamska et al. (submitted) might favor overestimated Eddington ratios for ERQs. Their results are based on measurements of the rest-frame B-band luminosities ($L_B$) of ten ERQ host galaxies observed with the Hubble Space Telescope, which they then use to estimate black hole masses from the $L_B$--$M_{BH}$ scaling relations in normal galaxies \citep{Kormendy1995, Kormendy2013}. Those scaling relations might not apply to ERQs, e.g., if ERQs are characteristically a young quasar population residing in young host galaxies. Even if the scaling relations do apply in principle to ERQs, they might produce overestimates of the Eddington ratios in ERQs. For example, if the measured values of $L_B$ in ERQ host galaxies are significantly affected by dust extinction across the galaxies (as they are toward the central quasars), the scaling relations would underestimate the true black hole masses overestimate the Eddington ratios in ERQs. Another potential bias is that ERQs are necessarily very luminous to be included as extremely red objects in the BOSS quasar catalog \citep{Hamann2017}. Selecting luminous quasars from the overall quasar population tends to favor larger black hole masses, which could then favor larger values of $M_{BH}$ for a given $L_B$ compared to the average or median in $L_B$--$M_{BH}$ scaling relations. This would again mean that the scaling relations applied to ERQs would underestimate the black hole masses and overestimate the Eddington ratios. 

It is not possible to quantify these various uncertainties and potential biases without additional data and detailed analysis. That work is beyond the scope of our current study. We favor the results as shown in Figure~\ref{vel_Edd}, with a nominal expected extinction correction applied to the ERQ points, indicating that ERQs do not have anomalously large Eddington ratios compared to normal/blue quasars at similar redshifts and luminosities. 

\begin{figure*}
\includegraphics[width=\textwidth]{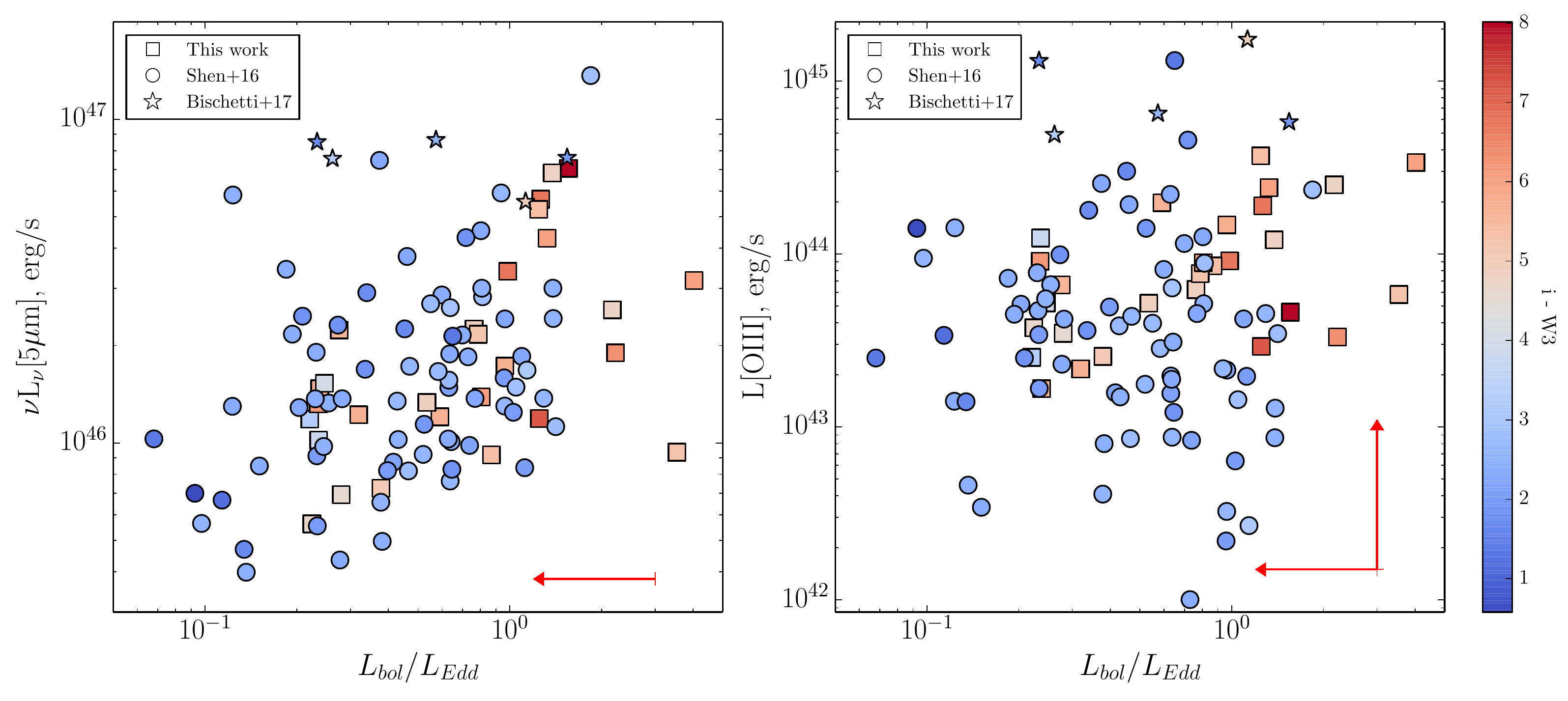}

 \caption{Mid-infrared luminosity (left) and [O{\sevensize III}] luminosity (right) as a function of the Eddington ratio for the quasars in this study (squares), 1.5 $< z <$ 3.5 luminous quasars from \citet{Shen2016} (circles) and the five $z \approx$ 2.3 - 3.5 luminous quasars from \citet{Bischetti2017} (stars). All the symbols are color-coded indicating their i$-$W3. The red arrows represent the typical extinction correction for ERQs.} 
 \label{L5_LO3_Edd}
\end{figure*}

Figure~\ref{L5_LO3_Edd} compares the quasar mid-infrared luminosities (left panel) and [O{\sevensize III}] luminosities (right panel) to their  Eddington ratios. There is a weak dependence of Eddington ratio on infrared luminosity, as expected if the infrared luminosity is tied to the bolometric luminosity and more luminous quasars naturally favor larger Eddington ratios as a selection effect. There is perhaps a similar weak trend for larger Eddington ratios with larger $L$([O{\sevensize III}]) if we consider only the ERQs (right panel). However, there is no correlation with [O{\sevensize III}] luminosity with the luminous blue quasars included.

\section{Discussion}
\label{discussion}

The main result of our study is that ERQs routinely exhibit [O{\sevensize III}]~$\lambda$4959,5007 emission lines with unprecedented velocity widths, with $w_{90}$ ranging between 2053 and 7227 km s$^{-1}$, and unprecedented outflow speeds, with v$_{98}$ ranging from 1992 to 6702 km s$^{-1}$ (Section~\ref{analysis} and Tables~\ref{tab:fits} and \ref{tab_derived}). Comparisons to previous work on [O{\sevensize III}]  outflows (Section~\ref{results}) reveal that ERQs have the broadest and most blueshifted [O{\sevensize III}] lines ever reported. Figs~\ref{L5_LO3_vel}$-$\ref{L5_LO3_Edd} show directly the different [O{\sevensize III}] line widths and velocities measured in ERQs compared to normal blue quasars roughly matched in luminosity. The median values of $w_{90}$ and v$_{98}$ measured in ERQs are 4050 and 3664 km s$^{-1}$, respectively, compared to luminous blue quasars with 1567 and 1309 km s$^{-1}$ in the \citet{Shen2016} sample or 2143 and 1783 km s$^{-1}$ in the \citet{Bischetti2017} sample. These differences in the [O{\sevensize III}] kinematics are clearly related to i$-$W3 color and {\it{not}} the quasar luminosities. If we consider only the reddest ERQs in our sample, with i$-$W3 $>$ 5.6, the median values of $w_{90}$ and v$_{98}$ are even more extreme at 4659 and 4296 km s$^{-1}$, respectively. Another important result is that the faster [O{\sevensize III}] outflows in ERQs are not driven by larger Eddington ratios nor radio-loudness. 

In the following subsections, we use our results above to estimate the energetics (Section~\ref{energetics}) and discuss possible acceleration mechanisms (Section~\ref{mechanism}) for such powerful [O{\sevensize III}] outflows. We conclude with a brief discussion of the implications for quasar feedback and how ERQs might fit into quasar/galaxy evolution schemes (Section~\ref{implications}).

\subsection{Energetics}
\label{energetics}

The physical properties of the ionized outflow can be constrained through the observational parameters of the [O{\sevensize III}] emission line. We use the [O{\sevensize III}] luminosity as a mass tracer with the understanding that it is also sensitive to unknown parameters like the gas temperature, metallicity, ionisation, and electron density. \citet{Cano2012} show that, under nominal ionization and excitation conditions, an order of magnitude estimate of the kinetic power in the [O{\sevensize III}]-emitting gas is given by
\begin{ceqn}
\begin{equation}
\label{E_kin}
 \dot{E}_{k} = 5.17 \times 10^{43} \frac{C \,L_{44}([{\rm O{\scriptstyle III}}]) \, {\rm v}^3_3}{\langle n_{e3} \rangle \, 10^{[{\rm O/H}]} \, R_{out}} \,\, {\rm erg \, s^{-1}}
\end{equation}
\end{ceqn}
where $n_{e3}$ is the electron density in units of 1000 cm$^{-3}$, $C$ = $ \langle n_e \rangle^2 / \langle n^2_e \rangle$ is the condensation factor (i.e., measure the homogeneity of the ionized gas distribution), $L_{44}$([O{\sevensize III}]) is the [O{\sevensize III}] luminosity in units of 10$^{44}$ erg s$^{-1}$, v$_3$ is the outflow velocity in units of 1000 km s$^{-1}$, 10$\rm^{[O/H]}$ is the metallicity in units of solar metallicity and $R_{out}$ is the radius of the outflowing region, in units of kpc. 

We compute $ \dot{E}_{k}$ for the ERQs in our sample according to Eq.~\ref{E_kin}. We assume $C$ $\approx$ 1, [O/H] $\sim$0 (solar metallicity). 
For the estimate of the outflow velocity we use v$_{98}$, inferred from the [O{\sevensize III}] profile. The [O{\sevensize III}] emission line profile in ERQs is a complex blend. The narrower [O{\sevensize III}] components in the fits are relatively broad and in many cases blueshifted relative to $z_{best}$ (Section~\ref{fitting}), and there is no evidence they are linked to the gas in dynamical equilibrium with the host galaxy. Therefore, we use L([O{\sevensize III}]) values of the entire [O{\sevensize III}] $\lambda$5007 line.

$n_e$ is often measured from the emission-line ratio [S{\sevensize II}] $\lambda$6716/$\lambda$6731 which is sensitive to $n_e$. Typical [S{\sevensize II}]-based measures of electron density vary from n$_e \sim$ 100$-$1000 cm$^{-3}$ (e.g. \citealp{Osterbrock2006}). We choose a value of $n_e$ = 200 cm$^{-3}$.  This parameter choice is consistent with similar studies of ionized outflows (e.g. \citealp{Bischetti2017}). For a more typical density of the NLR $n_e$ =10$^3$ cm$^{-3}$ \citep{Netzer2004, Baskin2005} the radius of the [O{\sevensize III}] $\lambda$5007 emitting region is $R_{out}$ $\sim$3 kpc, for a mean ERQ luminosity. However, we assume $R_{out}$ = 1 kpc. This choice is justified by recent observations of two ERQs in our sample performed with W.M. Keck Observatory OSIRIS integral field spectrograph (IFS) with adaptive optics. A preliminary data reduction indicates that the [O{\sevensize III}] emission is spatially unresolved in both sources on scales $\leq$ 1.2 kpc (Perrotta et al. in prep).

\citet{Zakamska2016} used an approach to estimate the kinetic energy of the ionized outflow based on the use of recombination lines \citep{Nesvadba2006}. To this purpose, they assumed the [O{\sevensize III}]/H$\beta$ ratio to be close to its standard value of 10 \citep{Dopita2002} in the extended emission-line region.
The use of this method to derive the kinetic energy associated to the ionized outflow (for additional details see \citealp{Zakamska2016}) yields comparable results.

The left panel of Figure~\ref{Ekin_Lbol} plots the kinetic power of the ionized outflows for the ERQs in this study (squares) and the luminous quasars in the samples from \citet{Shen2016} (circles) and \citet{Bischetti2017} (stars), as function of $L_{bol}$. We compute $ \dot{E}_{k}$ for the luminosity-matched blue quasars consistently to our sample.
In order to give an idea of the potential impact of the uncertainties affecting $\dot{E}_{k}$, we report error bars in Figure~\ref{Ekin_Lbol} (bottom right) estimated by varying $n_e$ (solid line) and $R$ (dashed line). In the first case, the upper bound correspond to the assumption of $n_e$ = 100 cm$^{-3}$ (as in \citealp{Liu2013, Harrison2014, Brusa2015}), while the lower bound correspond to $n_e$ = 1000 cm$^{-3}$ (i.e. the typical value for the NLR, \citealp{Peterson1997}). For what concerns uncertainties in the extension of the outflows, we consider $R_{out}$ = 0.5 kpc as a lower limit and $R_{out}$ = 7 kpc (e.g. \citealp{Bischetti2017}) an upper limit. 

The right panel of Figure~\ref{Ekin_Lbol} shows how the result changes if we adopt $\langle {\rm v}_r\rangle$ as the outflow velocity.
Table~\ref{tab_derived} shows that these flux-weighted outflow velocities $\langle {\rm v}_r\rangle$ are generally $\sim$10 to $\sim$30 percent smaller than the maximum measured velocities v$_{98}$. This has a significant effect on the kinetic power because of the v$^3$ dependence in Eq.~\ref{E_kin}. In principle, the flux-weighted velocities $\langle {\rm v}_r\rangle$ are a better indicator of the speed at which most of the gas is moving and, therefore, a better indicator of the kinetic energy. However, $\langle {\rm v}_r\rangle$ and v$_{98}$ can both be underestimates if there are orientation effects, e.g., if the outflows are axisymmetric and the flow axes are not aimed at the observer. Thus using the larger v$_{98}$ values in the kinetic energy formula might provide a first-order correction for these effects.  

The values of $\dot{E}_{k}$ shown for the blue quasars in Figure~\ref{Ekin_Lbol} \citep[from][]{Shen2016, Bischetti2017} are probably too large for two reasons. First, our use of $L$([O{\sevensize III}]) measured from the full [O{\sevensize III}] $\lambda$5007 line profiles overestimates $\dot{E}_{k}$ for quasars in which a distinct narrow [O{\sevensize III}] component is clearly present and not participating in an outflow. This occurs for most of the quasars in the \cite{Shen2016} sample and inspection of the [O{\sevensize III}] line profiles suggests that our derived $\dot{E}_{k}$ are too large factors of $\sim$2. Second, the actual radii of NLRs in luminous blue quasars are generally larger than our adopted $R_{out}$ = 1 kpc by factors of several \citep{Netzer2004, Baskin2005}, indicating that the true $\dot{E}_{k}$ values should be several times lower (see Eq.~\ref{E_kin}). For example, the [O{\sevensize III}] outflows in two of the WISSH quasars have been spatially resolved to have radial sizes of roughly $R_{out}$ = 7 kpc \citep{Bischetti2017}, which indicates that their $\dot{E}_{k}$ values should be $\sim$0.84 dex lower than shown in Figure~\ref{Ekin_Lbol}. 

In contrast, our $\dot{E}_{k}$ estimates for the ERQs in Figure~\ref{Ekin_Lbol}  are almost certainly too small by factors of several because we use \textit{observed} [O{\sevensize III}] luminosities without extinction corrections. As discussed in Section~\ref{radio}, the amount of extinction at 5007\AA\ in the ERQs is typically $\sim$2 magnitudes (or a factor or 6.3, based on comparisons between the typical SEDs of ERQs and normal blue quasars). Therefore, we expect the $\dot{E}_{k}$ values shown for the ERQs in Figure~\ref{Ekin_Lbol} to be too small by similar factors of 3 to 10 \citep[see also][]{Zakamska2016}. The red arrow in Figure~\ref{Ekin_Lbol} shows the effects of typical extinction correction for ERQs (see Section~\ref{luminosity}).

Figure~\ref{Ekin_Lbol} shows that, even with these systemic biases, the median kinetic power of ERQs is 1.5 dex larger than the median $\dot{E}_{k}$ of blue quasars with comparable luminosity. Correcting for the systemic biases makes the differences substantially larger. The main reasons for this are much larger [O{\sevensize III}] outflow velocities (Figure~\ref{L5_LO3_vel}) and, secondarily, larger [O{\sevensize III}] luminosities (after correcting for dust extinction the median L([O{\sevensize III}]) of ERQs is 1.06 dex larger than that of the other samples) in ERQs.

\subsection{Acceleration mechanism}
\label{mechanism}

In the previous Sections, we have seen that [O{\sevensize III}] outflows in ERQs exhibit extreme velocities unlike any known quasar population. What can be the mechanism of acceleration for such [O{\sevensize III}] outflows?
Various theoretical models have been developed over the last two decades to produce powerful large-scale outflows (e.g. \citealp{Murray2005, King2011, Faucher2012, Zubovas2012, McKinney2014, Thompson2015}). 

One possibility is that ultra-fast nuclear winds (with initial velocity v$_{in}$ $\geq$ 10,000 km s$^{-1}$), usually detected through UV and X-ray absorption features (e.g. \citealp{Weymann1981, Tombesi2015}), cause shocks interacting with the surrounding ISM. If the hot shocked gas does not cool efficiently (i.e. energy-conserving outflow), an expanding hot gas bubble can do work on the ambient gas and drive large-scale outflows (e.g. \citealp{King2011, Faucher2012, Zubovas2012}). This mechanism requires the hot shocked gas to be reasonably well confined to build-up enough pressure and be able to drive powerful galaxy-wide outflows. In this scenario, the velocity of the swept-up material, v$_s$, is generally $\ll$ v$_{in}$. Such shocked wind bubbles can decelerate to a few 100 km s$^{-1}$ (or less). Maximum values of v$_s$ $\sim$1000 km s$^{-1}$ are in good agreement with measurements of outflows in local ultra-luminous infrared galaxies \citep[e.g.][]{Rupke2011, Sturm2011}.

In our case, shocks can either contribute to or govern the [O{\sevensize III}] line excitation. Energetically it is challenging to channel so much energy into [O{\sevensize III}] via shocks \citep[e.g.][]{King2011, Faucher2012}. However, ERQs are remarkable objects with line properties related to extreme physical conditions. Therefore, the possibility of shocks should be re-examined in future analysis.

Another possibility is that outflows are accelerated by radiation pressure on dust grains. The strong correlation between [O{\sevensize III}] properties and  i$-$W3 color seen in Section~\ref{results} suggests that this mechanism may play a major role in accelerating large-scale gas to such high velocities.
This process has been investigated analytically (e.g. \citealp{Murray2005, Thompson2015, Ishibashi2015, Ishibashi2016}), in radiative transfer calculations (e.g. \citealp{Proga2004,  Krumholz2012, Krumholz2013, Bieri2017}) and through cosmological simulations \citep{Debuhr2011, Debuhr2012, Costa2018}. In all cases, IR multi-scattering has been reported as crucial to guarantee enough momentum can be transferred to the surrounding gas.
The optical and UV radiation emitted by quasars is absorbed and re-emitted at IR wavelengths before escaping the galactic nucleus. If the dust opacity is important also at IR frequencies (i.e. IR optical depth $\rm \tau_{IR}$ $>$ 1), and the dusty envelope has a large covering fraction as seen from the central emission source, instead of streaming out, the reprocessed IR photons will scatter multiple times before escaping, thus multiplying the momentum transfer and producing a very efficient coupling between the quasar radiation and the surrounding ISM.
In this scenario, the radiation force can exceed $L$$\rm_{AGN}$/$c$ at most by a factor of $\rm \tau_{IR}$, and approaches this maximum value if the radiation is efficiently confined by optically thick gas. In particular, \citet{Costa2018} show that the IR trapping is efficient as long as the optically thick gas has a high covering fraction and the IR diffusion times are short in comparison to the hydrodynamic response time.

\citet{Hamann2017} present an extensive discussion to explain the unusual SEDs of ERQs, which are surprisingly flat across the rest-frame UV given their red UV to mid-IR colors \citep[see Figure 16 in][]{Hamann2017}. Interestingly, one possibility is that patchy obscuration by small dusty clouds with typical $\sim$90 per cent covering fractions could produce the observed UV extinctions without substantial UV reddening. Another possibility is that the continuum light from the quasar is scattered into our line of sight by the surrounding medium. \citet{Alexandroff2018} found that the rest-frame UV continua of two ERQs have a polarization between 10 and 15 per cent.

\citet{Costa2018} include trapped IR radiation pressure in cosmological radiation-hydrodynamic simulations. They show that this mechanism has the ability to generate large-scale outflows, clear the galactic nucleus and generate low density channels through which the radiation field can escape. Since only the central regions of massive halos are optically thick in the IR, radiation pressure affects the star formation rate more efficiently within the innermost 1 kpc. However, these simulations are also not able to reproduce the typical outflow velocities of ERQs. 
Additional factors to boost the outflow speeds reached in these models might be needed, such as higher IR optical depths, different feedback mechanisms, or higher resolutions to accurately describe the observed outflows.

Interestingly, observations of two ERQs performed with OSIRIS/Keck revealed that [O{\sevensize III}] emission appears compact ($<$ 1.2 kpc, see Section~\ref{energetics}). If this result holds generally for ERQs, it would support a scenario where trapped IR radiation produces more compact and higher-speed [O{\sevensize III}] emission. However, the geometry of the dust obscuration in the host galaxies is not clear, making difficult to estimate the extension of the [O{\sevensize III}] emission. We will explore the size and morphology of [O{\sevensize III}] outflows in ERQs in a forthcoming paper.
In reality, hot shocks due to ultra-fast nuclear winds and trapped IR radiation pressure may co-exist and contribute to the dynamics of the outflows in quasars to somewhat different degrees.

\subsection{Implications for Quasar Feedback}
\label{implications}

Dust-obscured and red quasars provide important tests of galaxy/quasar evolution. They are expected to be young, appearing during the brief blowout/transition phase between the initial dusty starbursts and later normal blue quasars \citep{Urrutia2008, Glikman2012, Assef2015, Banerji2015, Banerji2017}. 

ERQs have sky densities a few percent of luminous blue quasars consistent with a short-lived phase of quasar evolution. They may represent the short-lived ``blowout'' evolution stage that precedes the much more common (and more evolved) luminous blue quasar phase. 
A possible evolutionary scenario is that [O{\sevensize III}] outflows attain their large velocities during a fully obscured phase, when they are well confined by the ambient medium, and can build-up of pressure within the innermost $\sim$1 kpc as the quasar continues injecting energy and momentum into this region via the wind. When the pressure force exceeds the inertia of the surrounding ISM quasar feedback can clear the nucleus and reveal the SMBH as an optical quasar. In the subsequent optical quasar phase, the outflows may decelerate while shock-heating the surrounding ISM, and expanding to larger scales.

ERQs show [O{\sevensize III}] emission lines with $w_{90}$ reaching $\sim$7200 km s$^{-1}$, and maximum outflow speeds v$_{98}$ up to 6700 km s$^{-1}$. These extreme outflow signatures, unprecedented in other luminous quasar samples \citep[Figure~\ref{compare_others}, also][]{Zakamska2014, Shen2016}, correlate with red color strongly (see Figure~\ref{scatter_vel}). This important result suggests that the dust content in the host galaxy may play a crucial role in coupling the energy released by the central accreting SMBH to the ISM efficiently, and explaining the clear distinction of [O{\sevensize III}] widths between ERQs and blue quasars matched in luminosity. 

Our target selection gives priority not only to quasars with extreme red color, but also with REW(C{\sevensize IV}) $>$ 100\AA. However, our sample contains six ERQs with REW(C{\sevensize IV}) $<$ 100\AA\,and they all show [O{\sevensize III}] widths and velocities at least 2 times larger than the median values of luminosity-matched blue quasar samples. We will explore the possible relation between BLR winds (traced by C{\sevensize IV}) and [O{\sevensize III}] winds in a forthcoming paper (Hamann et al. in prep). 

\begin{figure*}
 \includegraphics[width=\columnwidth]{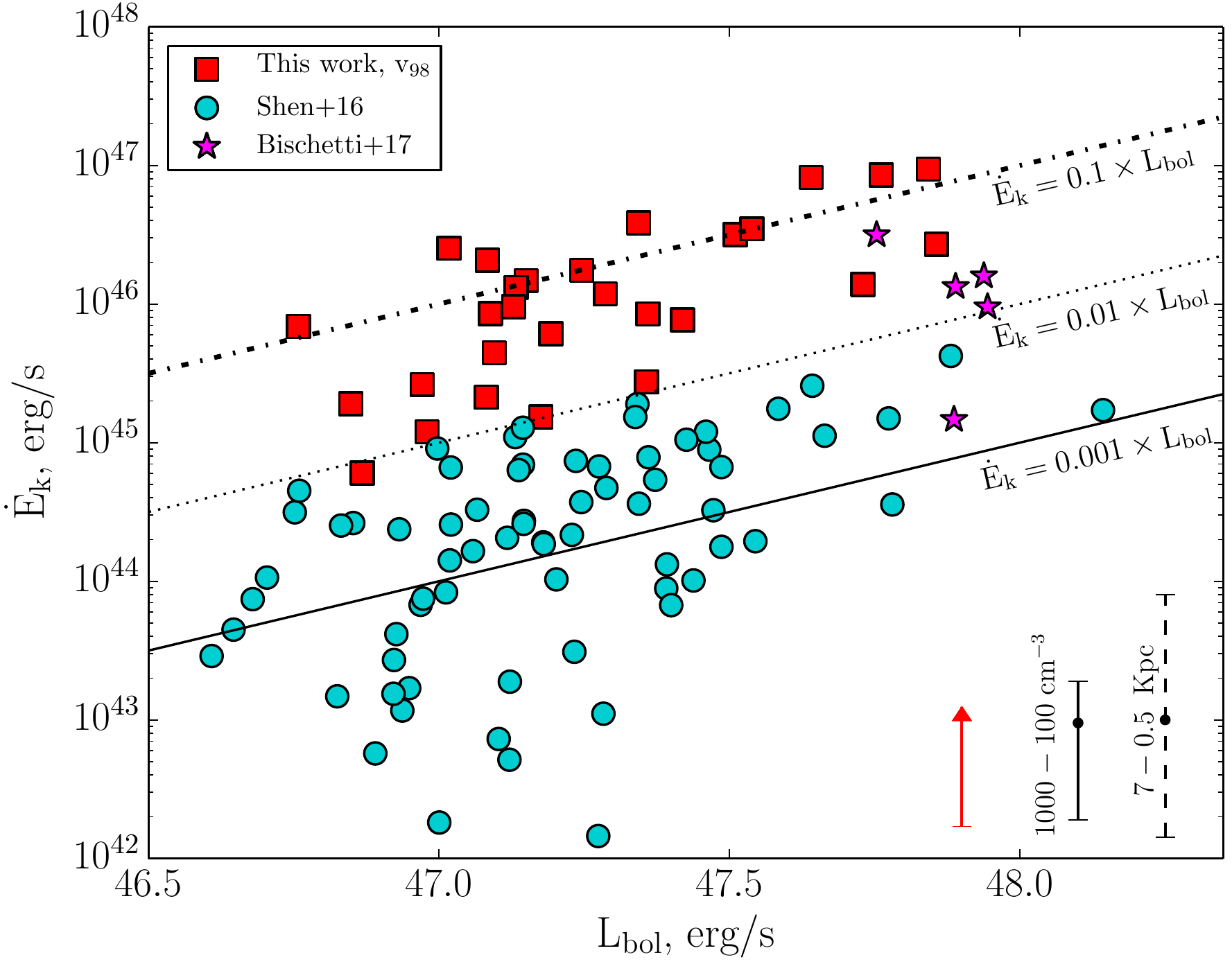}
 \includegraphics[width=\columnwidth]{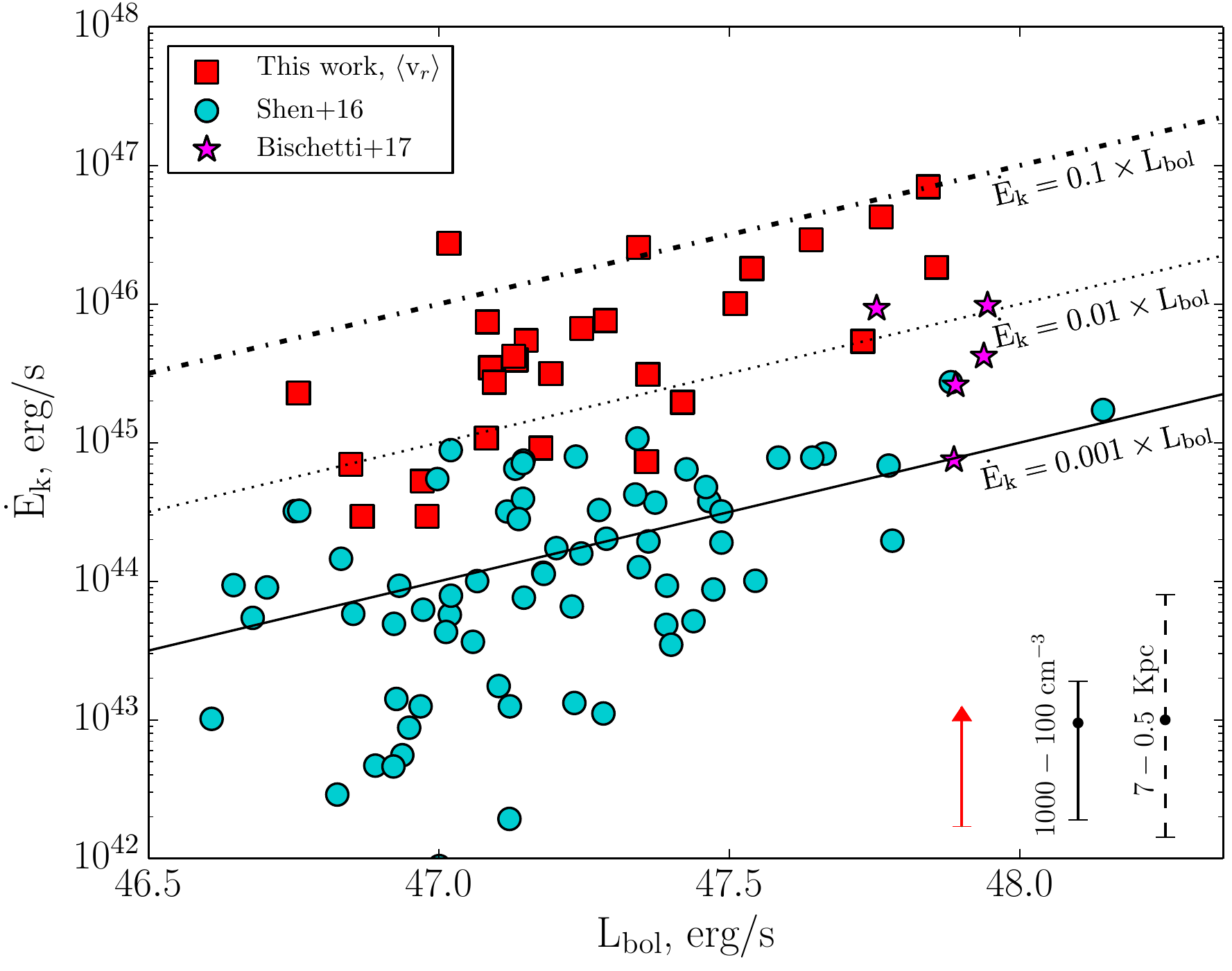}

 \caption{Kinetic power of the ionized outflows as a function of $L_{bol}$. $\dot{E}_{k}$ is calculated adopting v$_{98}$ (left panel) and $\langle$v$_r \rangle$ (right panel). Values obtained for the ERQs (squares)
are compared with 1.5 $< z <$ 3.5 luminous quasars from \citet{Shen2016} (circles) and the five $z \approx$ 2.3$-$3.5 luminous quasars from \citet{Bischetti2017} (stars). Error bars (bottom right) are calculated as described in Section~\ref{energetics}. The dashed-dotted, dotted, and solid lines represent outflow kinetic powers that are 10 per cent, 1 per cent and 0.1 per cent of the quasar luminosity, respectively. }
 \label{Ekin_Lbol}
\end{figure*}

The dashed, dotted, and solid lines in Fig.~\ref{Ekin_Lbol} represent outflow kinetic powers that are 10 per cent, 1 per cent and 0.1 per cent of the quasar bolometric luminosities, respectively. 
Various studies have shown that it is sufficient for an outflow to carry $\sim$5 per cent of the radiated energy to directly entrain most of the galactic gas at large radii, quenching the star formation in the host galaxy \citep[e.g.][]{Scannapieco2004, DiMatteo2005, Hopkins2005, Hopkins2008}. However, only $\sim$0.5 per cent of the luminosity is required to generate instabilities that shred the ISM clouds and mix them, efficiently increasing their cross-section. The enhancement in cross-section allows the quasar radiation to perturb more efficiently the clouds that were once too small and too dense to be affected by the radiation field \citep{Hopkins2010}. This effect can act dramatically on the cold gas, disrupting star formation in quasar host galaxies.

Figure~\ref{Ekin_Lbol} shows that ERQ outflow energy constitutes at least 3$-$5 per cent of quasar luminosity.
Such efficiencies are sufficient to regulate star formation in their host systems and prevent massive galaxies from growing too massive.

\section{Summary}
\label{summary}

ERQs are a remarkable population of heavily-reddened quasars at  redshift $z \sim$ 2 $-$ 3.4, discovered by \citet{Ross2015} and \citet{Hamann2017} in the BOSS and WISE surveys. ERQs show a suite of remarkable properties, unlike any known quasar population, that might all be tied to unusually powerful outflows during a brief young evolution stage \citep[Section~\ref{intro}, also][]{Ross2015, Zakamska2016, Hamann2017}. This study follows up on the discovery by \citet{Zakamska2016} of extreme [O{\sevensize III}] kinematics in the spectra of the reddest 4 ERQs. We carried out new IR observations with the aim of exploring the ubiquity of high-speed [O{\sevensize III}] outflows across the ERQ population. We selected 20 additional ERQs and 4 ERQ-like quasars to span a large range of reddenings and emission line properties. Our main results are the following:

\begin{itemize}
\item[1)] All 20 ERQs and 4 ERQ-like quasars routinely show powerful high-speed [O{\sevensize III}] outflows.
Their [O{\sevensize III}]~$\lambda$4959,5007 emission lines exhibit very broad and blueshifted profiles, with widths ($w_{90}$) ranging between 2053 and 7227 km s$^{-1}$, and exceptional outflows velocities (v$_{98}$) ranging from 1992 to 6702 km s$^{-1}$.

\item[2)] Comparisons to previous studies show that ERQs on average have [O{\sevensize III}] emission lines $\sim$0.43 dex broader than those of normal blue quasars matched in luminosity. Moreover, the maximum [O{\sevensize III}]-emitting gas velocities (v$_{98}$) in our sources are on average $\sim$0.47 dex larger than the typical values seen in blue quasars with similar luminosity (see Figure~\ref{L5_LO3_vel}).

\item[3)] There is a clear correlation between [O{\sevensize III}] outflow properties and i$-$W3 color and {\it{not}}  quasar luminosity (see Figures~\ref{L5_LO3_vel} and \ref{vel_Edd}). The redder the color, the broader the [O{\sevensize III}] emission line profile. This result provides a simple explanation of the differences in the [O{\sevensize III}] kinematics between ERQs and blue quasars with matched luminosity. Indeed, it suggests that the dust content in the host galaxy may play an important role in coupling the energy and momentum injected by the quasar to the surrounding ISM efficiently. See also \citet{Brusa2015} for similar conclusions in objects at lower luminosity.

\item[4)] Our study shows that the faster [O{\sevensize III}] outflows in ERQs are not tied to radio-loudness, nor larger Eddington ratios (see Figures~\ref{scatter_vel} and \ref{vel_Edd}). 

\item[5)] Our energetics estimates indicate that at least a few per cent of the ERQs bolometric luminosity is converted into the kinetic power of the ionized outflows. According to galaxy formation models, such efficiencies are in the range necessary to drive important feedback in host galaxies, regulating star formation and SMBH growth. Therefore, ERQs can drive strong feedback effect in host galaxies, and have a severe impact on their evolution (see Figure~\ref{Ekin_Lbol}).

We suggest that ERQs may represent a heavily-reddened quasar population caught during the short-lived ``blow-out'' phase of quasar feedback at the peak epoch of galaxy formation. Their powerful [O{\sevensize III}] winds have the potential to profoundly affect the evolution of the galaxies in which they occur.

\end{itemize}

\section*{Acknowledgements}
We thank the anonymous referee for his/hers useful suggestions that help to improve the quality of the manuscript. This work is based on observations made at the W.M. Keck Observatory, which is operated as a scientific partnership between the California Institute of Technology and the University of California; it was made possible by the generous support of the W.M. Keck Foundation. Data presented herein were partially obtained using the CIT Remote Observing Facility, and partially obtained using the UCI Remote Observing Facility, made possible by a generous gift from John and Ruth Ann Evans. 




\bibliographystyle{mnras}
\bibliography{ERQ_draft} 



\appendix
\section{Notes and Fits of Individual Targets}
\label{Appendix1}

J000610.67$+$121501.2 $-$ A ``perfect'' fit of the [O{\sevensize III}] would include 3 components for the [O{\sevensize III}] line. We prefer the simple 2 component fit, because it looks good for H$\beta$ and it captures all the essential properties of [O{\sevensize III}] in a way that is consistent with the other quasars. A 3rd component does not significantly change the strength and width of the [O{\sevensize III}]. The 2-component fit is also better because it yields a more conservative (lower) velocity for the blueshifted wing.

\begin{figure*}
 \includegraphics[width=0.8\columnwidth]{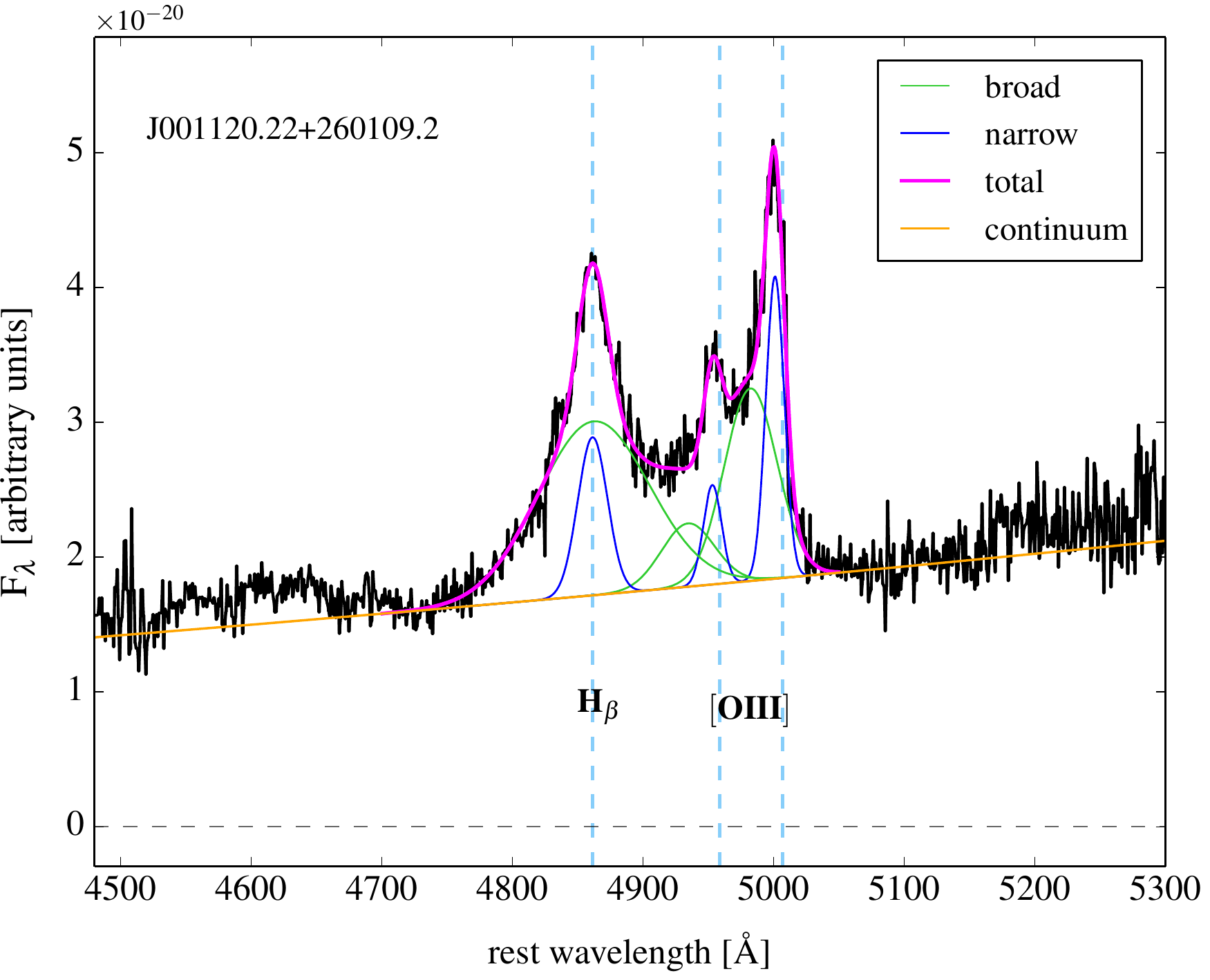}
 \includegraphics[width=0.8\columnwidth]{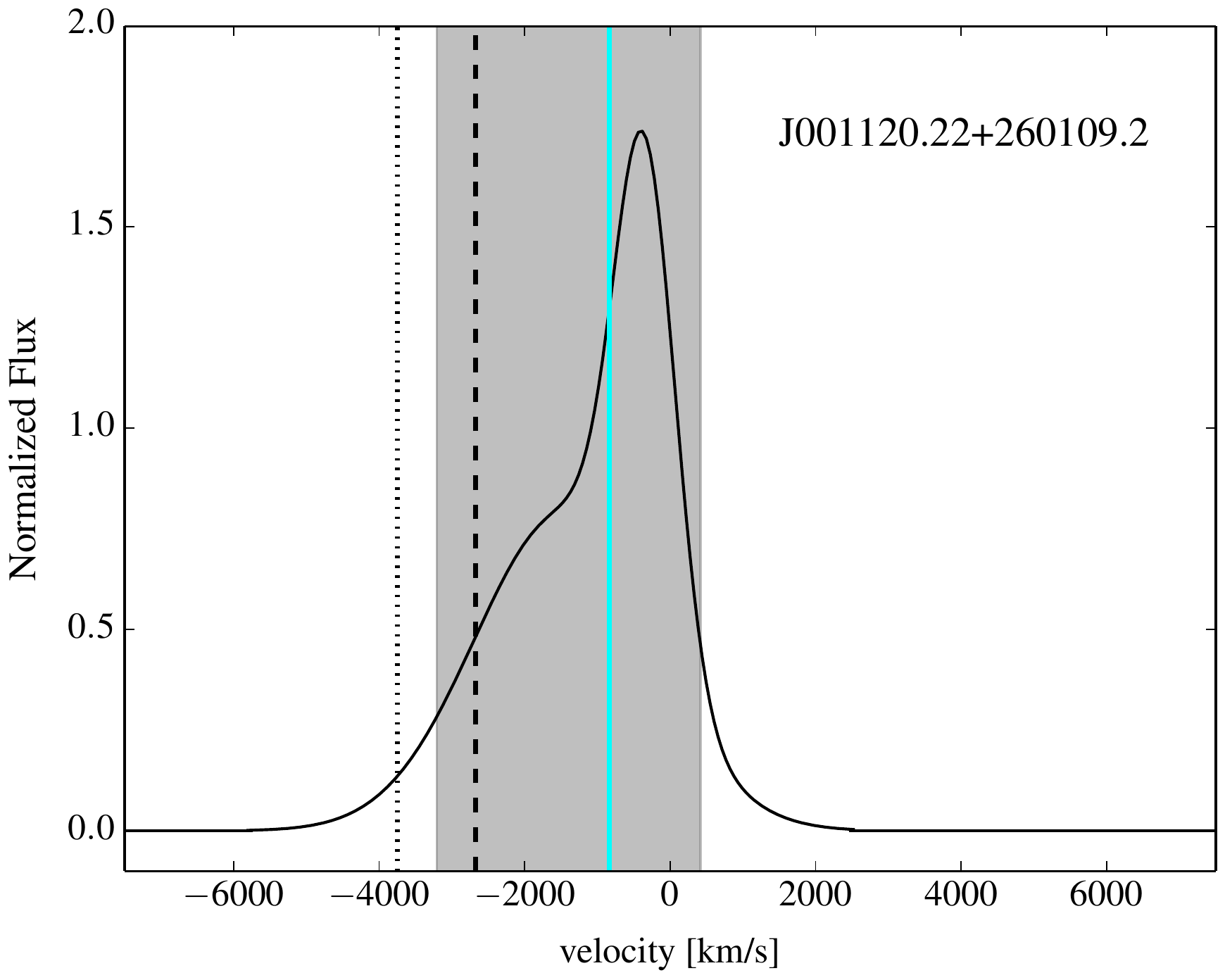}
 \includegraphics[width=0.8\columnwidth]{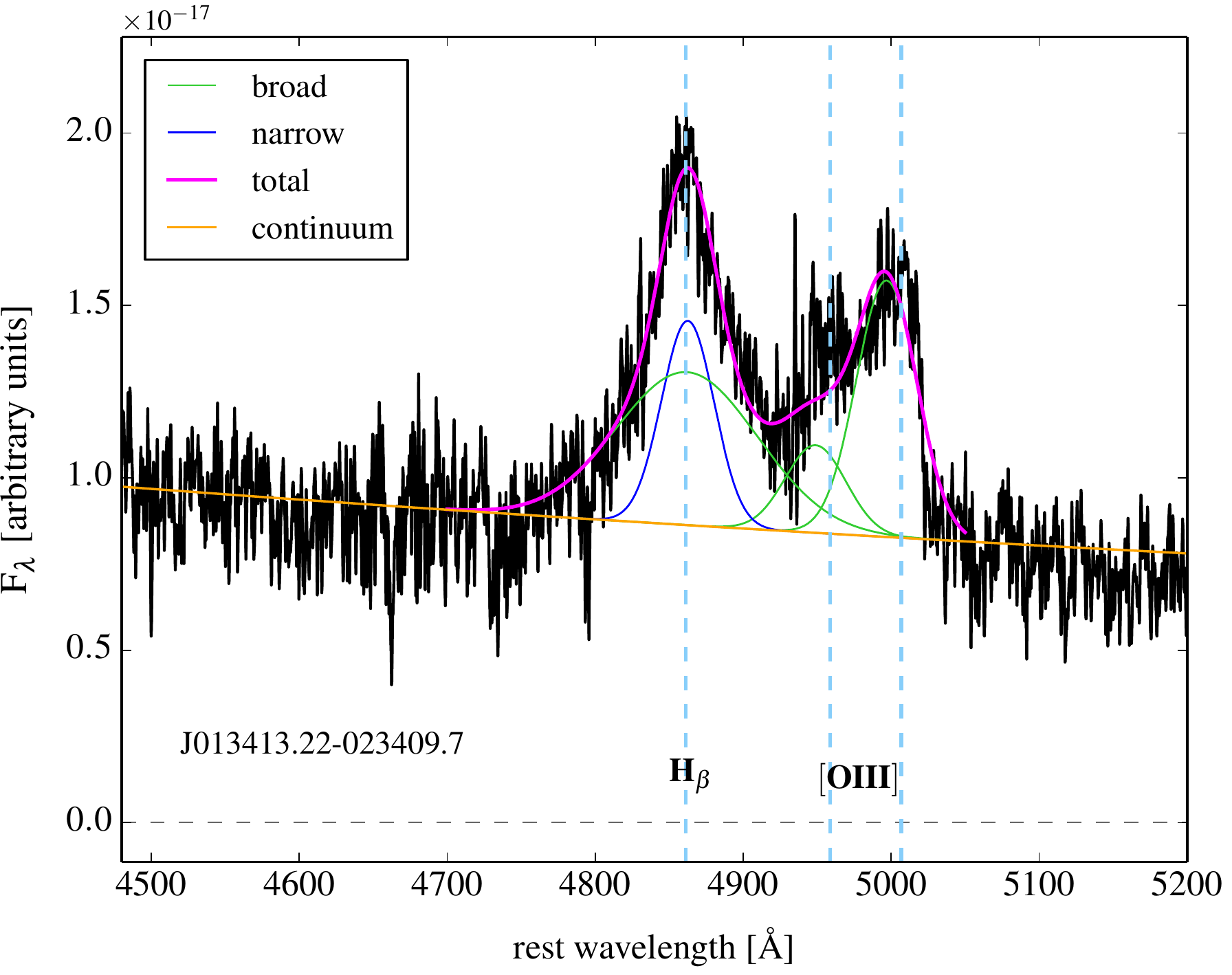}
 \includegraphics[width=0.8\columnwidth]{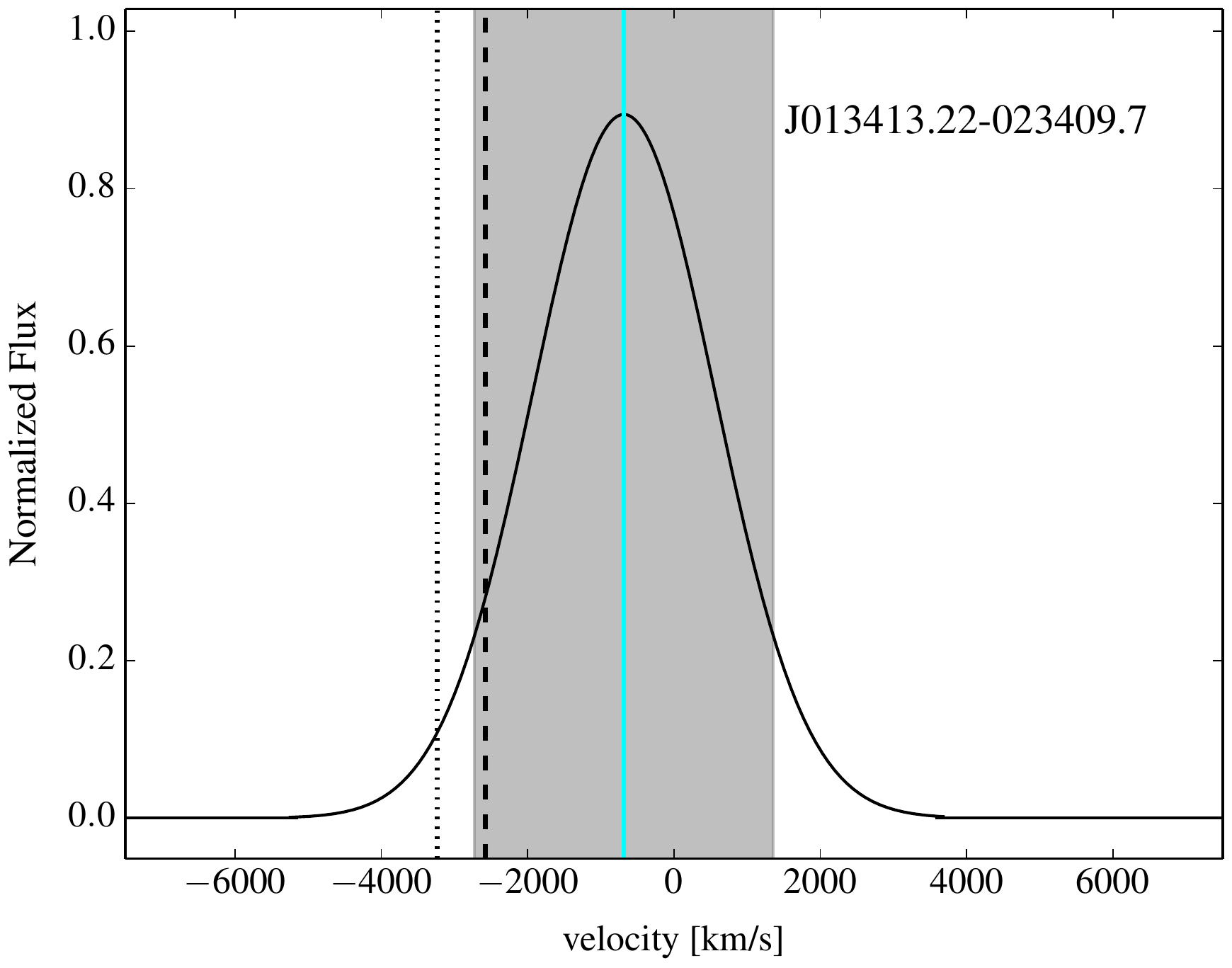}
 \includegraphics[width=0.8\columnwidth]{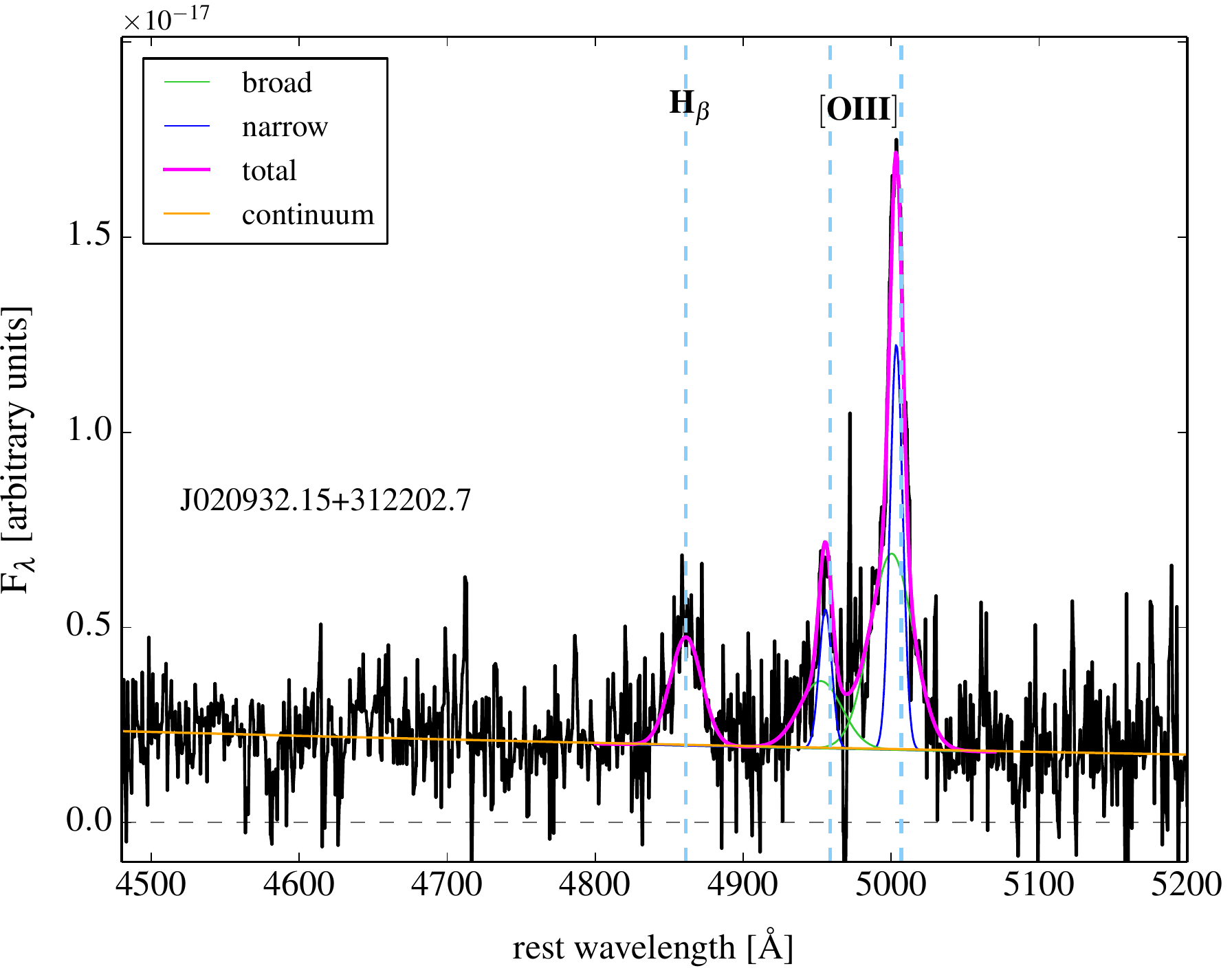}
 \includegraphics[width=0.8\columnwidth]{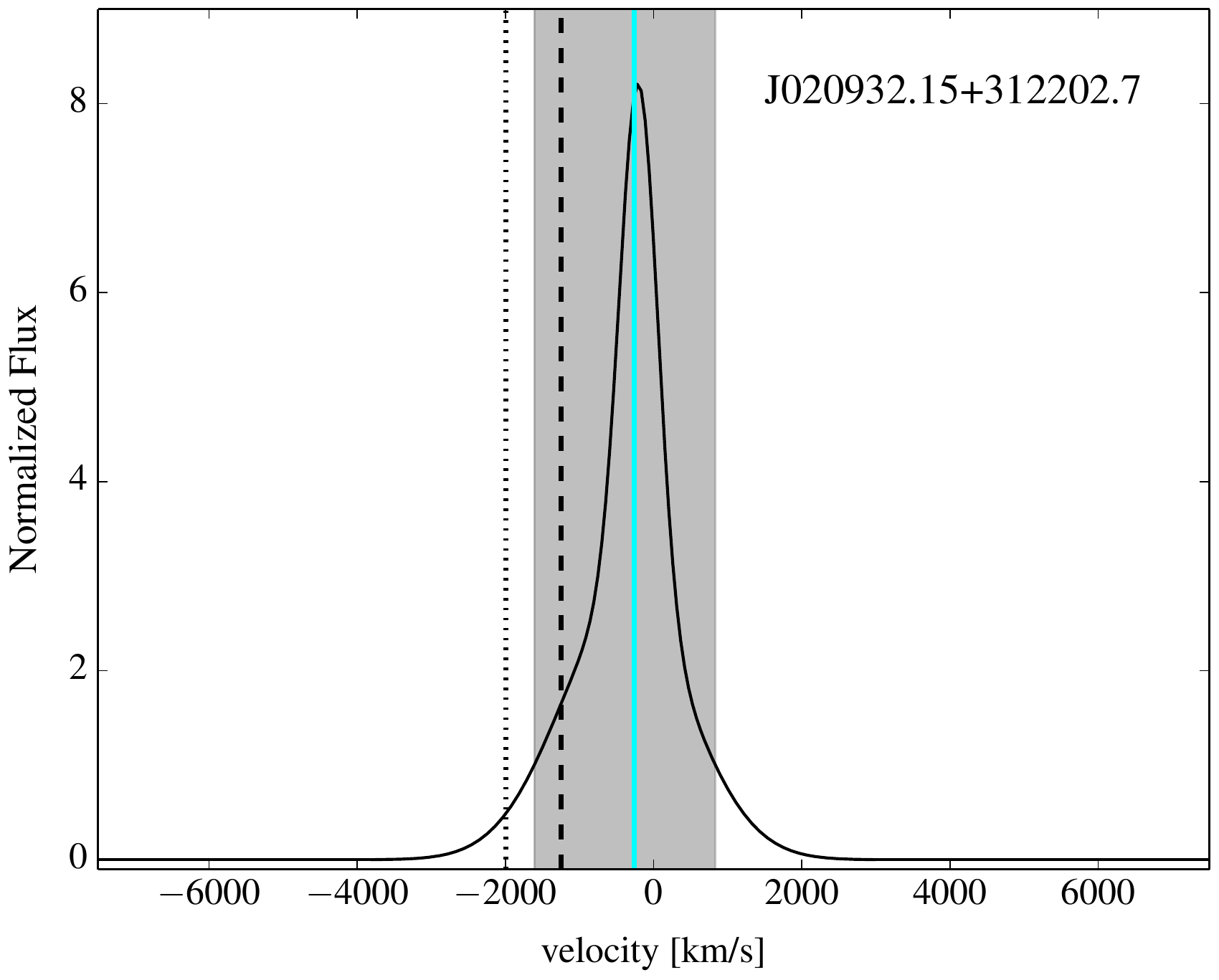}
 \includegraphics[width=0.8\columnwidth]{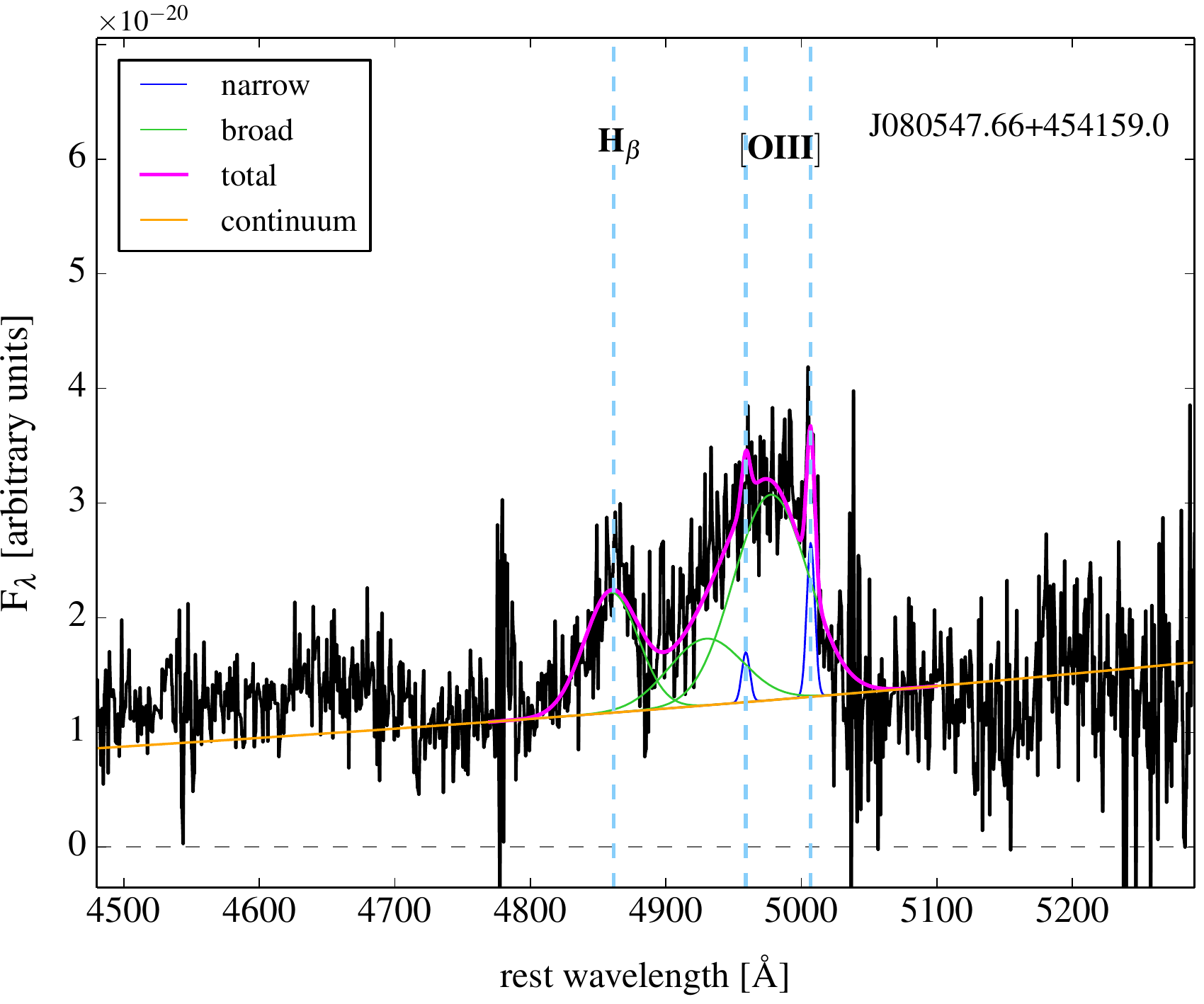}
 \includegraphics[width=0.8\columnwidth]{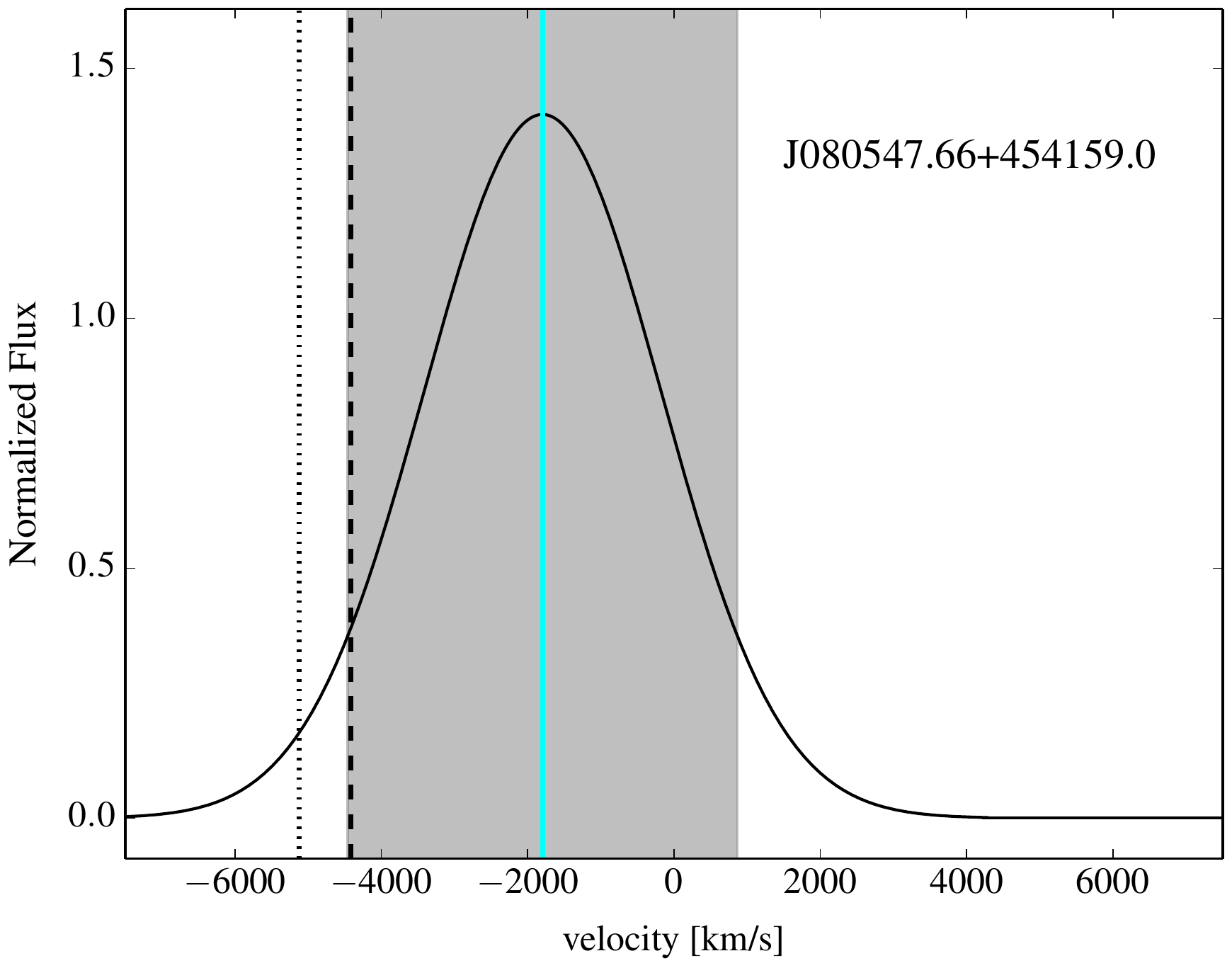}

 \caption{}
 \label{}
\end{figure*}

\begin{figure*}
\ContinuedFloat
 \includegraphics[width=0.8\columnwidth]{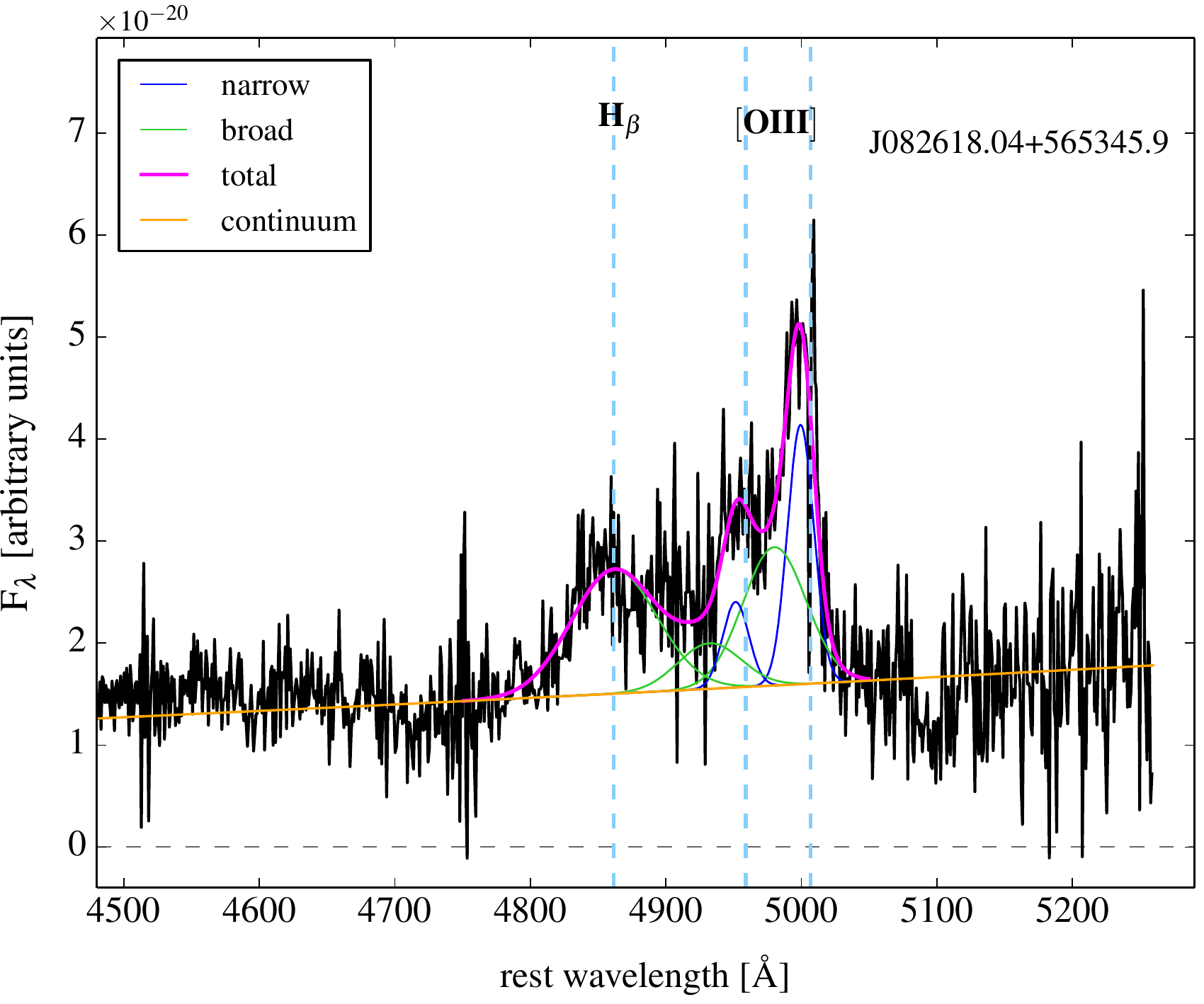}
 \includegraphics[width=0.8\columnwidth]{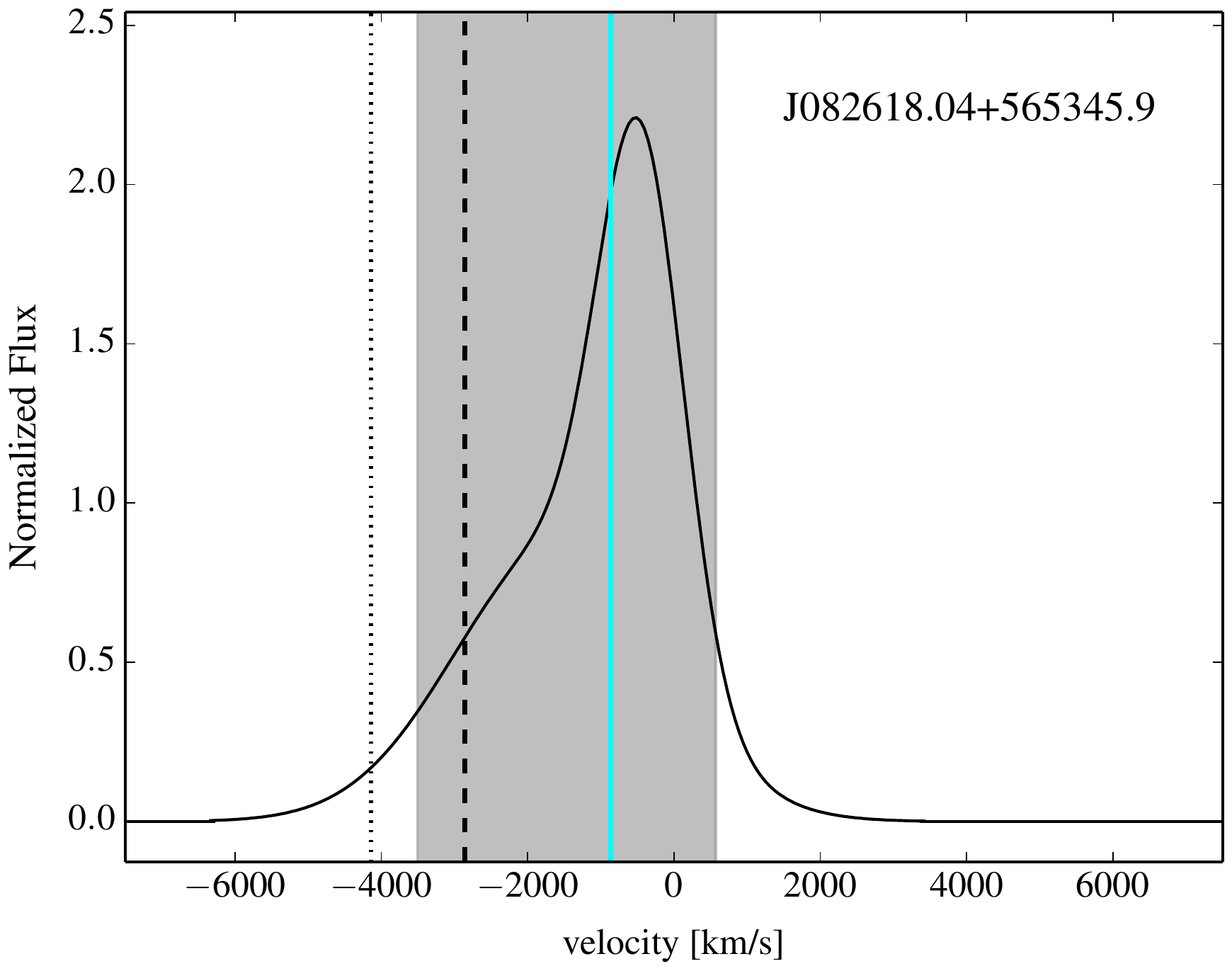}
 \includegraphics[width=0.8\columnwidth]{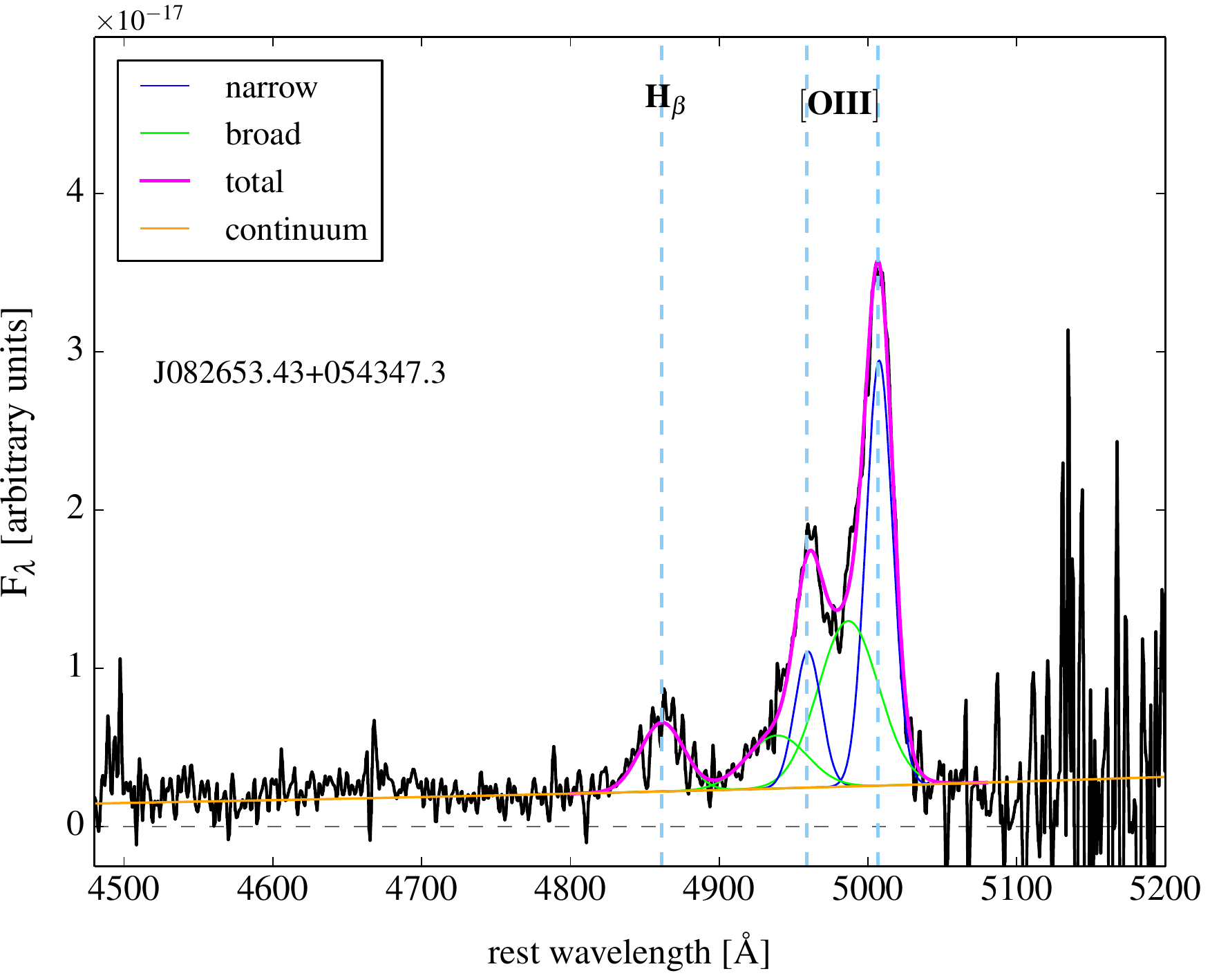}
 \includegraphics[width=0.8\columnwidth]{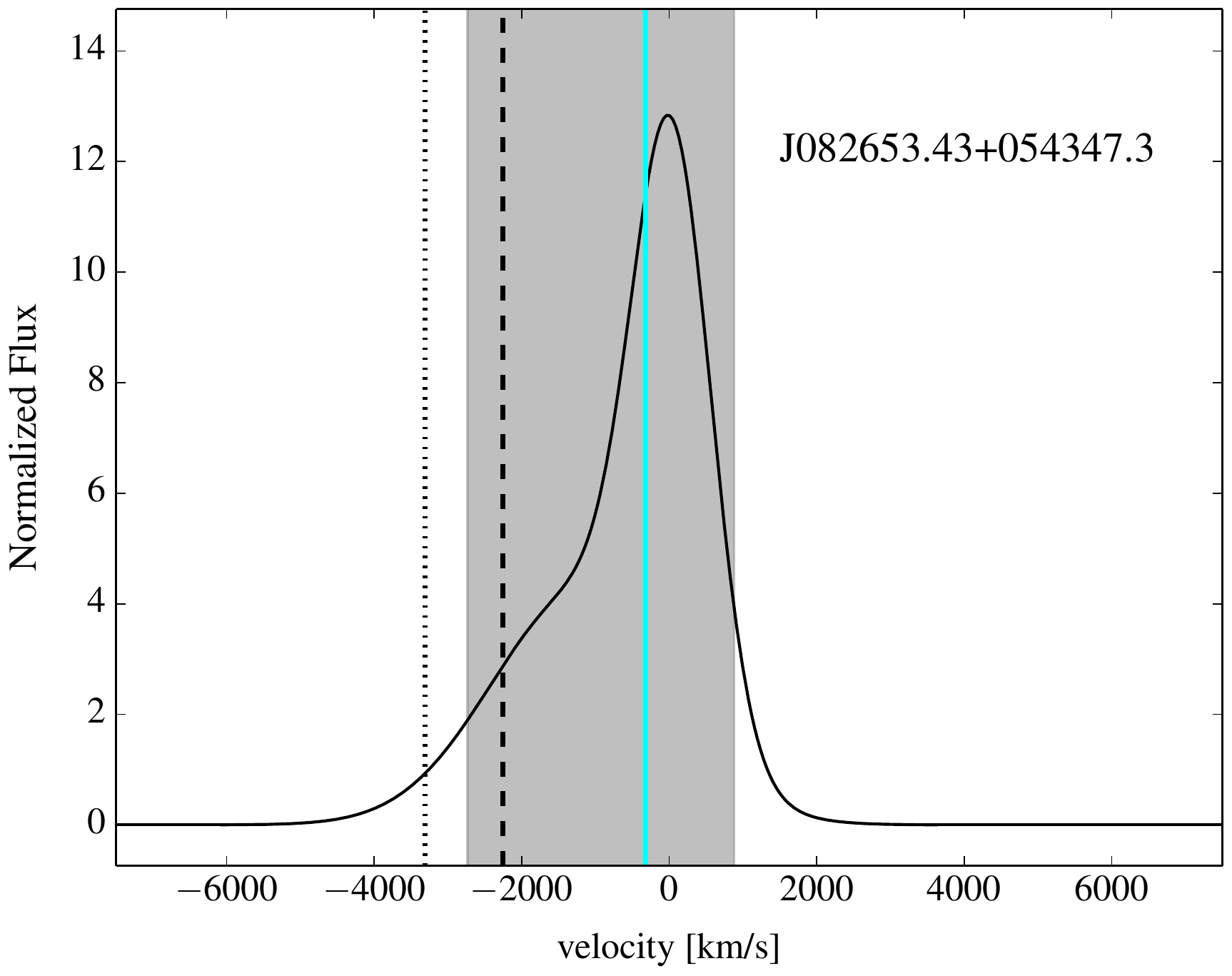}
 \includegraphics[width=0.8\columnwidth]{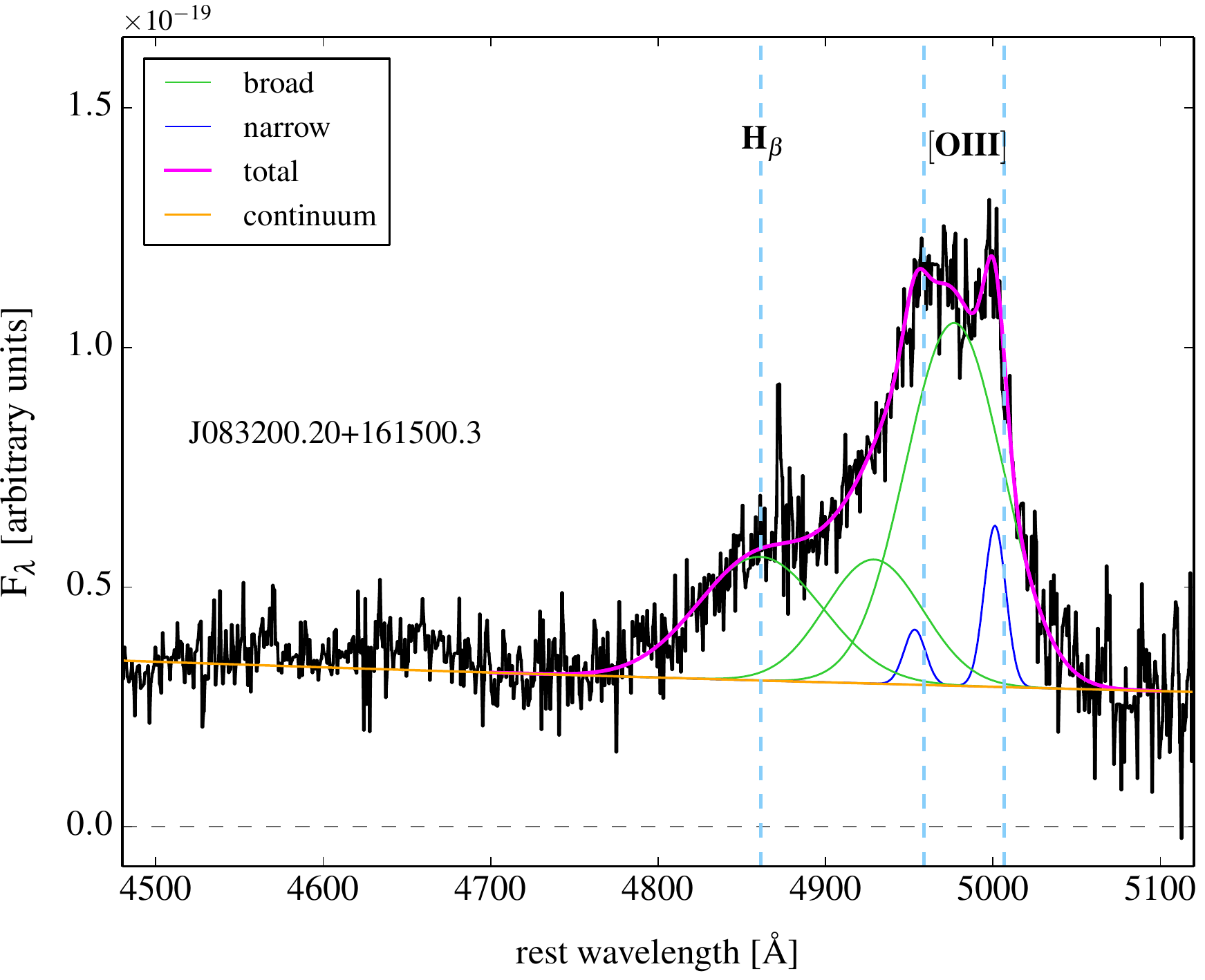}
 \includegraphics[width=0.8\columnwidth]{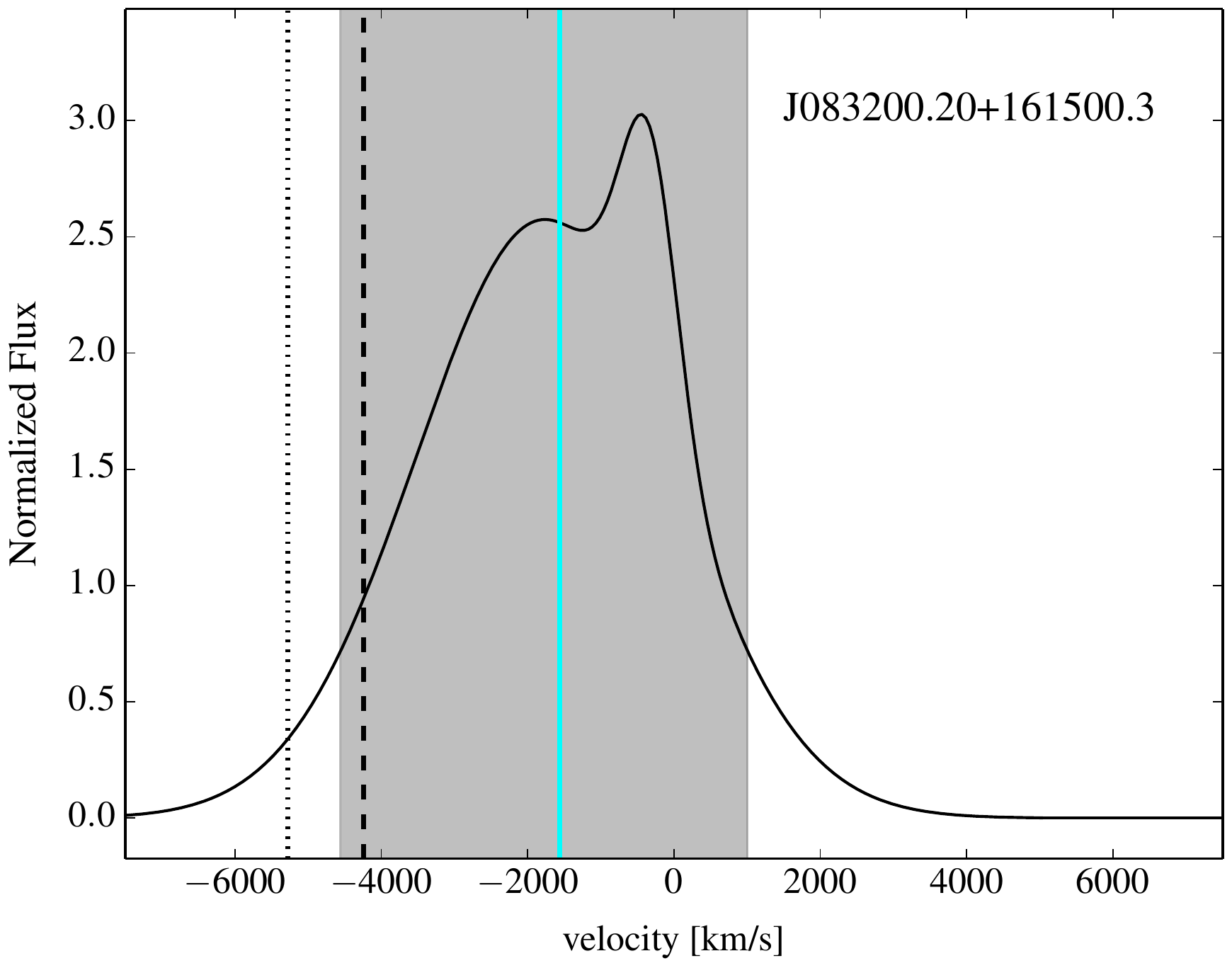}
 \includegraphics[width=0.8\columnwidth]{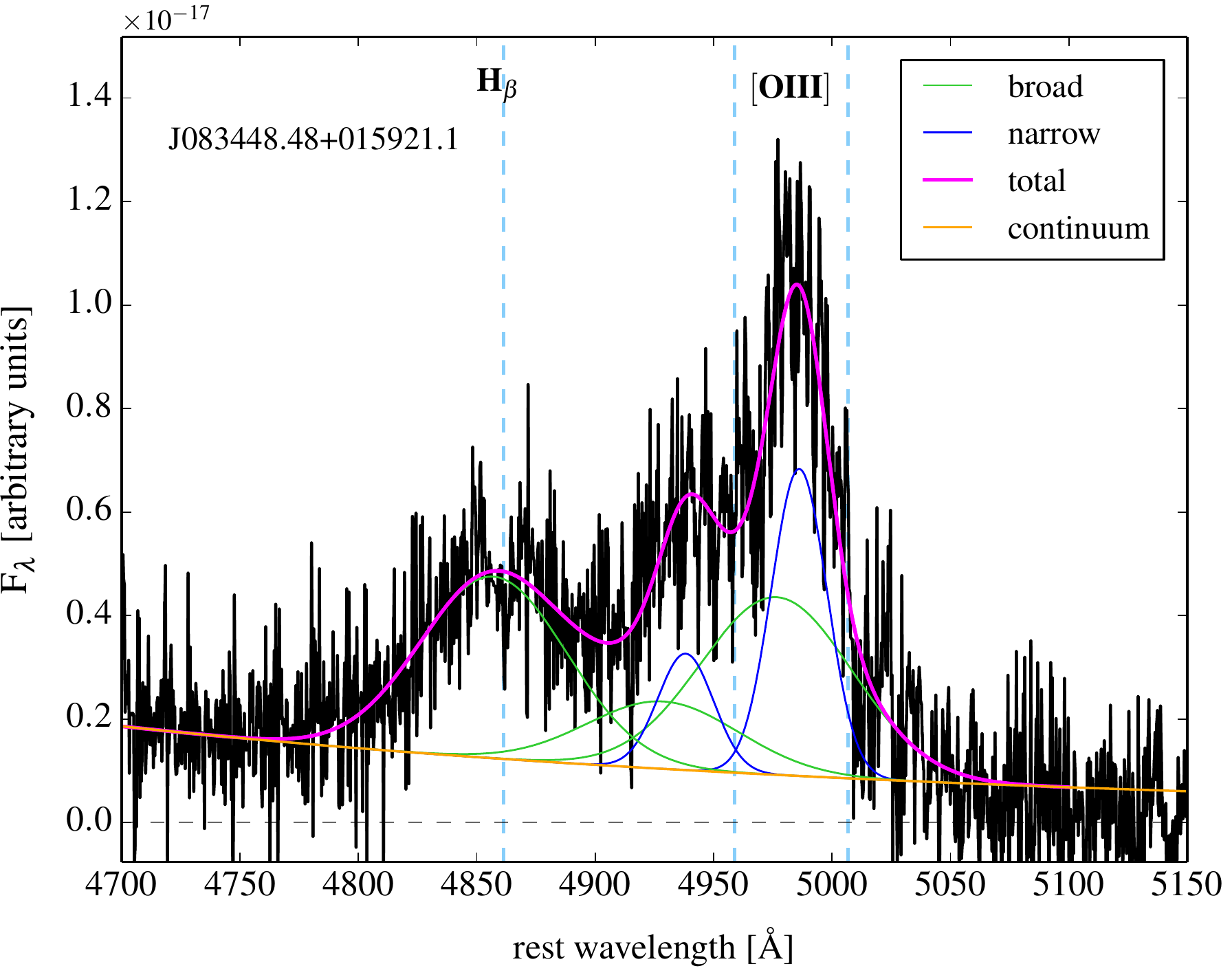}
 \includegraphics[width=0.8\columnwidth]{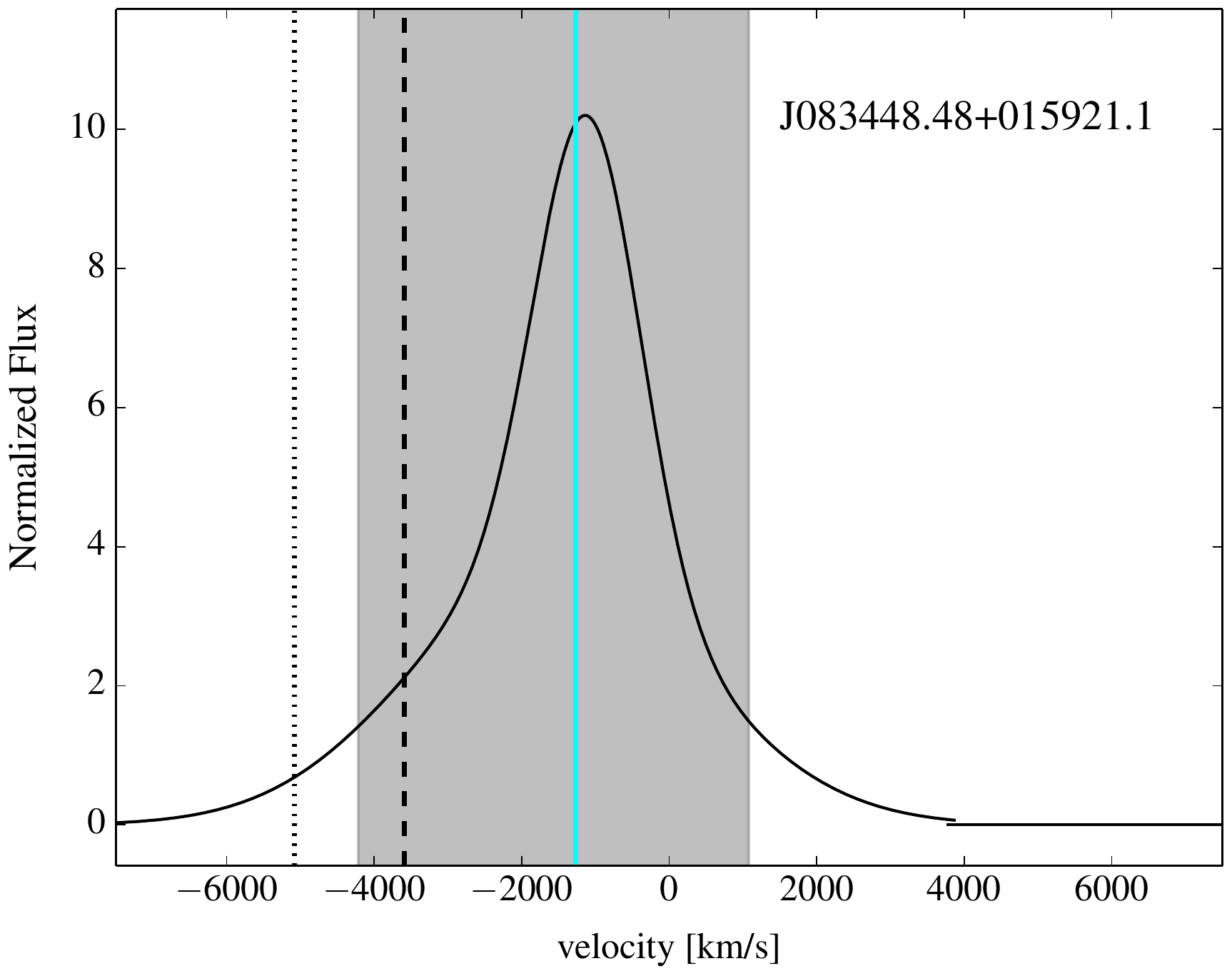}

 \caption{-- \bf continue}
 \label{}
\end{figure*}

\begin{figure*}
\ContinuedFloat
 \includegraphics[width=0.8\columnwidth]{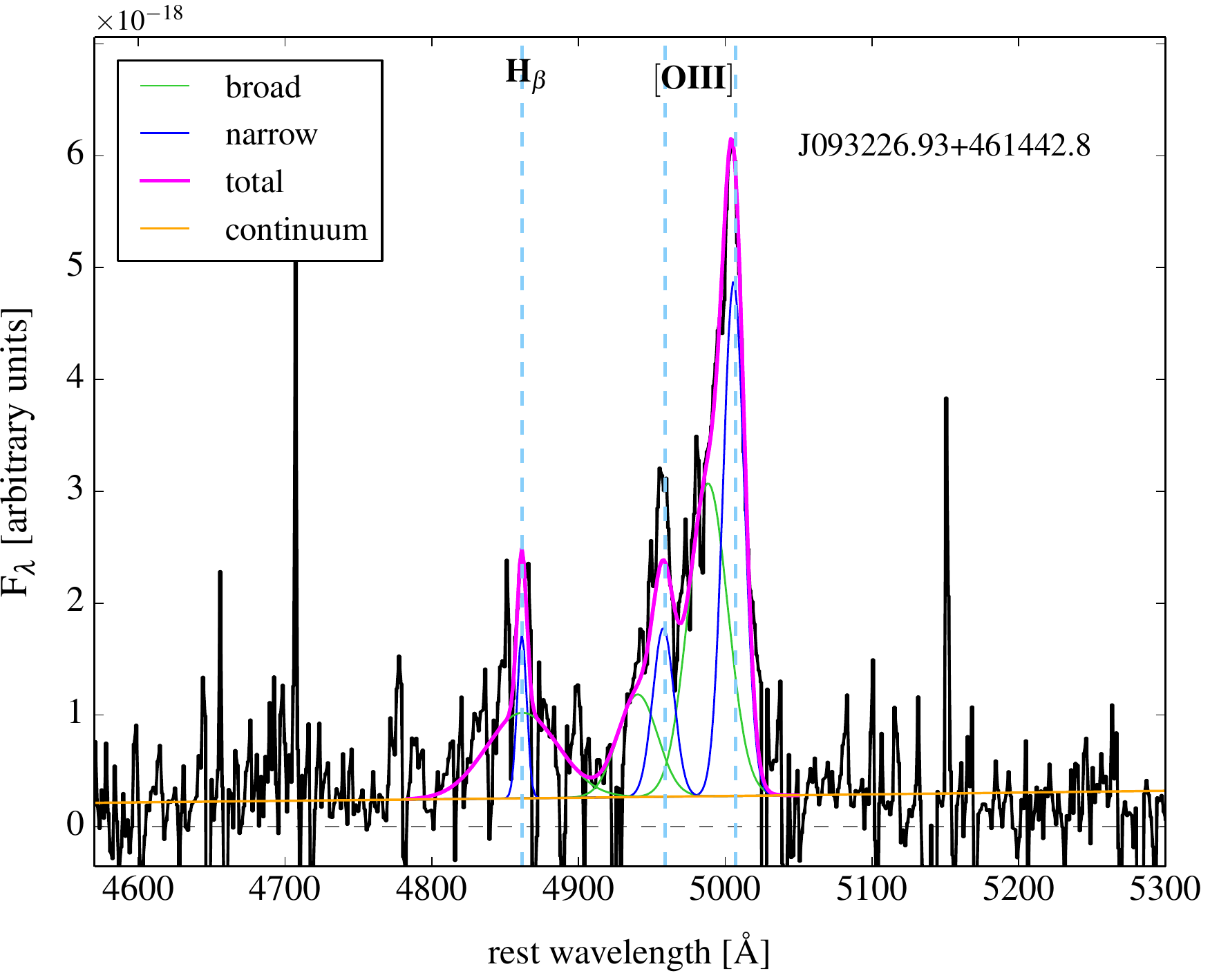}
 \includegraphics[width=0.8\columnwidth]{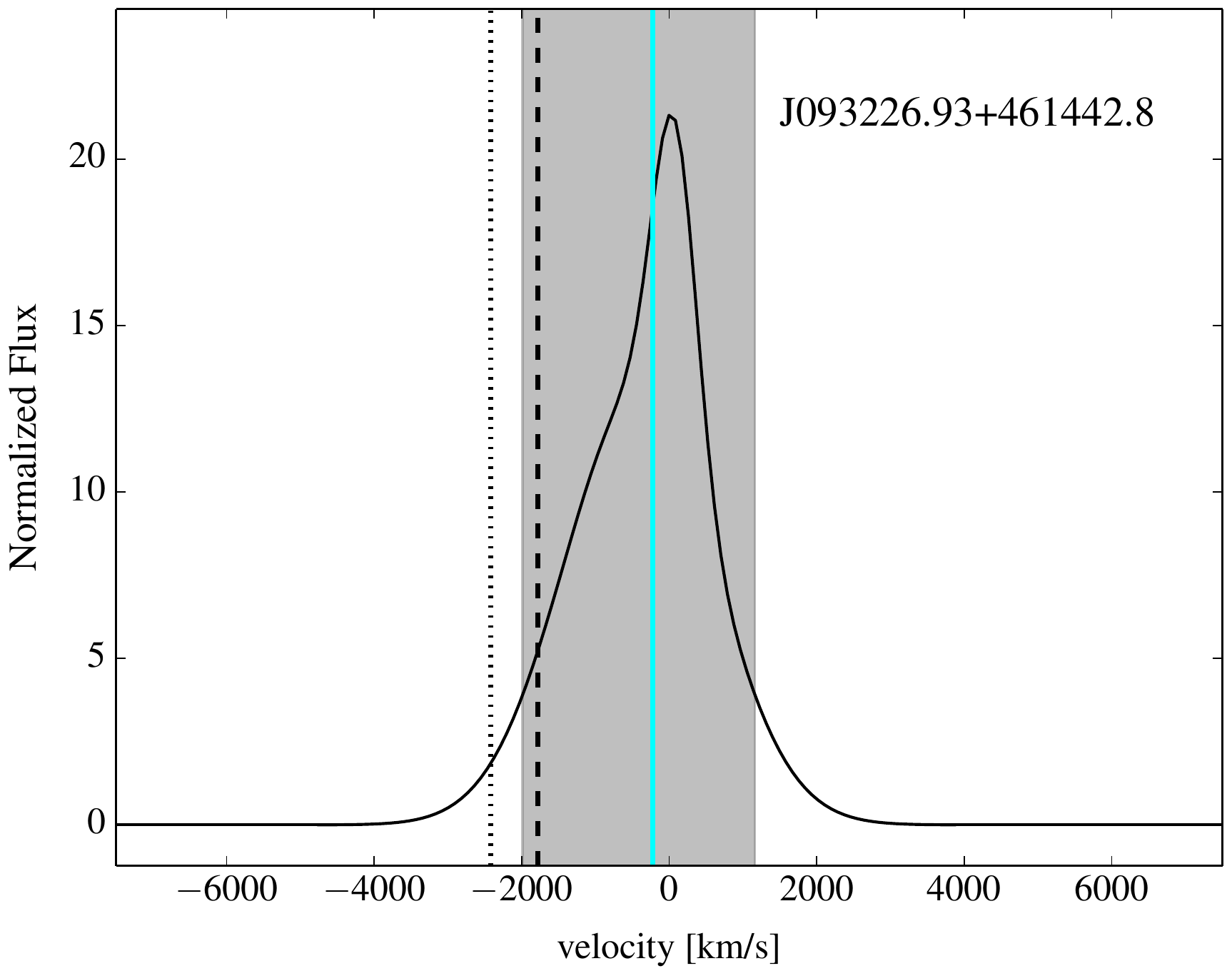}
 \includegraphics[width=0.8\columnwidth]{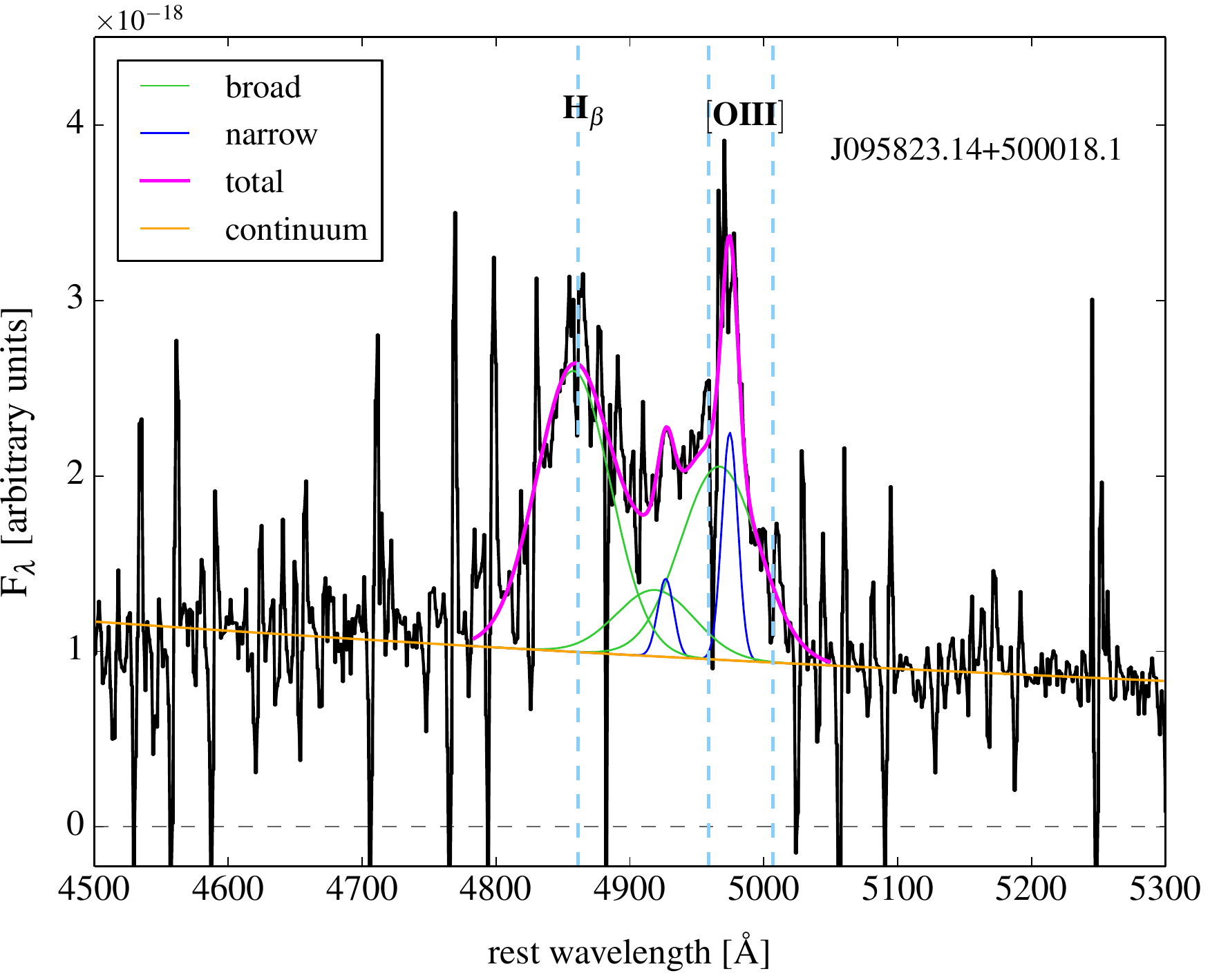}
 \includegraphics[width=0.8\columnwidth]{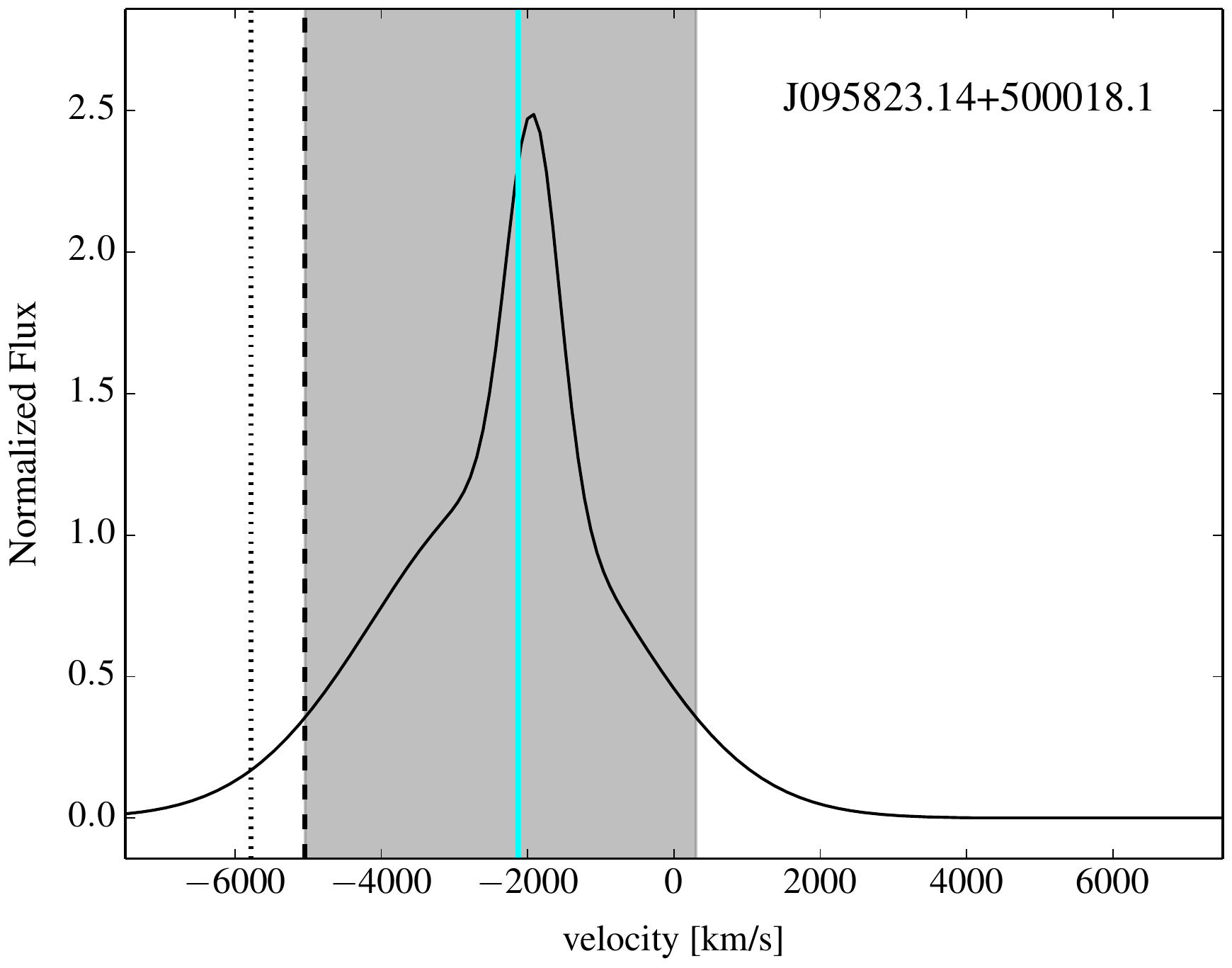}
 \includegraphics[width=0.8\columnwidth]{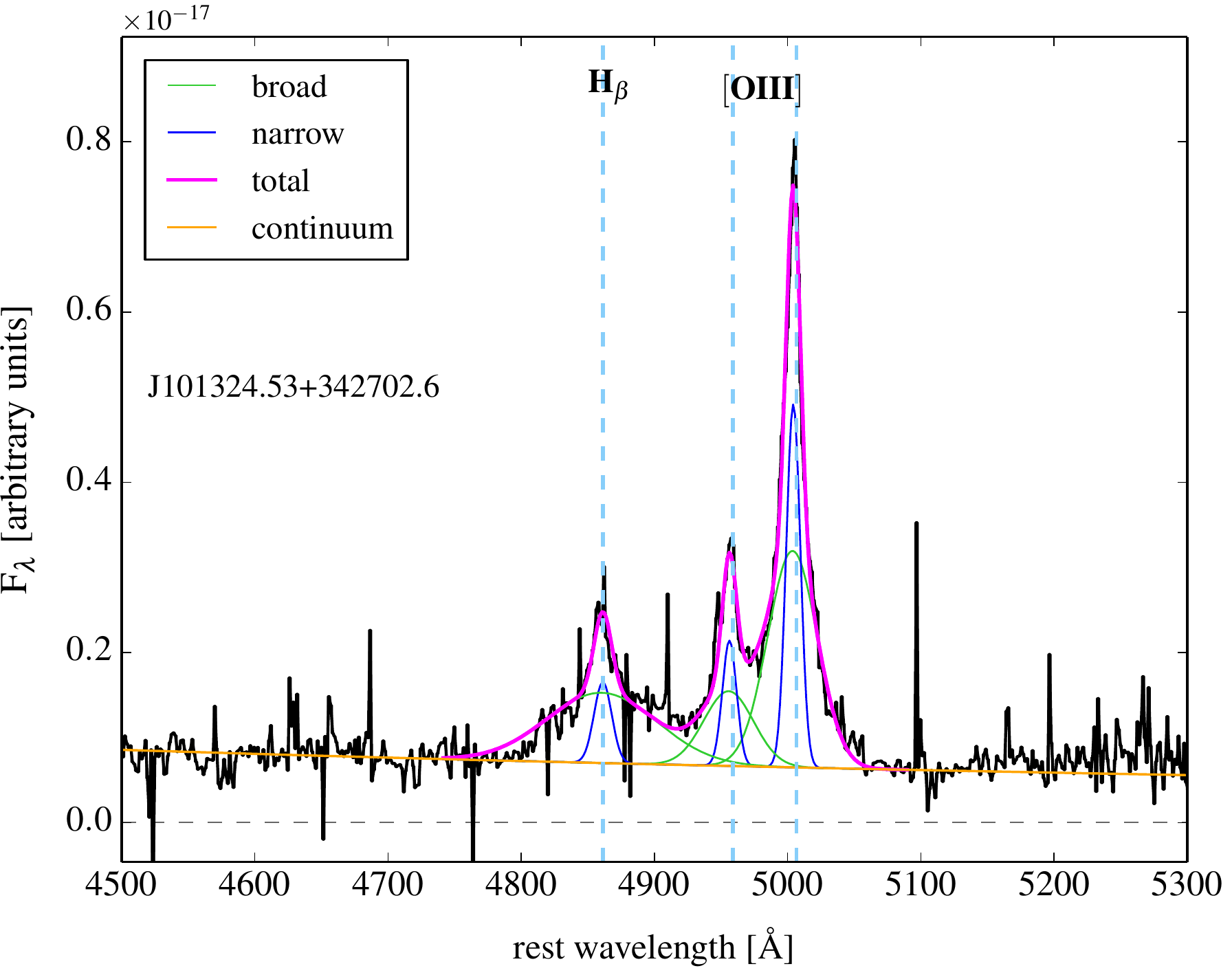}
 \includegraphics[width=0.8\columnwidth]{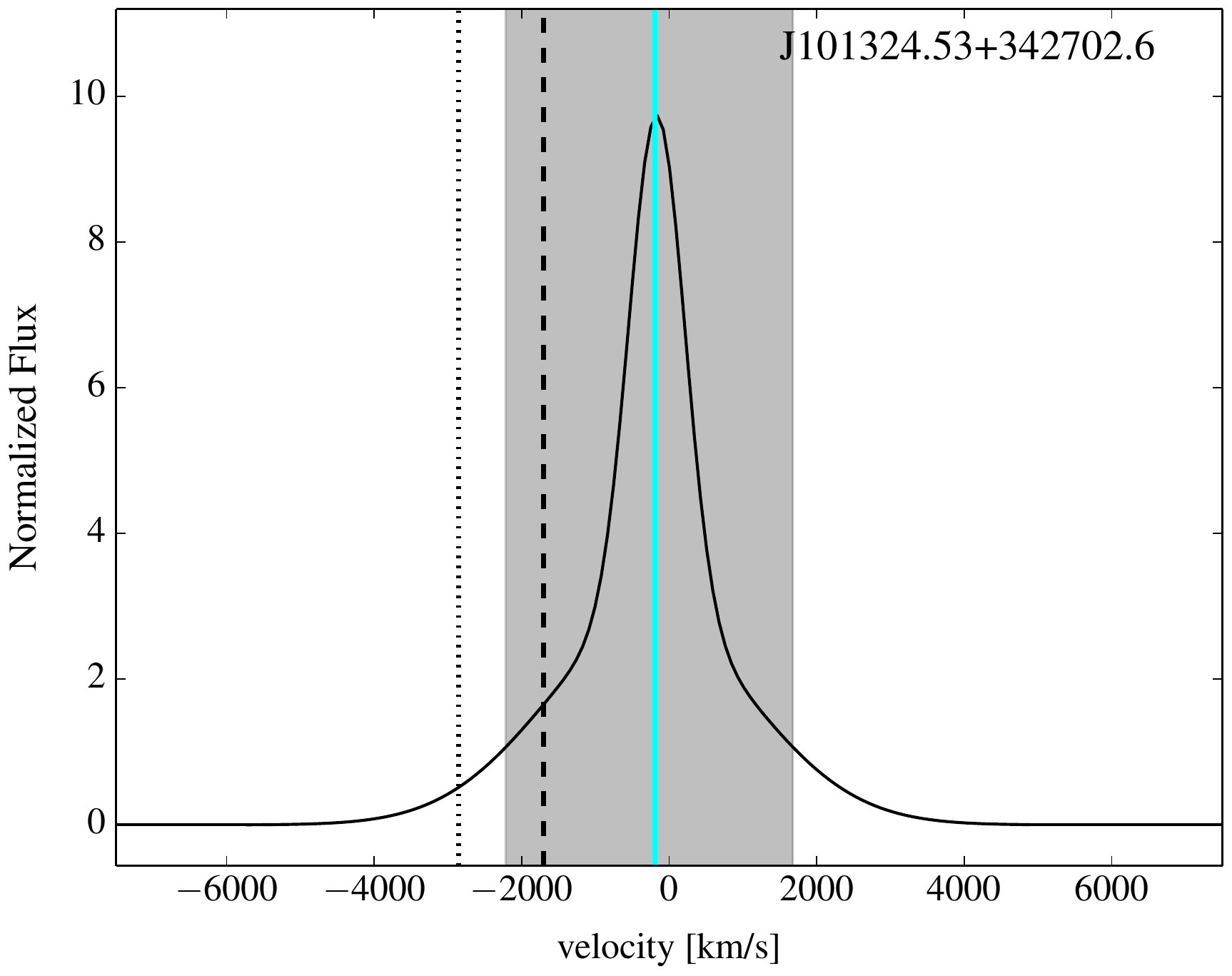}
 \includegraphics[width=0.8\columnwidth]{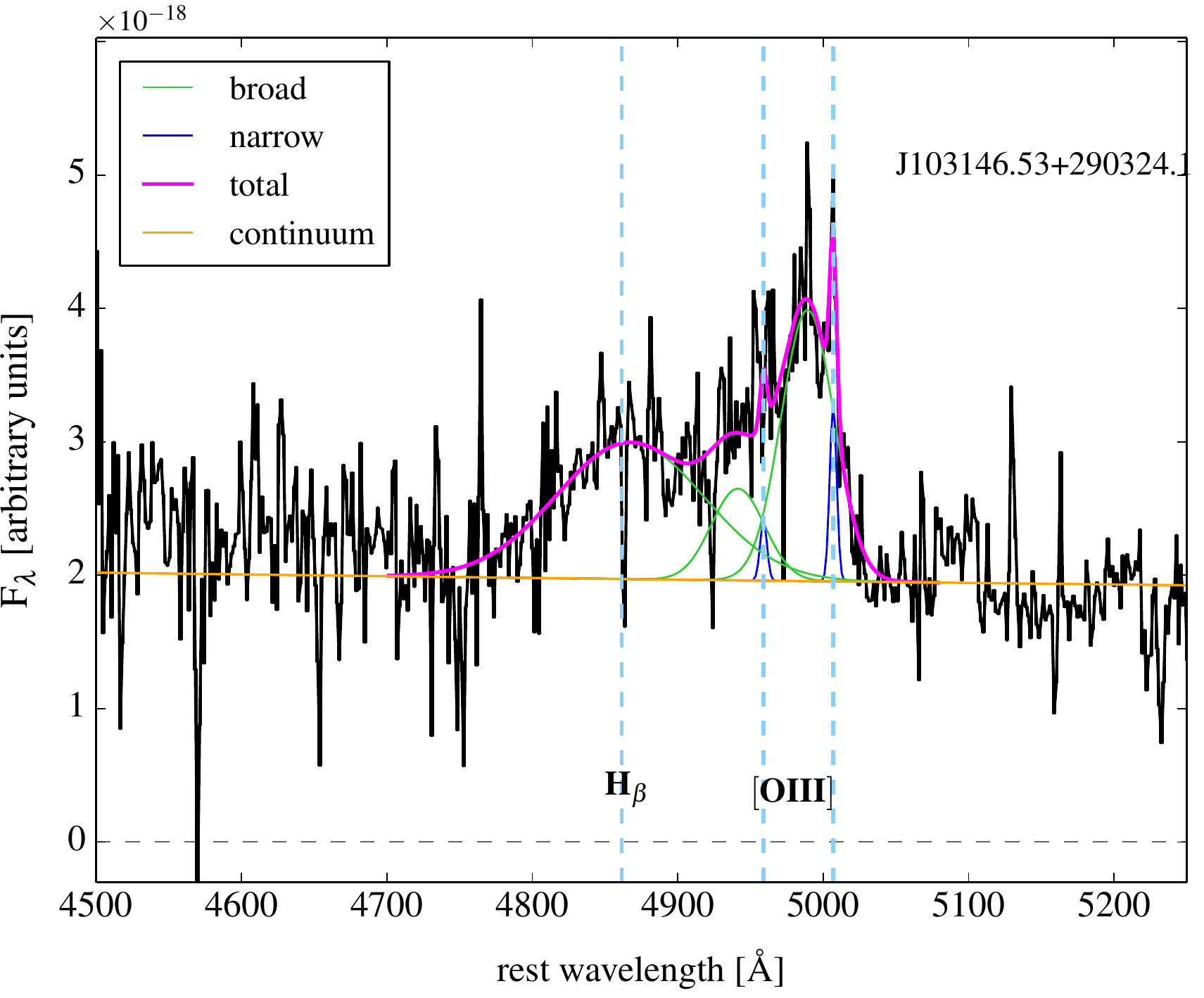}
 \includegraphics[width=0.8\columnwidth]{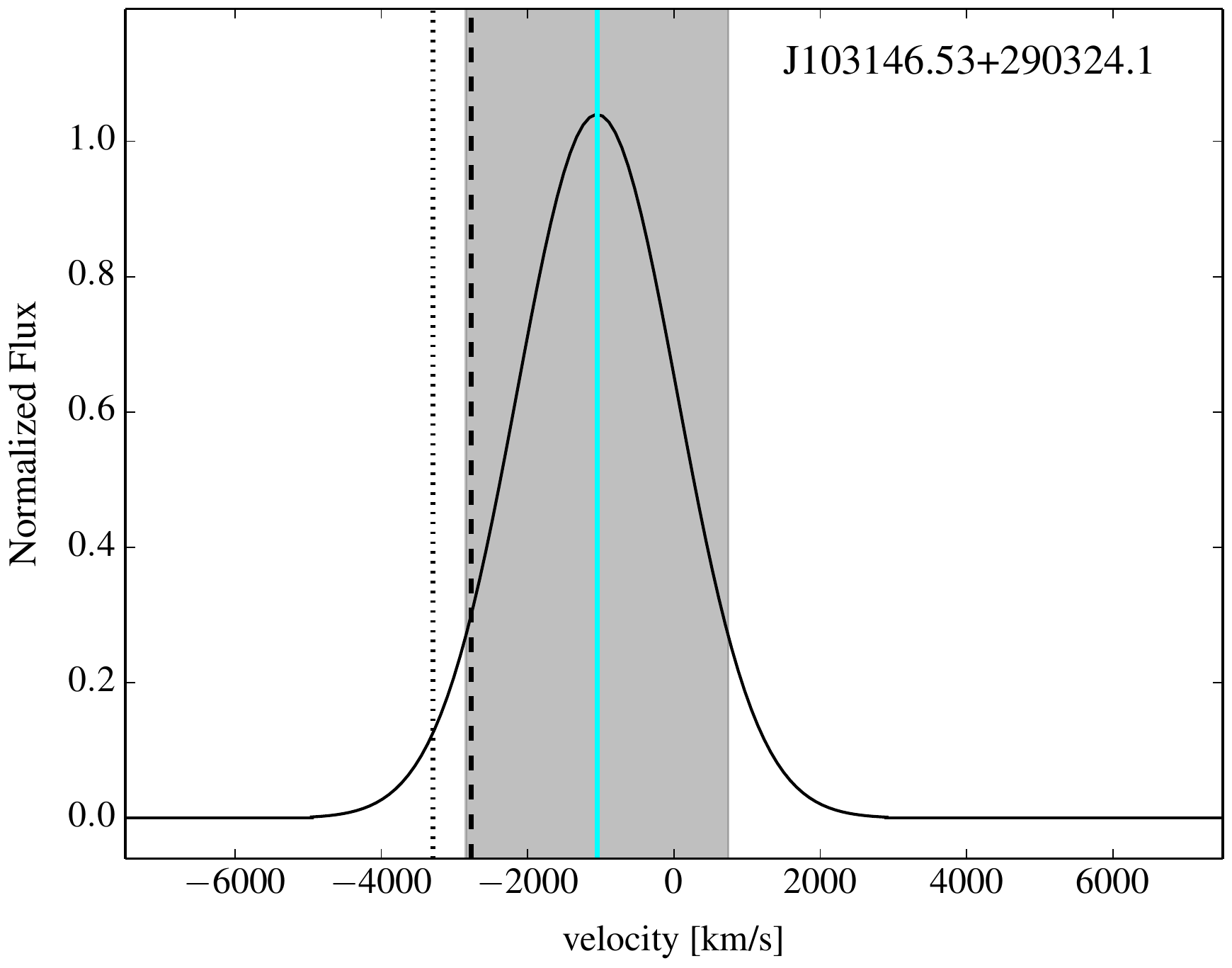}
 \caption{-- \bf continue}
 \label{}
\end{figure*}

\begin{figure*}
\ContinuedFloat

 \includegraphics[width=0.8\columnwidth]{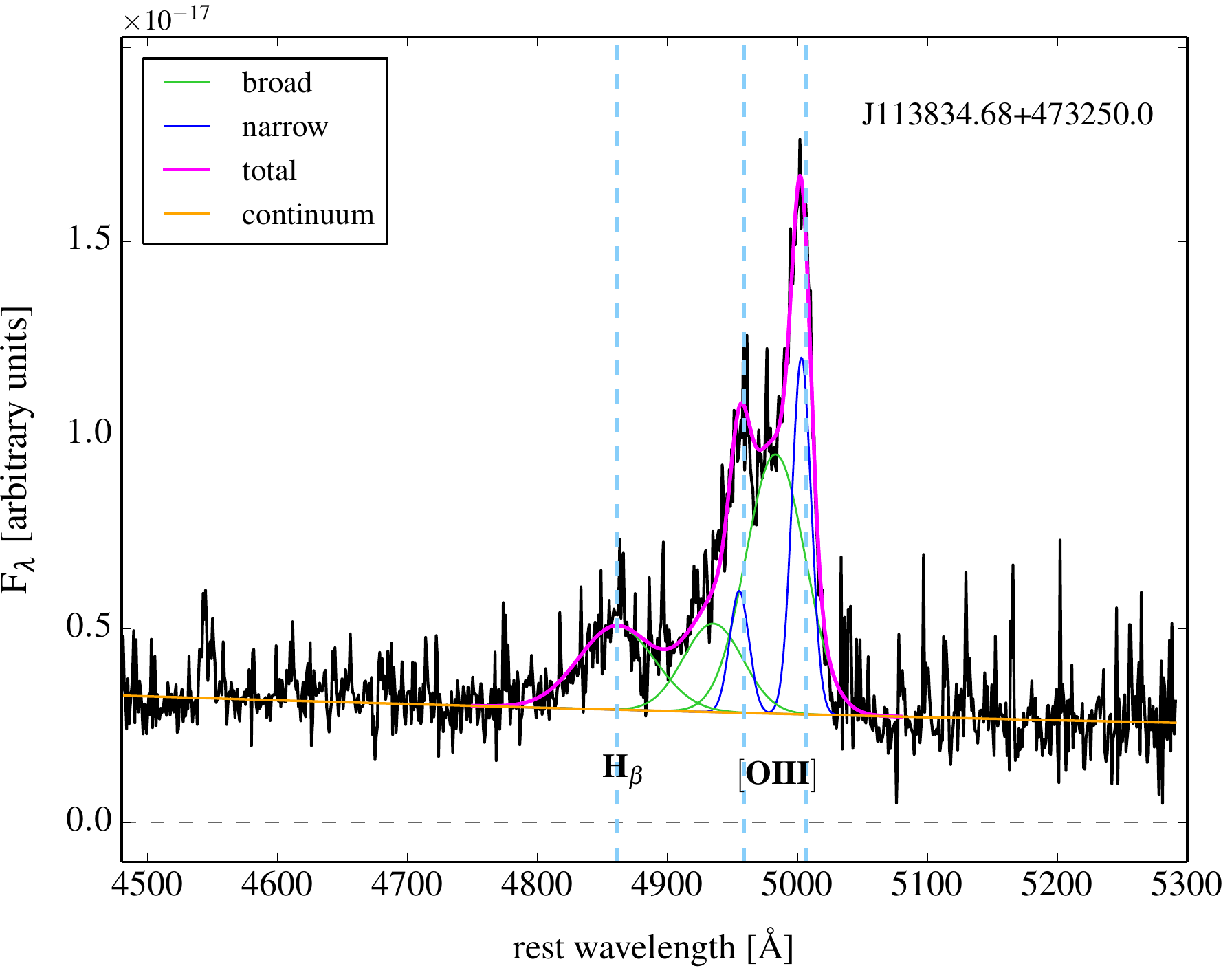}
 \includegraphics[width=0.8\columnwidth]{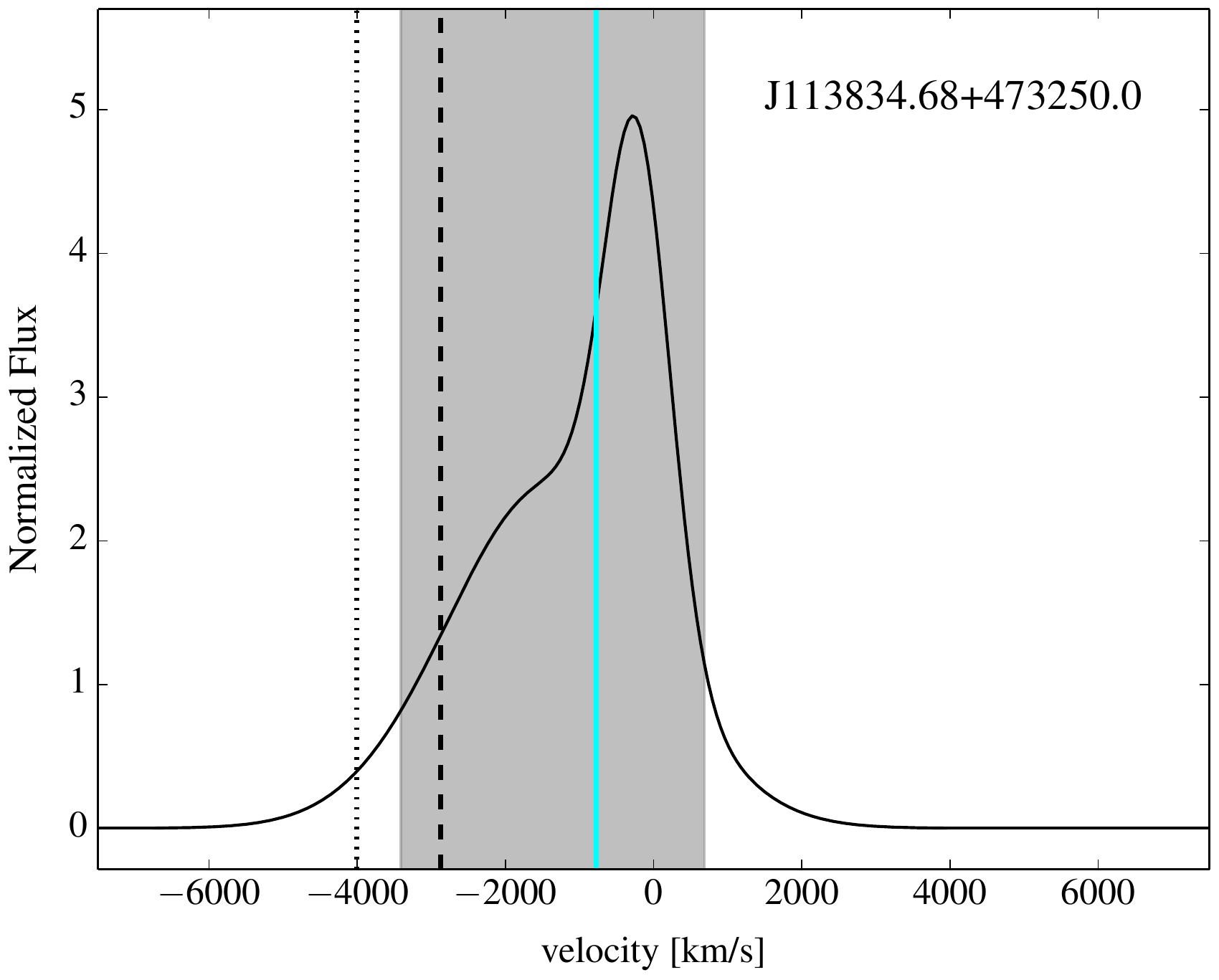}
 \includegraphics[width=0.8\columnwidth]{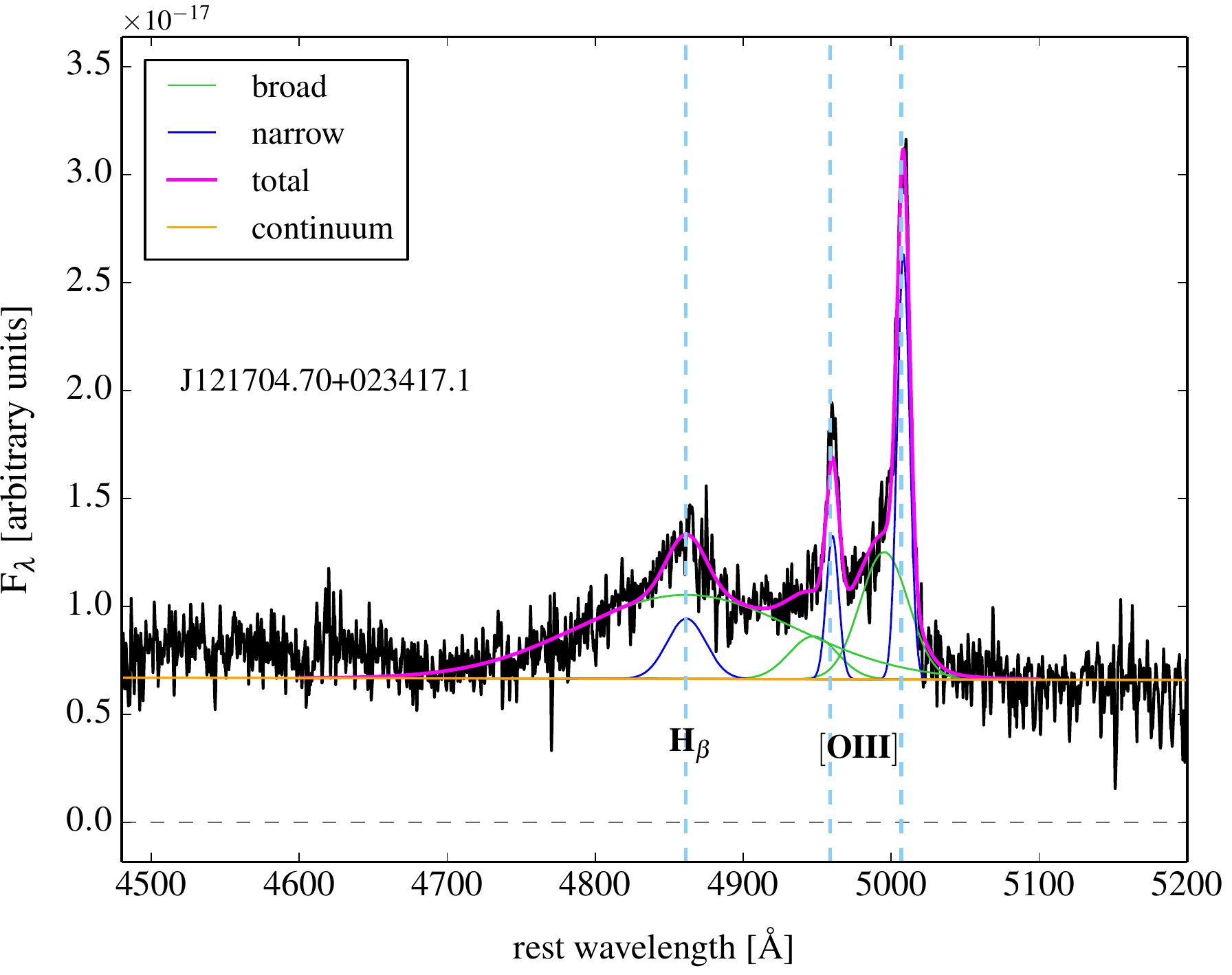}
 \includegraphics[width=0.8\columnwidth]{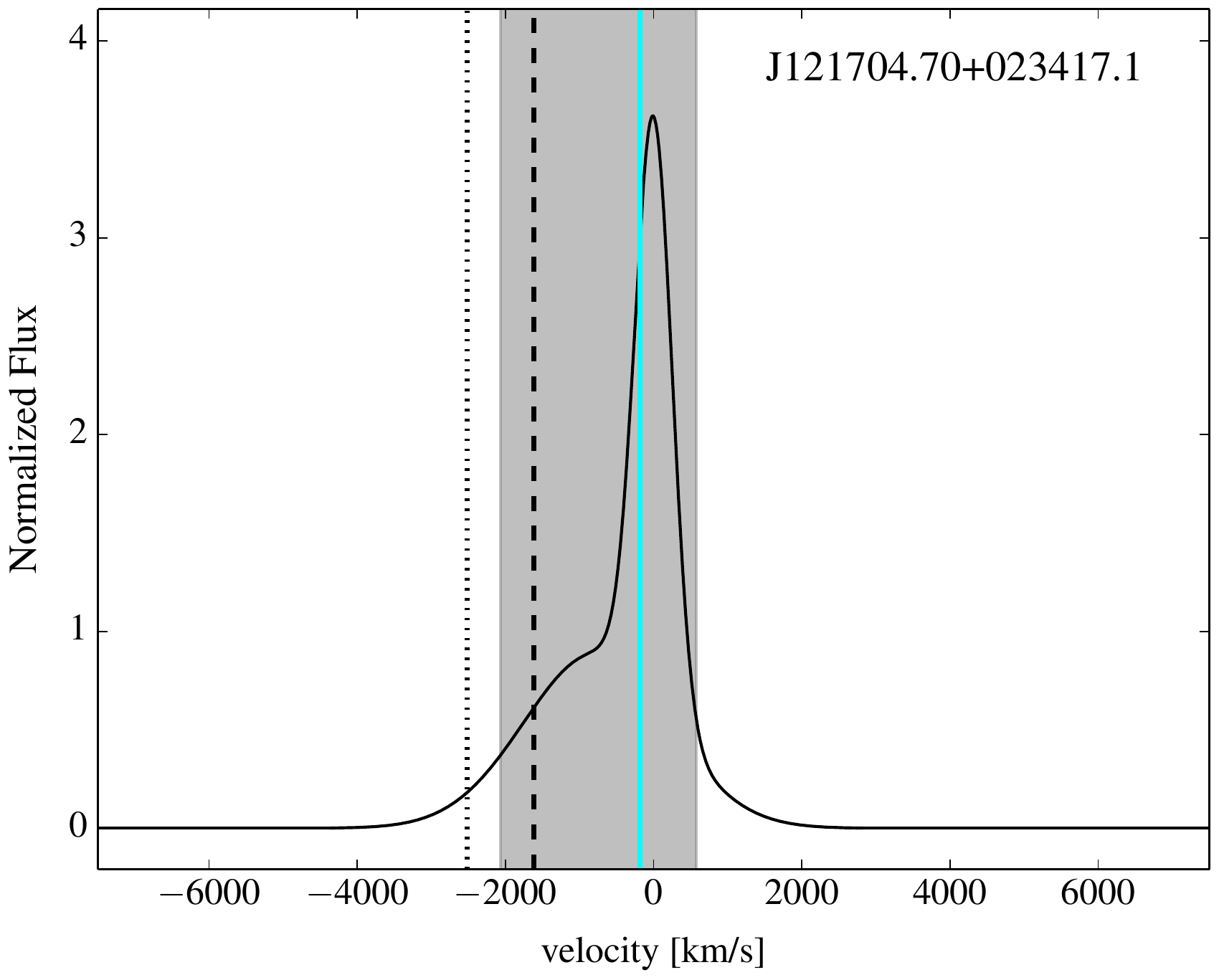}
 \includegraphics[width=0.8\columnwidth]{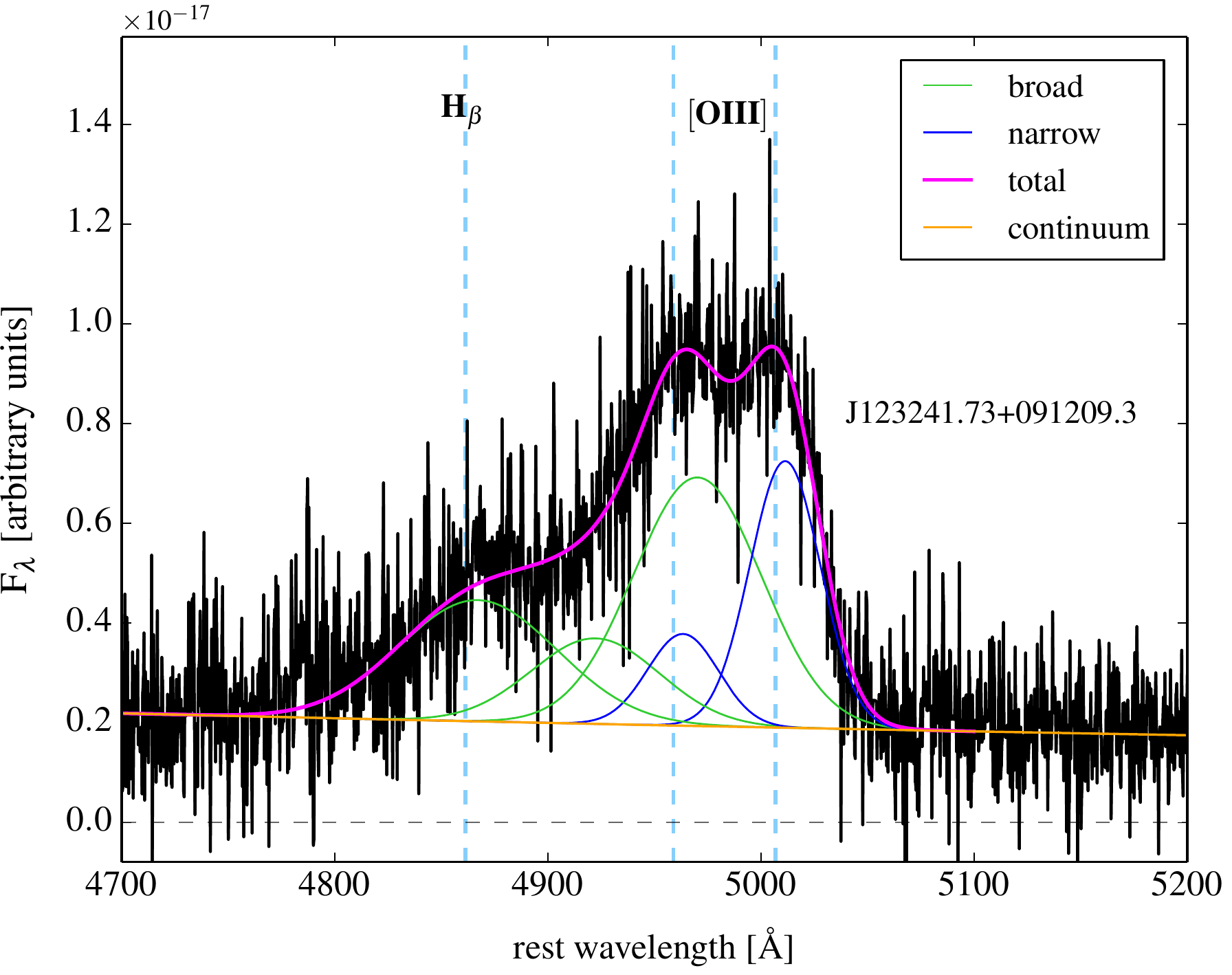}
 \includegraphics[width=0.8\columnwidth]{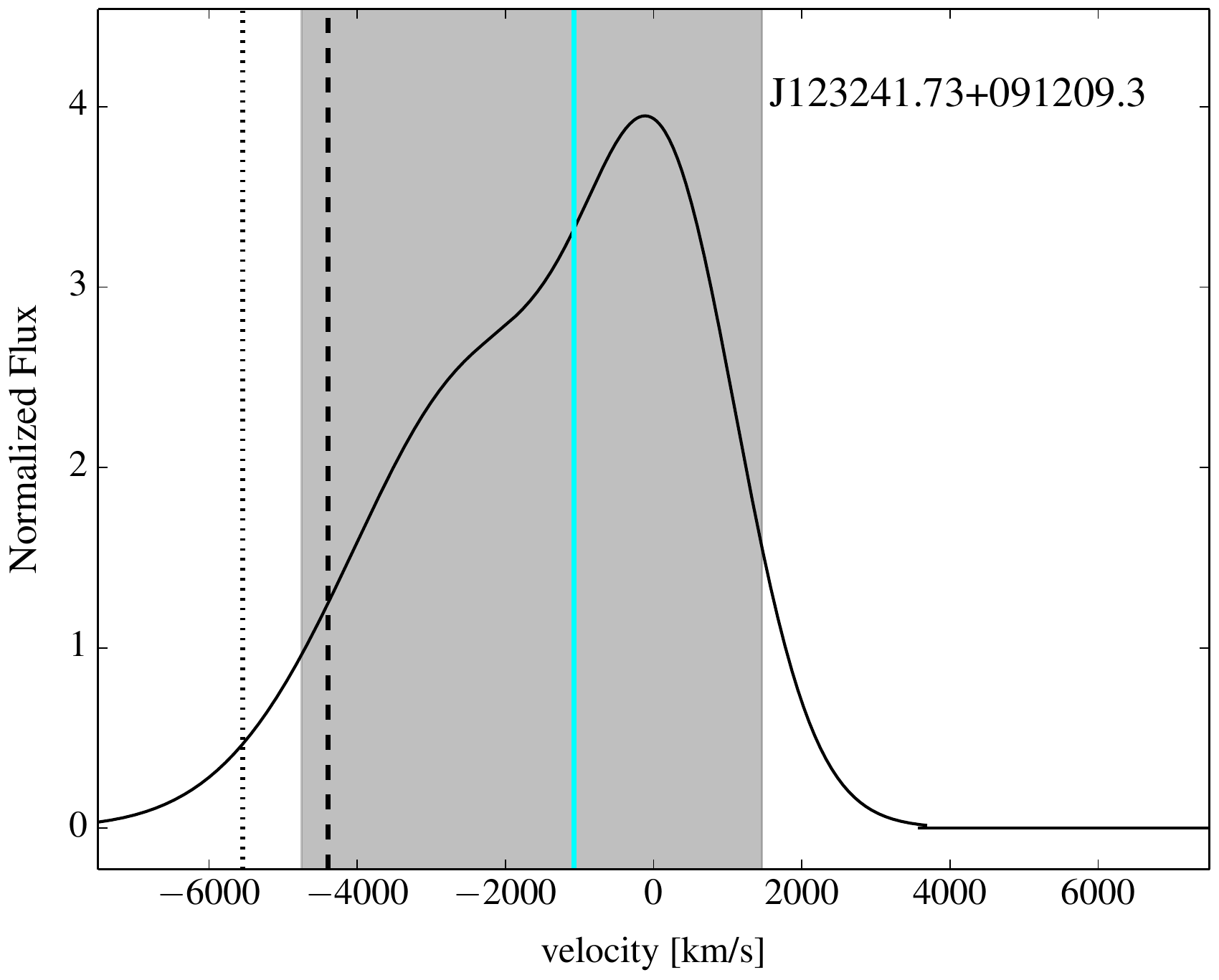}
 \includegraphics[width=0.8\columnwidth]{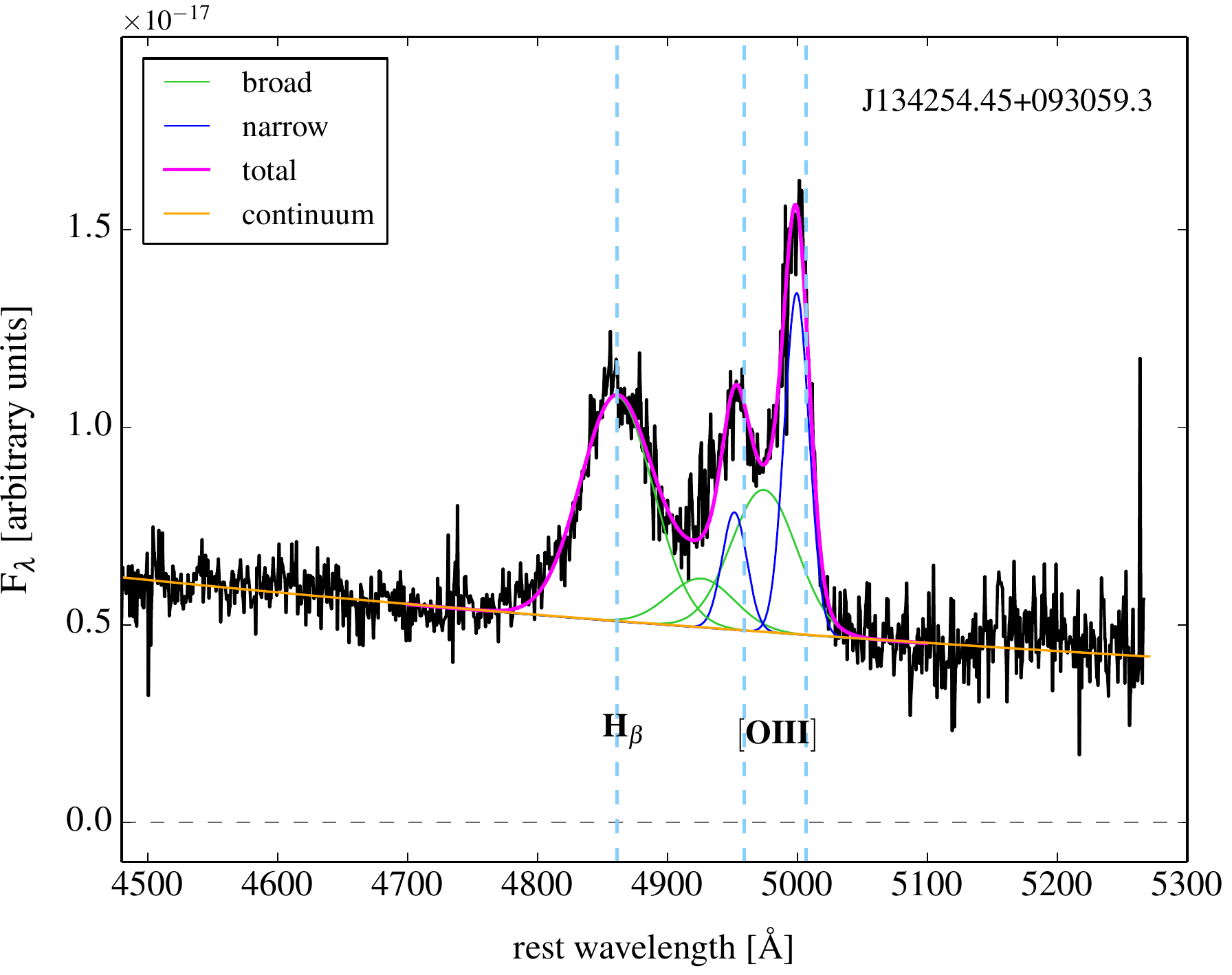}
 \includegraphics[width=0.8\columnwidth]{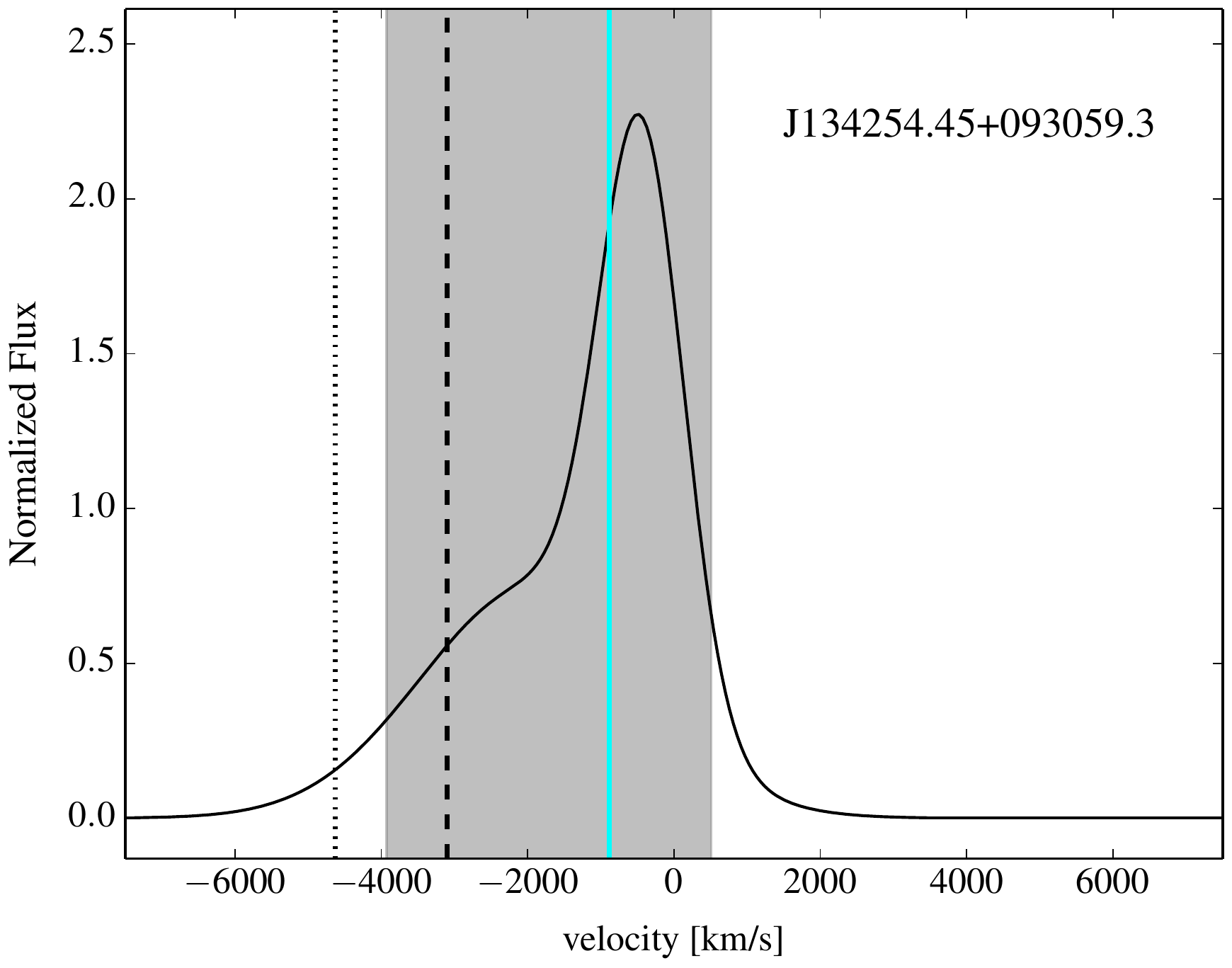}
 \caption{-- \bf continue}
 \label{}
\end{figure*}

\begin{figure*}
\ContinuedFloat

 \includegraphics[width=0.8\columnwidth]{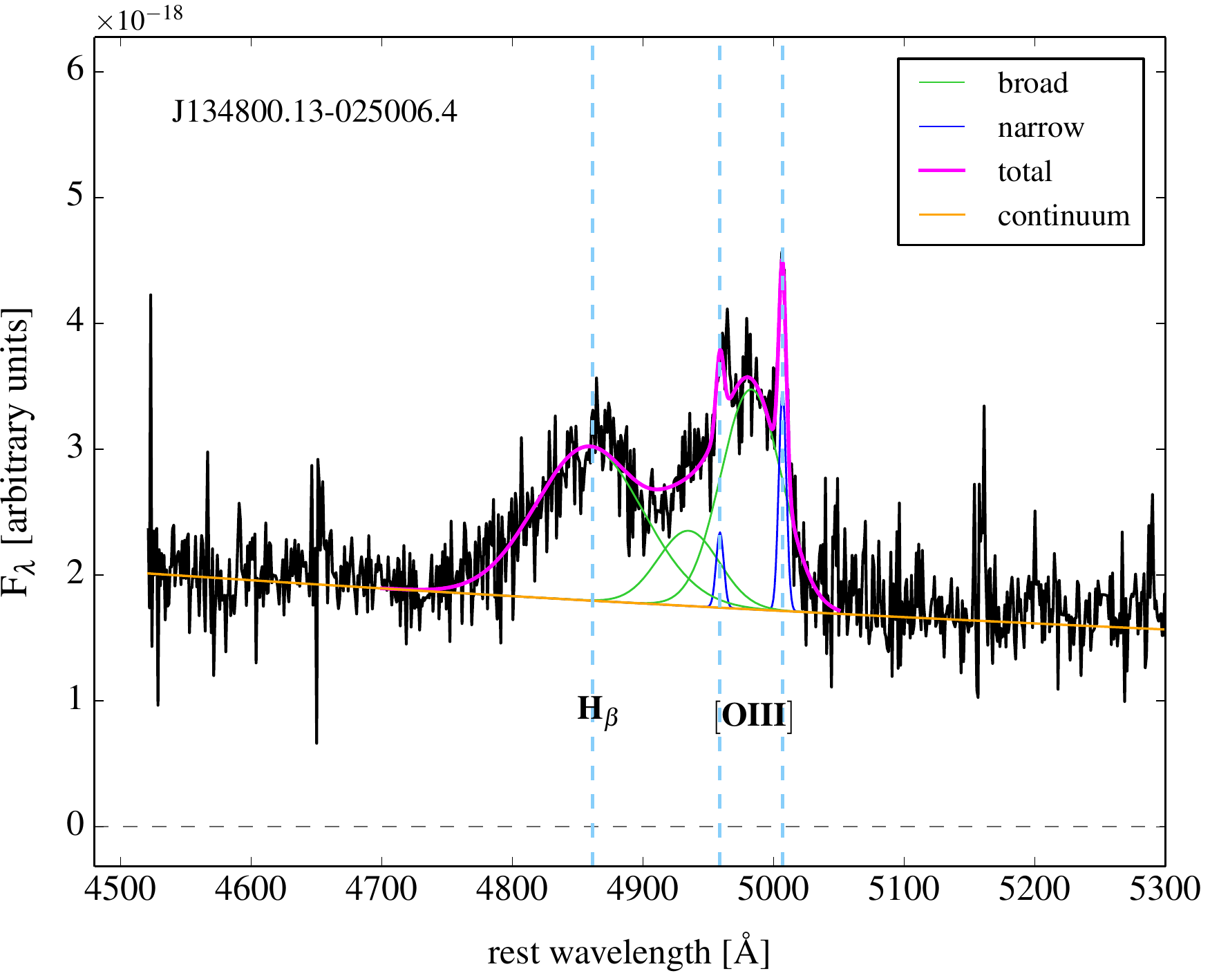}
 \includegraphics[width=0.8\columnwidth]{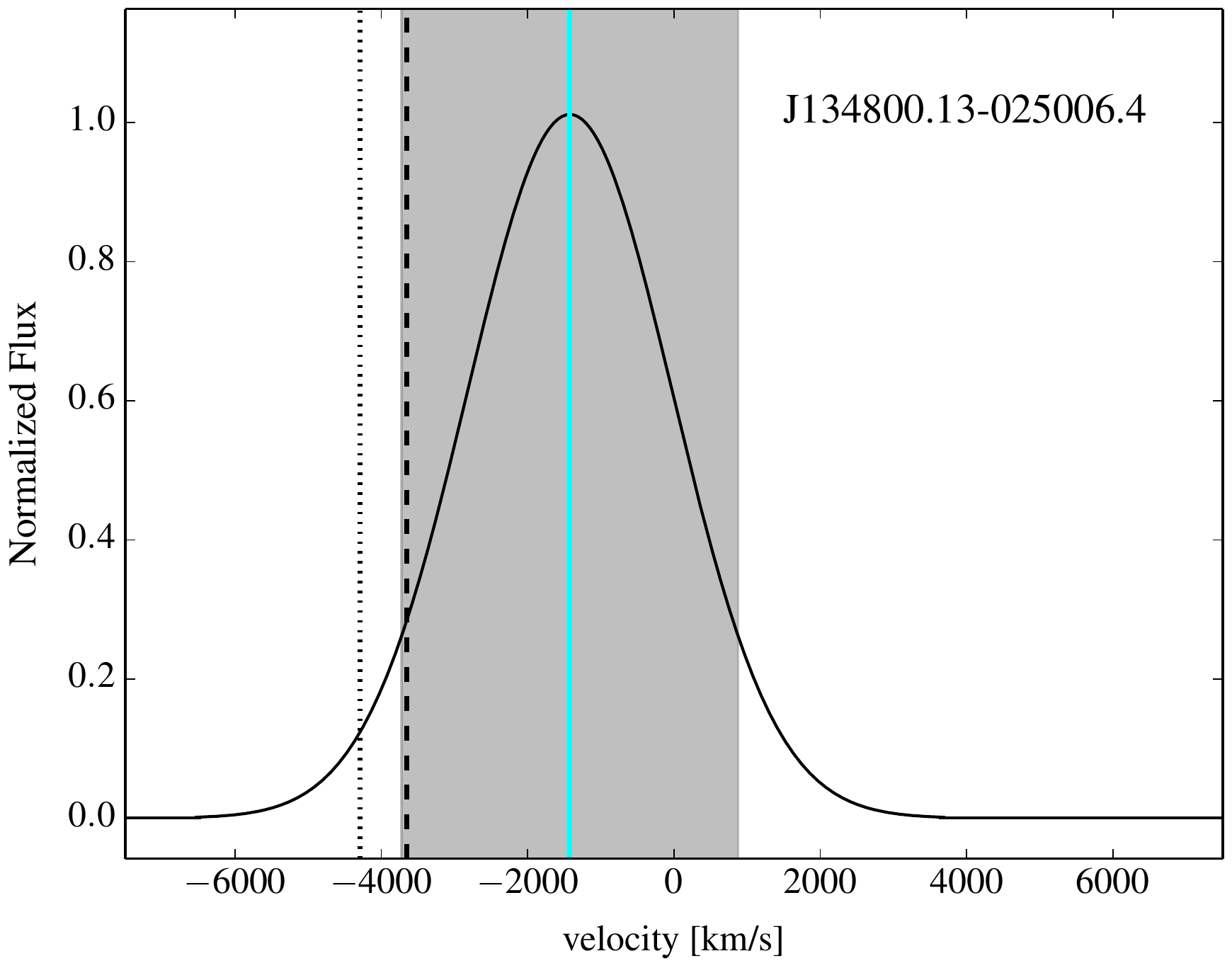}
 \includegraphics[width=0.8\columnwidth]{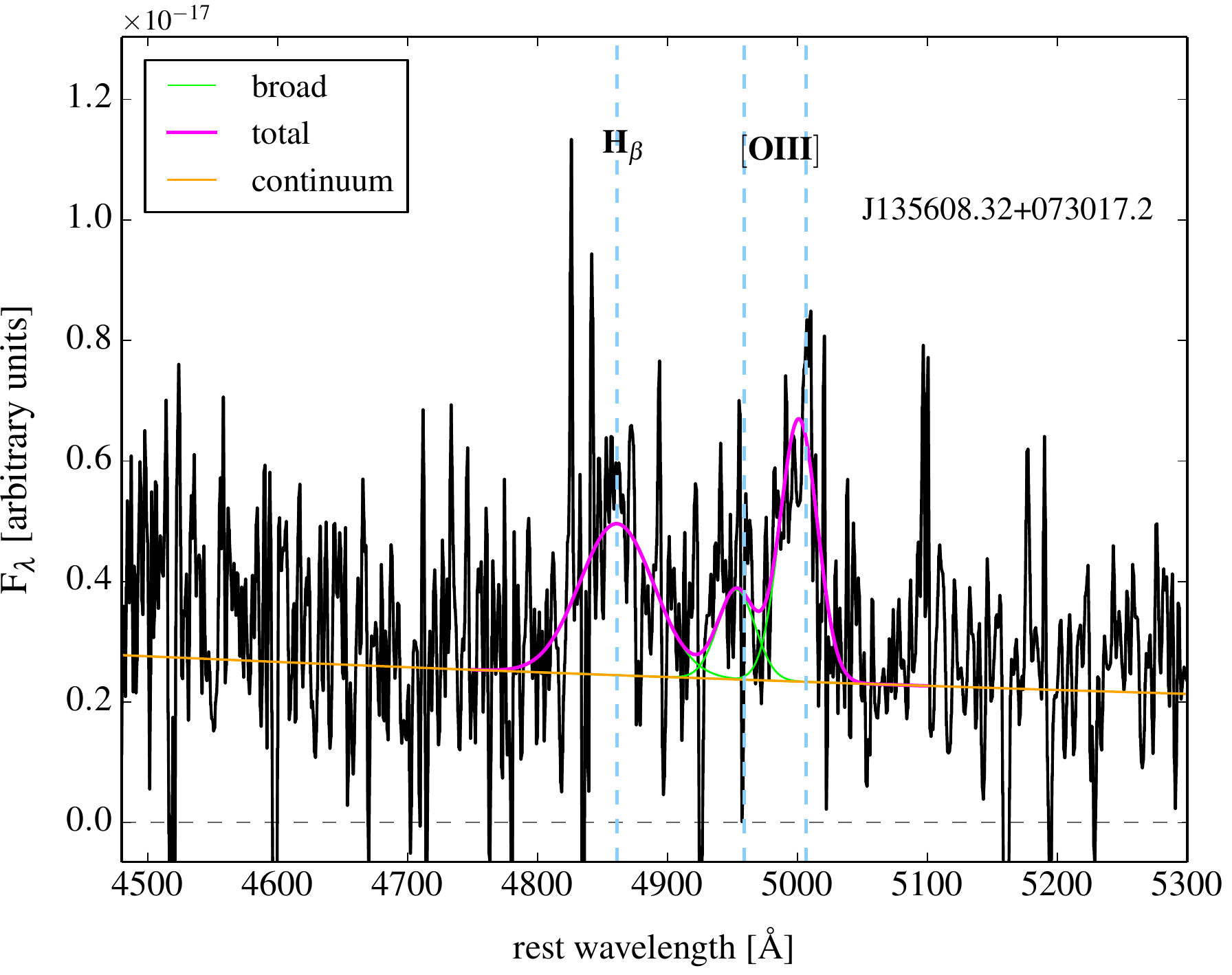}
 \includegraphics[width=0.8\columnwidth]{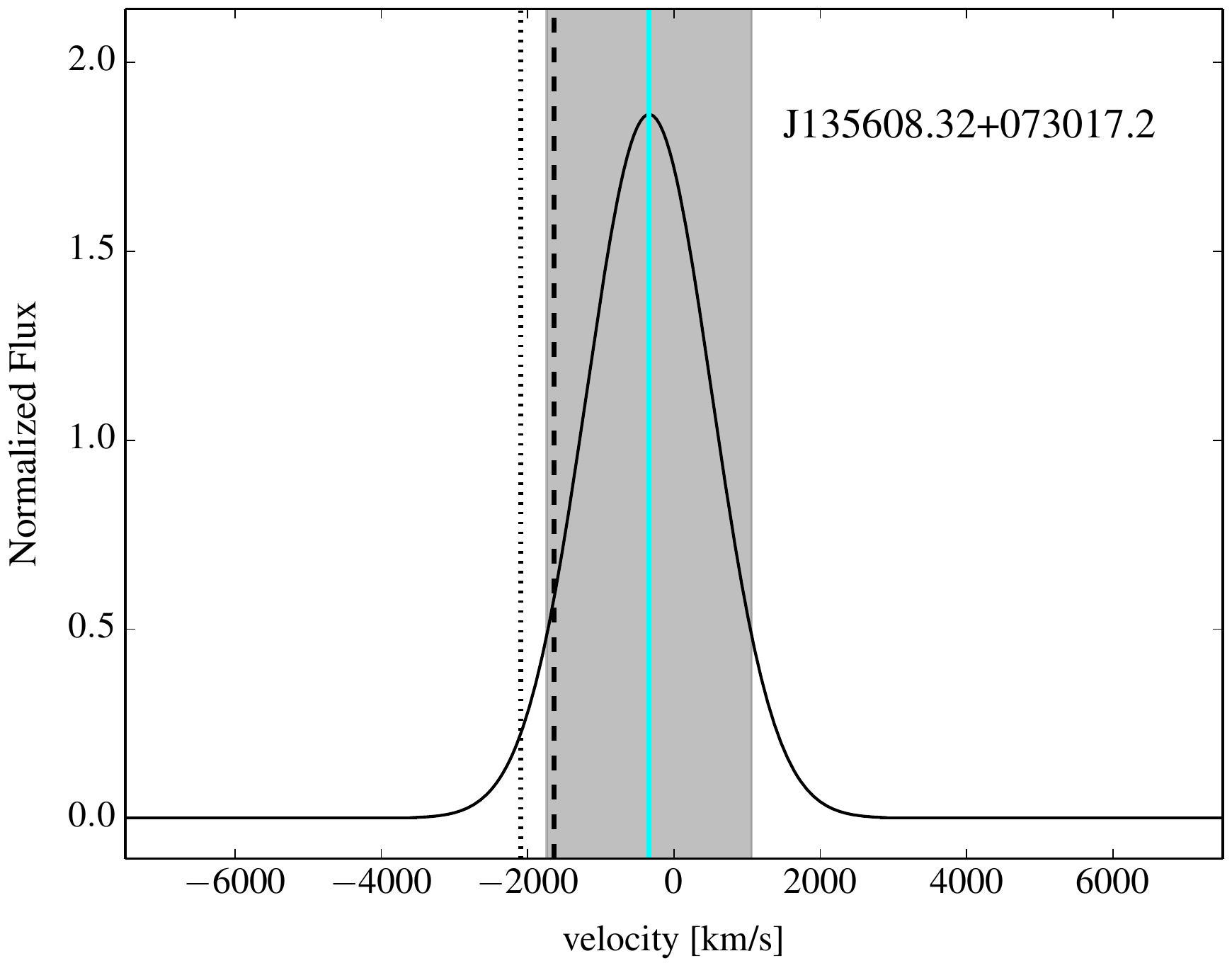}
 \includegraphics[width=0.8\columnwidth]{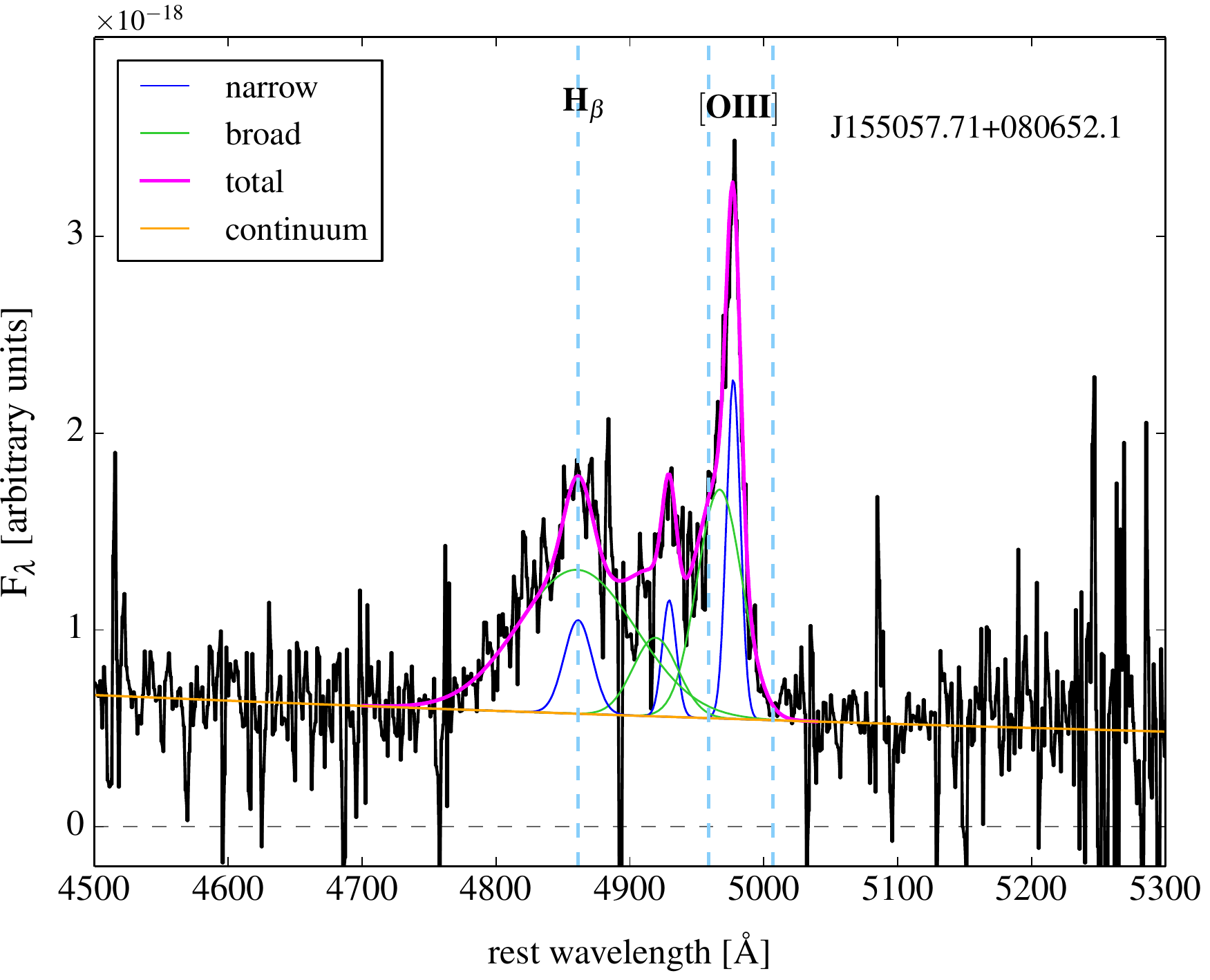}
 \includegraphics[width=0.8\columnwidth]{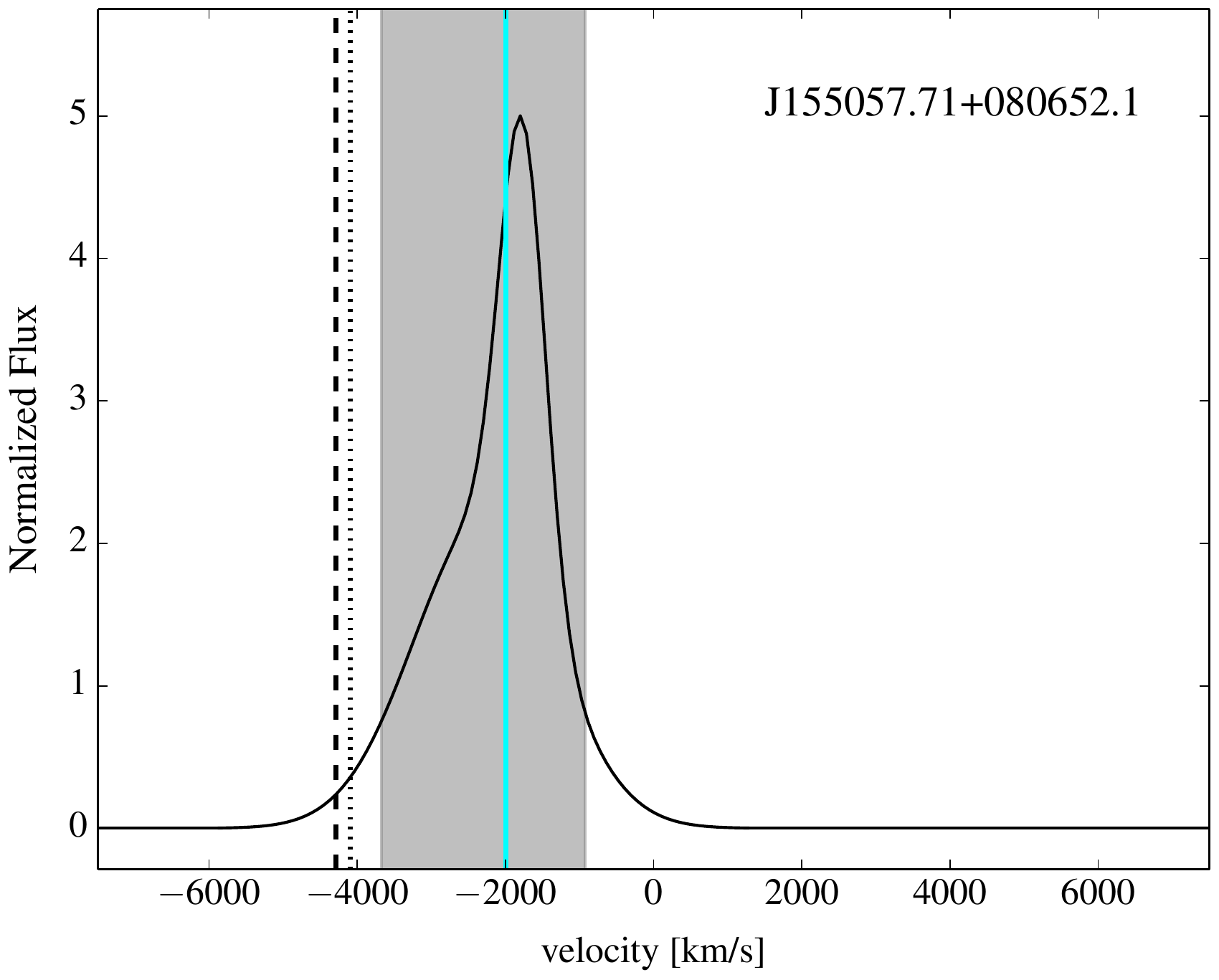}
 \includegraphics[width=0.8\columnwidth]{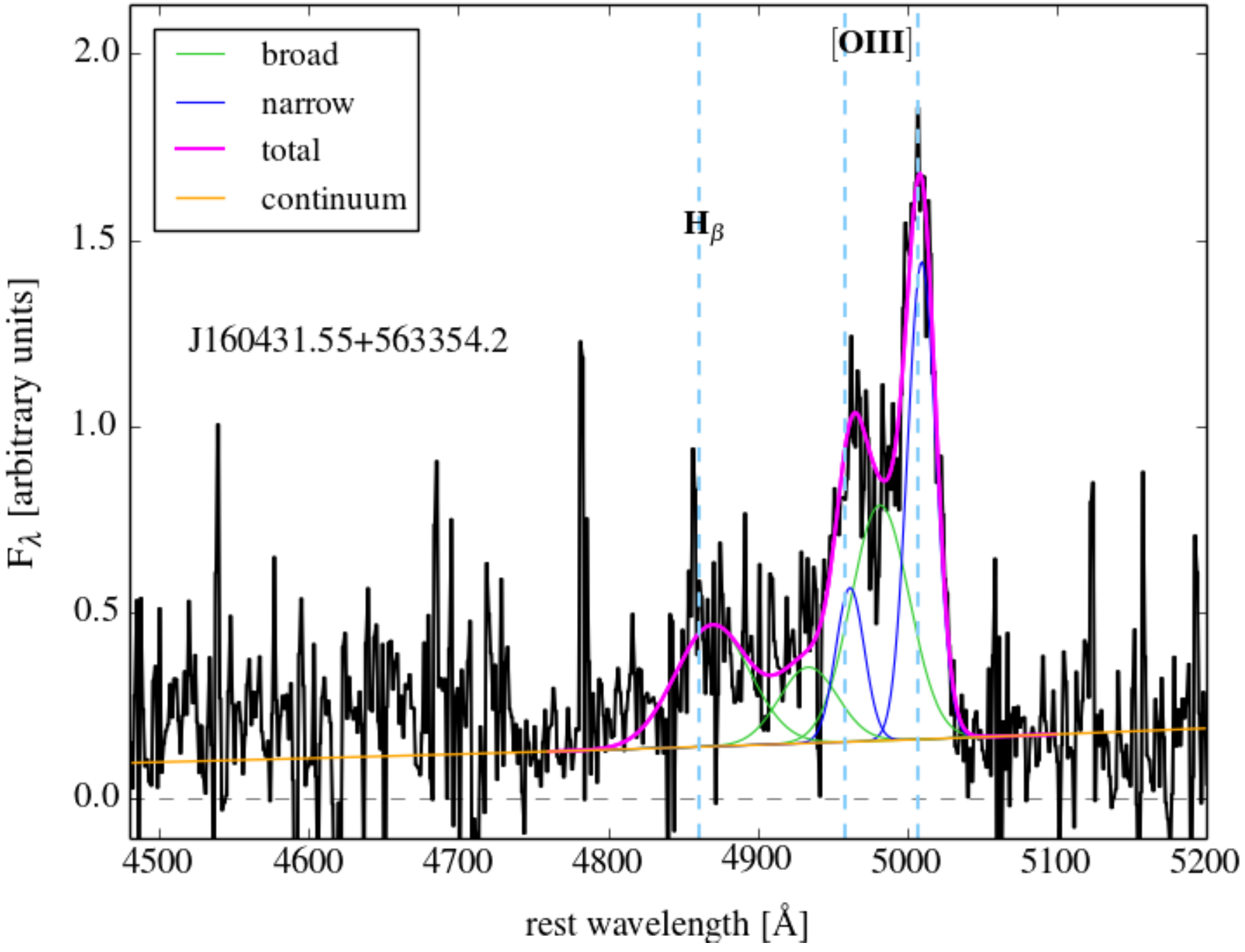}
 \includegraphics[width=0.8\columnwidth]{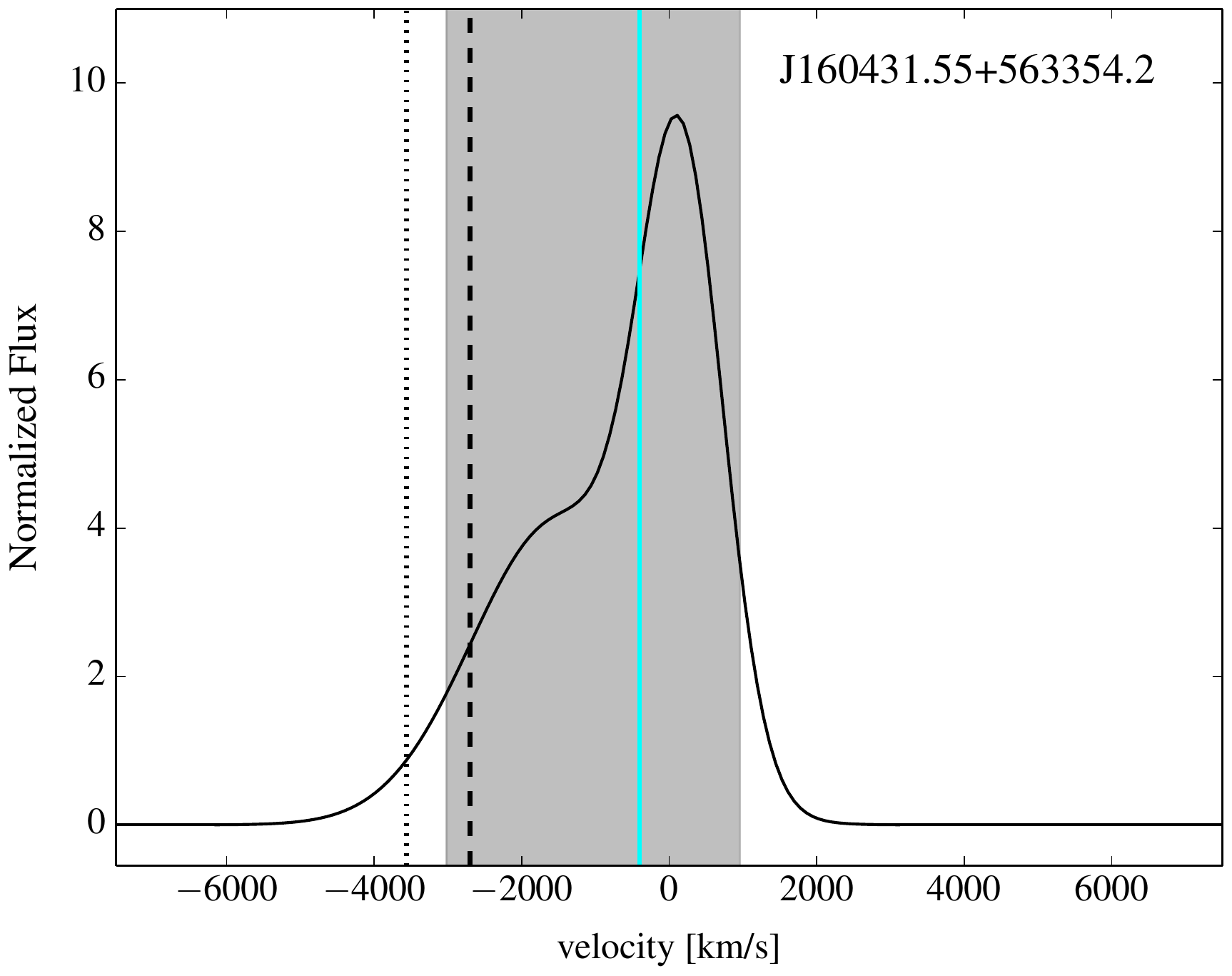}
 \caption{-- \bf continue}
 \label{}
\end{figure*}

\begin{figure*}
\ContinuedFloat

 \includegraphics[width=0.8\columnwidth]{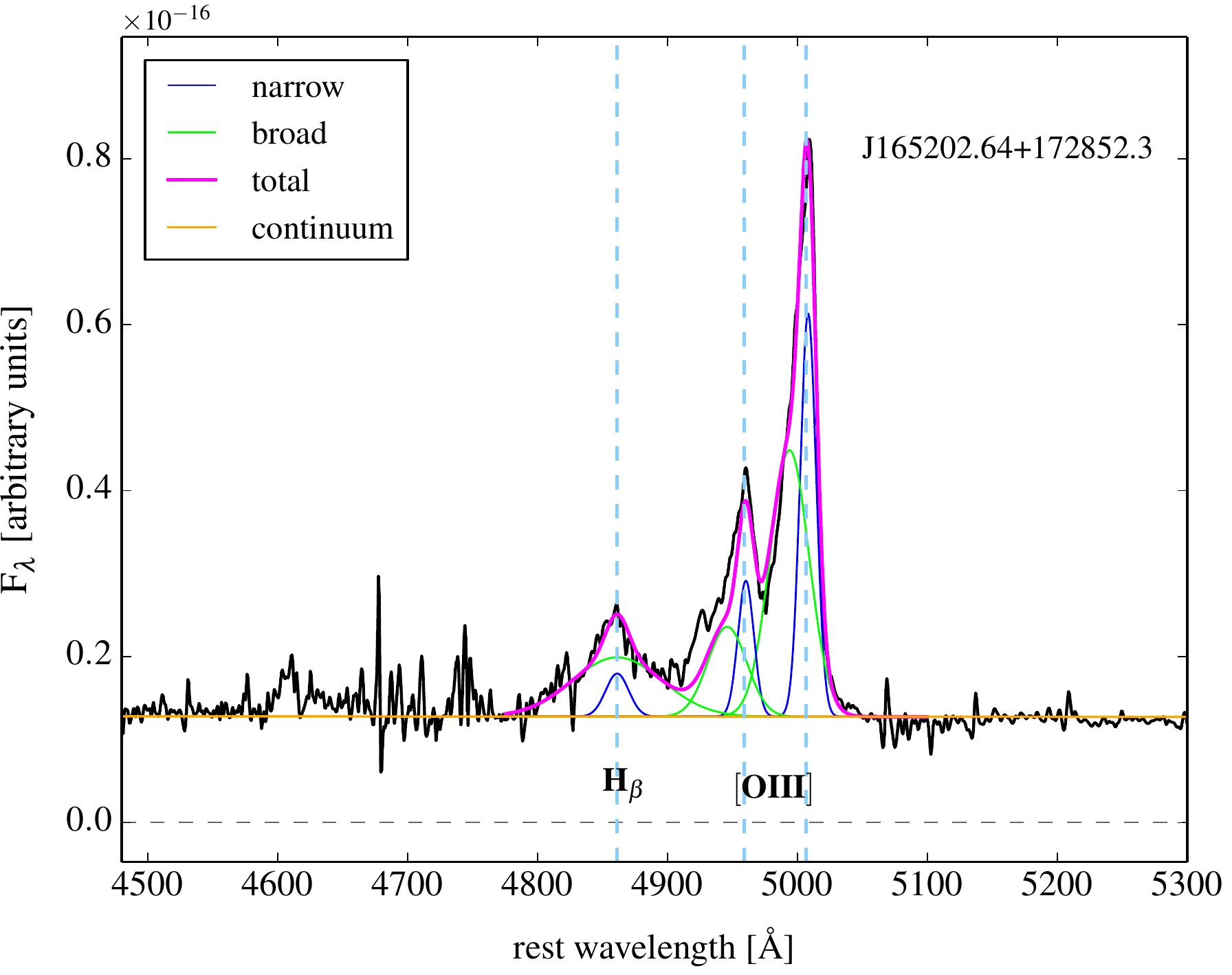}
 \includegraphics[width=0.8\columnwidth]{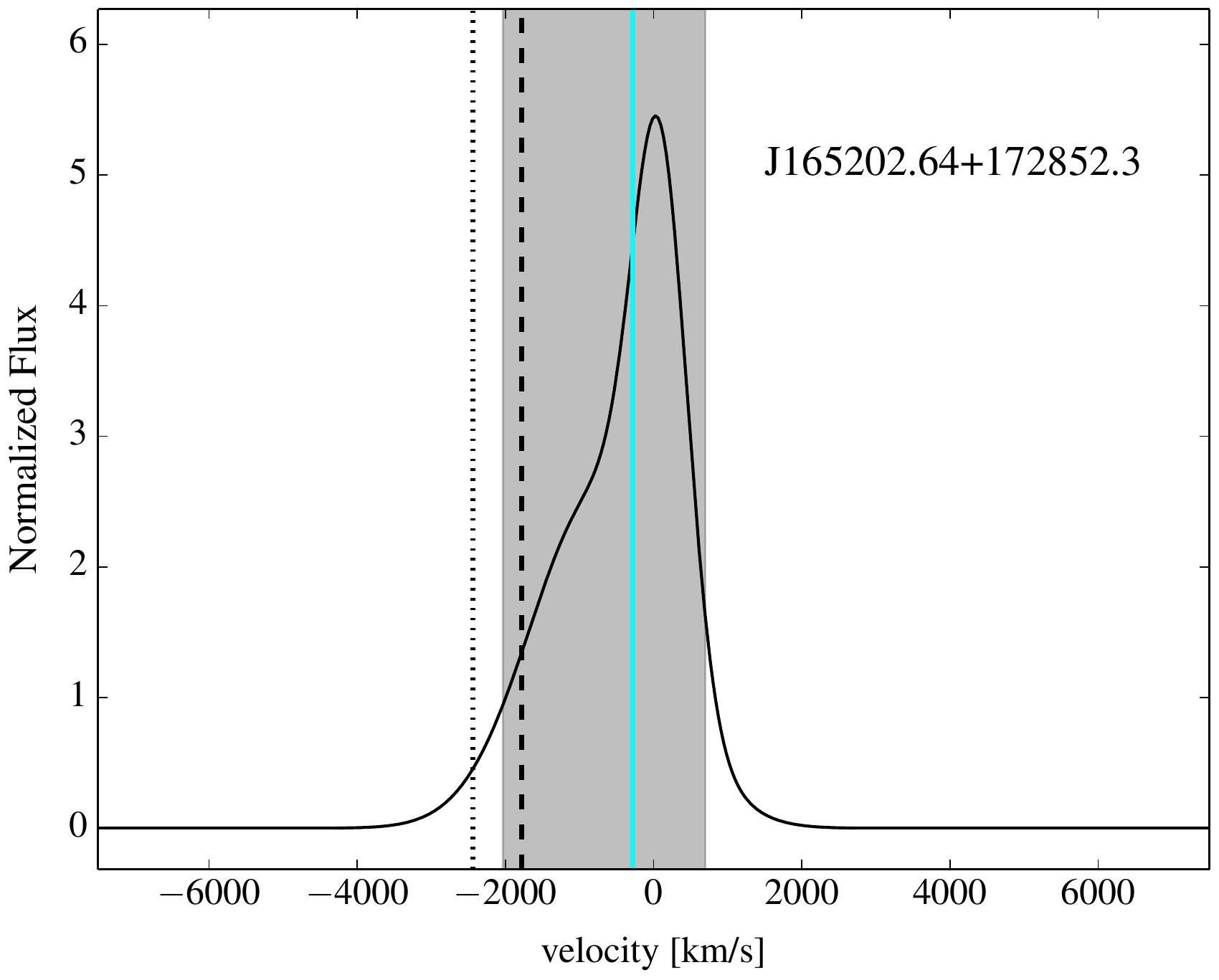}
 \includegraphics[width=0.8\columnwidth]{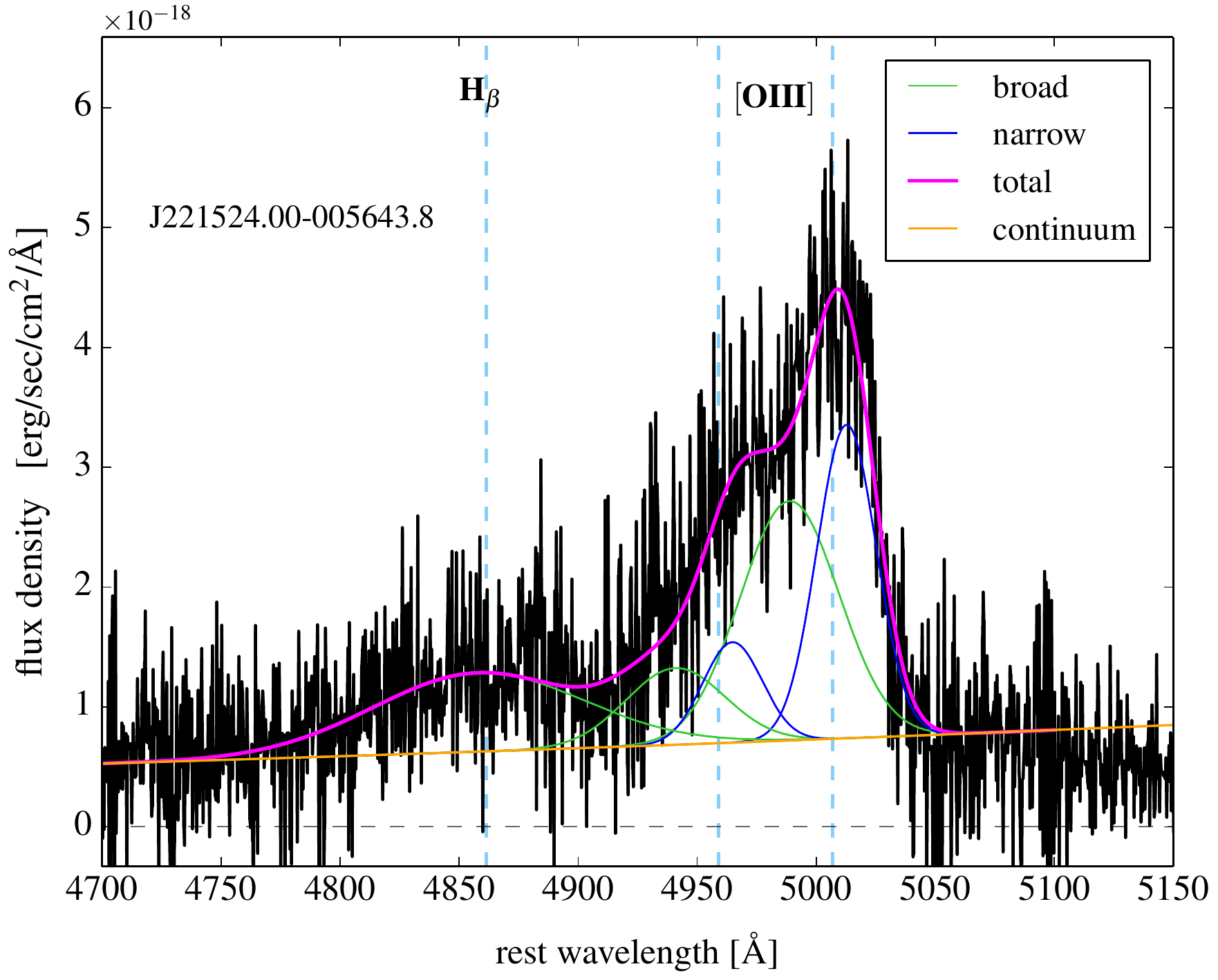}
 \includegraphics[width=0.8\columnwidth]{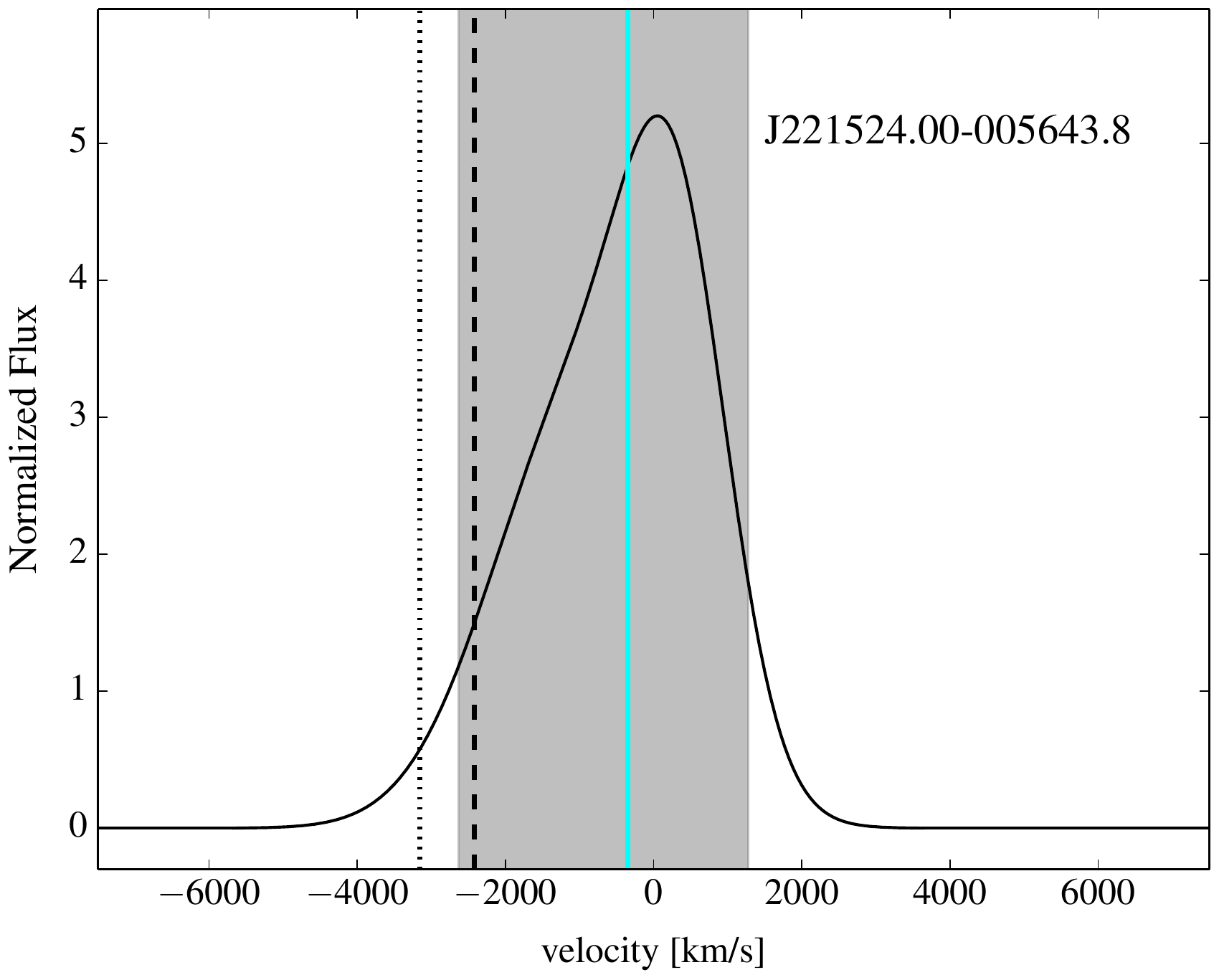}
  \includegraphics[width=0.8\columnwidth]{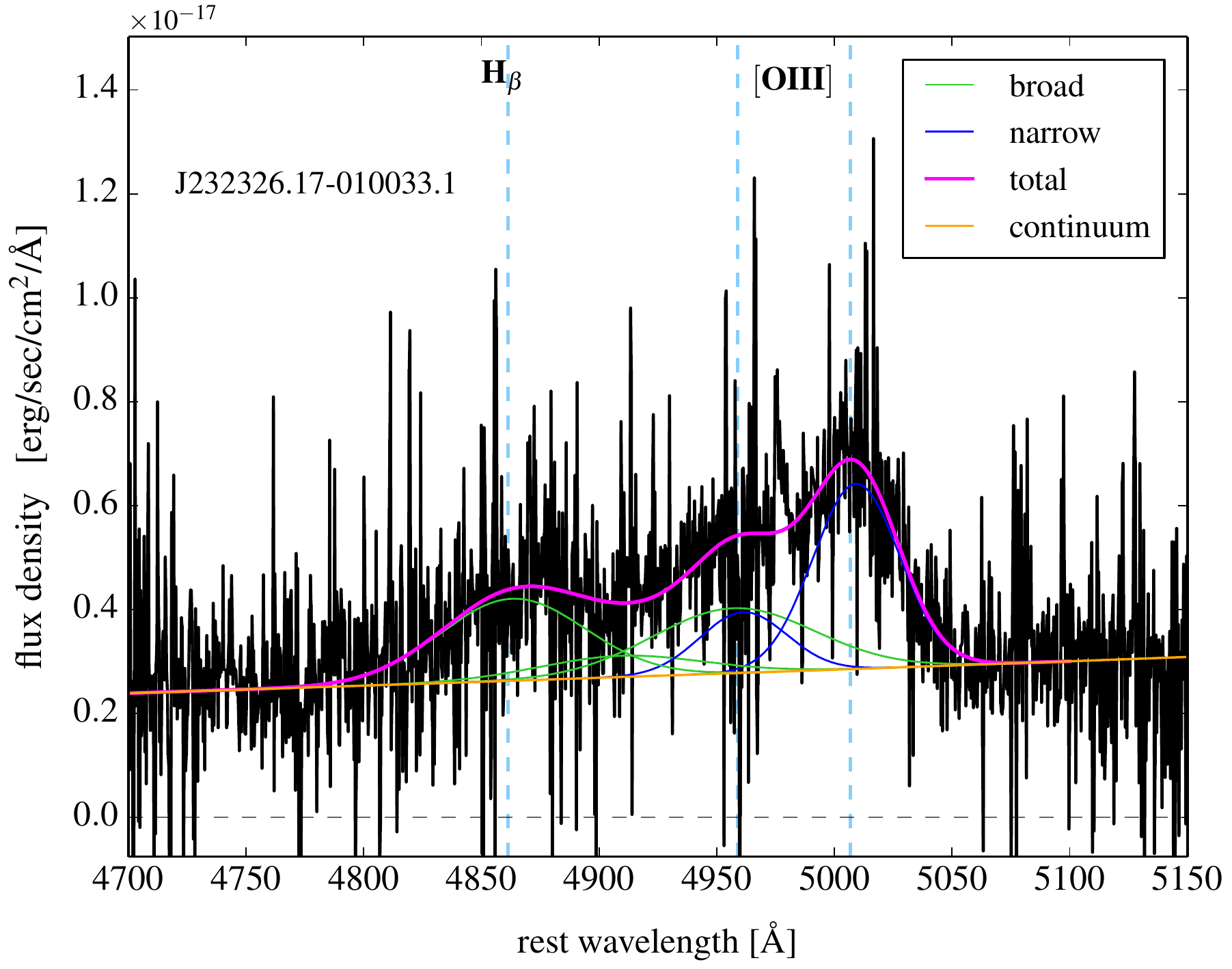}
 \includegraphics[width=0.8\columnwidth]{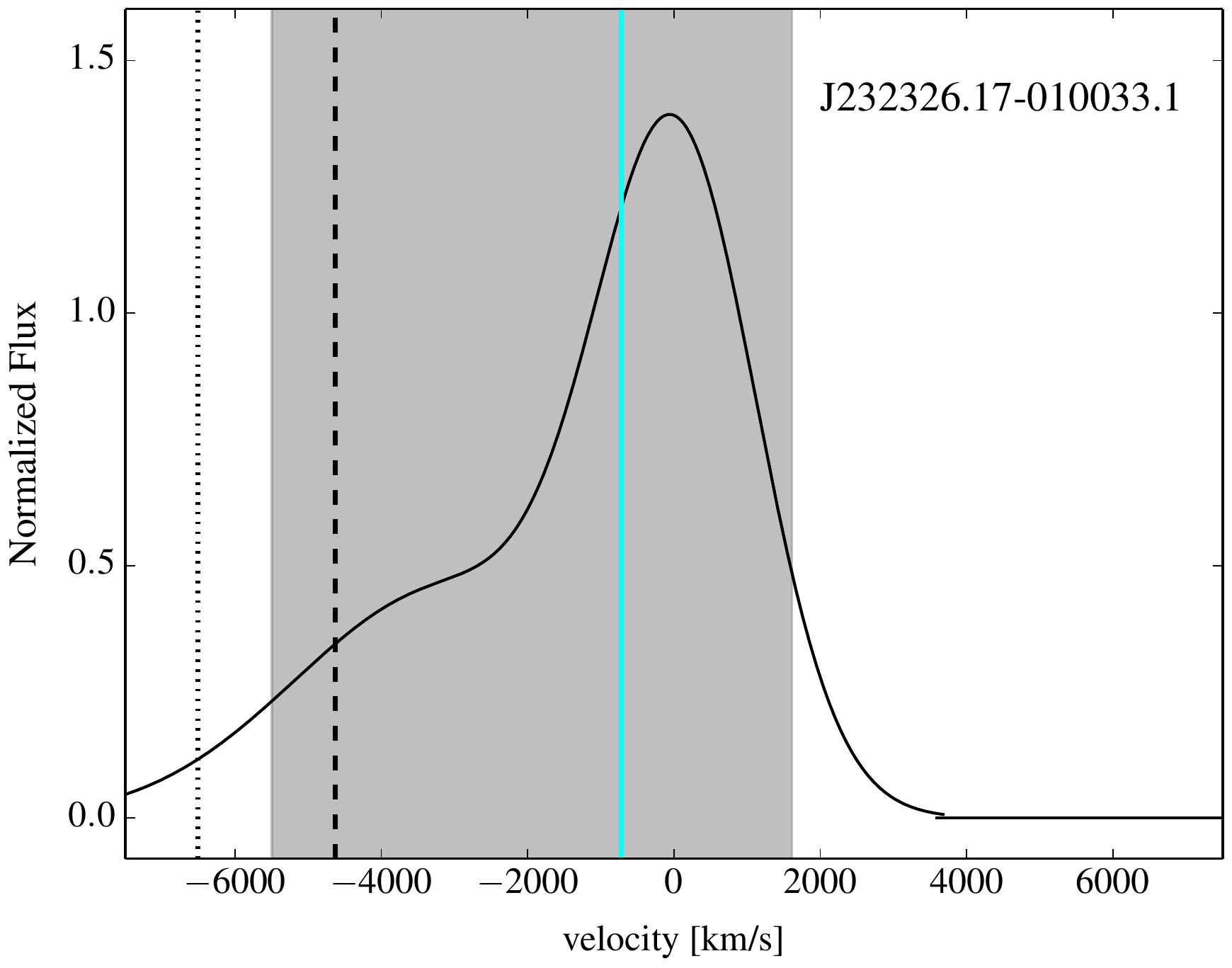}
 \includegraphics[width=0.8\columnwidth]{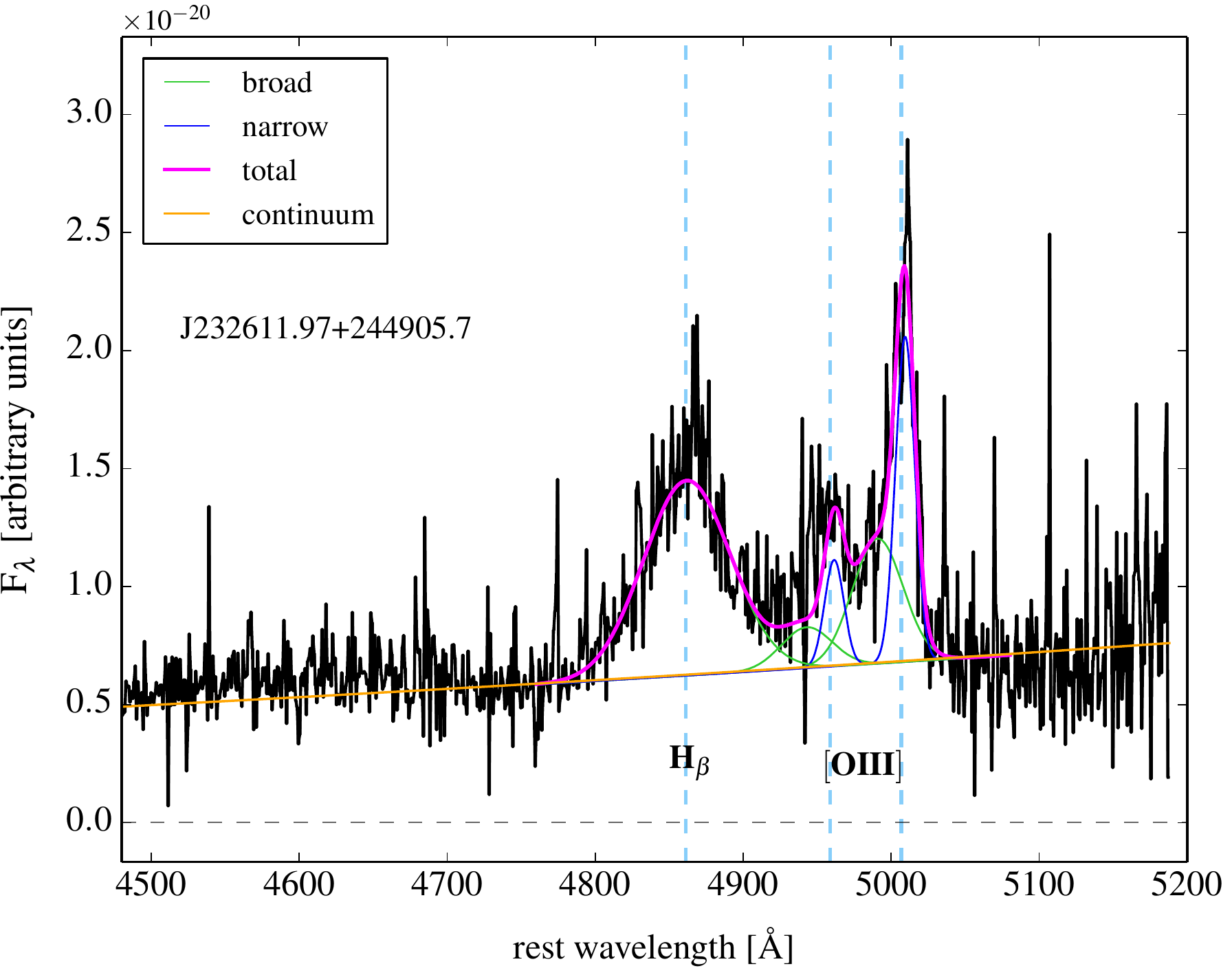}
 \includegraphics[width=0.8\columnwidth]{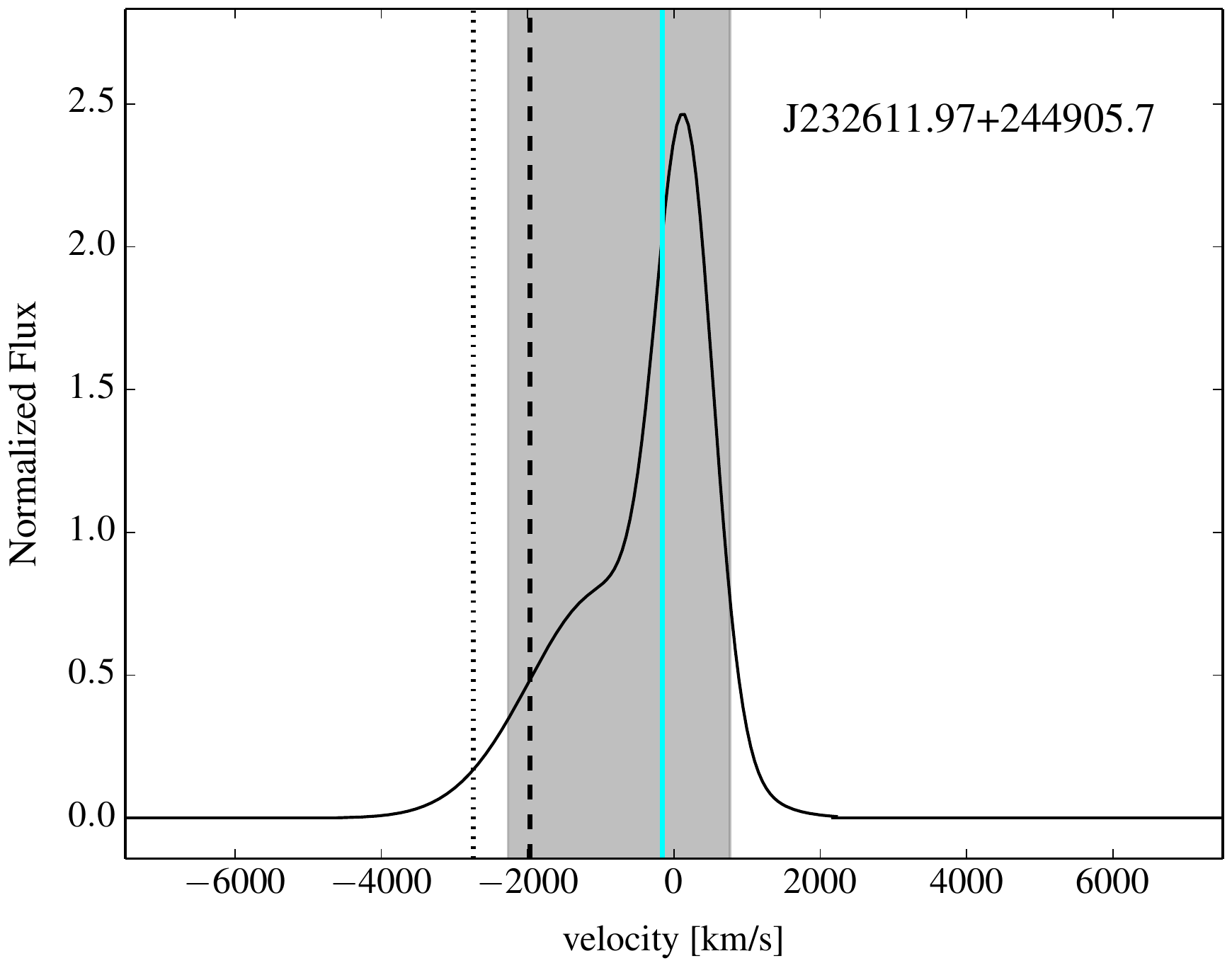}

 \caption{-- \bf continue}
 \label{}
\end{figure*}


\bsp	
\label{lastpage}
\end{document}